\pdfoutput=1
\documentclass[11pt]{article}

\usepackage[final]{acl}
\usepackage{times}
\usepackage{latexsym}

\usepackage[T1]{fontenc}
\usepackage[utf8]{inputenc}

\usepackage{microtype}

\usepackage{inconsolata}

\usepackage{graphicx}

\usepackage{hyperref}
\usepackage{booktabs}
\usepackage{graphicx}
\usepackage{multirow}
\usepackage[figuresright]{rotating}
\usepackage{colortbl}
\usepackage{siunitx}
\usepackage{subcaption}
\usepackage{array}
\usepackage{ragged2e}
\usepackage{xcolor}

\usepackage{xspace}
\newcommand{\airbench}{\textsc{AIR-Bench}\xspace}

\newcommand{\repolink}{\url{https://github.com/AIR-Bench/AIR-Bench}}
\newcommand{\leaderboardlink}{\url{https://huggingface.co/spaces/AIR-Bench/leaderboard}}
\newcommand{\hforglink}{\url{https://huggingface.co/AIR-Bench}}

\title{\airbench: Automated Heterogeneous Information Retrieval Benchmark}

\author{
 \textbf{Jianlyu Chen\textsuperscript{1,2,6}}~~
 \textbf{Nan Wang\textsuperscript{3}}~~
 \textbf{Chaofan Li\textsuperscript{2,4}}~~
 \textbf{Bo Wang\textsuperscript{3}}~~
\\
 \textbf{Shitao Xiao\textsuperscript{2}}~~
 \textbf{Han Xiao\textsuperscript{3}}~~
 \textbf{Hao Liao\textsuperscript{5}$^{*}$}~~
 \textbf{Defu Lian\textsuperscript{1,6}$^{*}$}~~
 \textbf{Zheng Liu\textsuperscript{2,7}\thanks{Corresponding authors}}
\\
 \textsuperscript{1}University of Science and Technology of China
\\
 \textsuperscript{2}Beijing Academy of Artificial Intelligence \ \ \
 \textsuperscript{3}Jina AI
\\
 \textsuperscript{4}Beijing University of Posts and Telecommunications \ \ \
 \textsuperscript{5}Shenzhen University
\\
 \textsuperscript{6}State Key Laboratory of Cognitive Intelligence
\ \ \textsuperscript{7}Hong Kong Polytechnic University\\
{\tt chenjianlv@mail.ustc.edu.cn} \ \ 
{\tt research@jina.ai}, \ \  {\tt haoliao@szu.edu.cn} \\
{\tt liandefu@ustc.edu.cn} \ \ \
{\tt zhengliu1026@gmail.com}
\\
}

\begin{document}
\maketitle
\begin{abstract}

Evaluation plays a crucial role in the advancement of information retrieval (IR) models.
However, current benchmarks, which are based on predefined domains and human-labeled data, face limitations in addressing evaluation needs for emerging domains both cost-effectively and efficiently.
To address this challenge, we propose the \textbf{A}utomated Heterogeneous \textbf{I}nformation \textbf{R}etrieval \textbf{Bench}mark (\textbf{\airbench}).
\airbench is distinguished by three key features: 1) Automated. The testing data in \airbench is automatically generated by large language models (LLMs) without human intervention. 2) Heterogeneous. The testing data in \airbench is generated with respect to diverse tasks, domains and languages. 3) Dynamic. The domains and languages covered by \airbench are constantly augmented to provide an increasingly comprehensive evaluation benchmark for community developers. 
We develop a reliable and robust data generation pipeline to automatically create diverse and high-quality evaluation datasets based on real-world corpora. Our findings demonstrate that the generated testing data in \airbench aligns well with human-labeled testing data, making \airbench a dependable benchmark for evaluating IR models.
The resources in \airbench are publicly available at \href{https://github.com/AIR-Bench/AIR-Bench}{\texttt{https://github.com/AIR-Bench/AIR-Bench}}.

\end{abstract}

\section{Introduction}

% Background
As information retrieval (IR) models grow in complexity and capability, the need for sophisticated evaluation techniques becomes increasingly critical.
In recent years, a series of milestone works have significantly advanced the field by introducing comprehensive evaluation datasets and benchmarks.
Early contributions to IR evaluation include MS MARCO~\cite{bajaj2016ms} and Natural Questions~\cite{kwiatkowski2019natural}, both designed for open-domain question answering (QA) tasks in English.
These datasets have been crucial in driving progress in monolingual IR systems and establishing baseline performance metrics.
Recognizing the importance of multilingual information retrieval, researchers developed Mr.TyDi~\cite{zhang-etal-2021-mr} and MIRACL~\cite{zhang2023miracl}.
These datasets cover ad hoc retrieval tasks in 11 and 18 languages, respectively, facilitating the development and evaluation of IR systems capable of handling diverse linguistic contexts.
More recently, the focus has shifted towards creating general-domain, zero-shot IR benchmarks. BEIR~\cite{thakur2021beir} and MTEB~\cite{muennighoff-etal-2023-mteb} represent this trend by aggregating multiple existing datasets from diverse tasks and domains.
These comprehensive benchmarks allow researchers to evaluate the generalization capabilities of IR models across various scenarios without task-specific fine-tuning.

% Issues
Despite their contributions, existing benchmarks are constrained to pre-defined domains and rely heavily on human-labeled data, making it challenging to efficiently address evaluation needs in emerging domains. With the emergence of powerful large language models (LLMs), several studies have explored their application for retrieval evaluation in retrieval-augmented generation (RAG) systems~\cite{es-etal-2024-ragas,saad-falcon-etal-2024-ares,salemi2024evaluating}, presenting a promising solution to this challenge. However, a comprehensive IR benchmark that addresses this limitation remains insufficiently developed.

% Moreover, the widespread use of in-domain training data during fine-tuning compromises the zero-shot evaluation intention of benchmarks like BEIR and MTEB, risking overfitting and poor real-world generalization. Additionally, the public availability of these datasets also raises concerns about inadvertent data leakage, where testing data might mistakenly incorporated into the training sets.

\begin{figure*}[!t]
\centering
\includegraphics[width=1.0\textwidth]{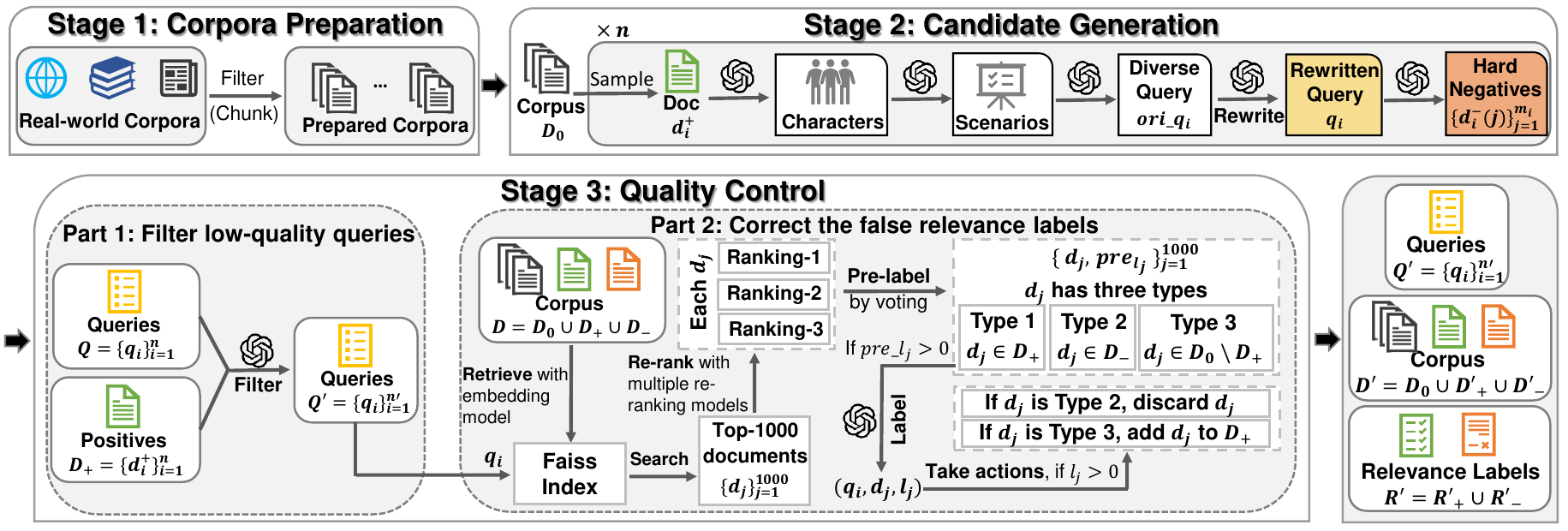}
\caption{The three-stage data generation pipeline of \airbench.}
\label{fig:generation_pipeline}
\vspace{-10pt}
\end{figure*}

% Features of AIR-Bench
In this work, we present the \textbf{A}utomated Heterogeneous \textbf{I}nformation \textbf{R}etrieval \textbf{Bench}mark (\textbf{\airbench}), which is characterized by three features:

\begin{enumerate}
    \item \textbf{Automated:} We develop a comprehensive data generation pipeline to automatically produce diverse and high-quality testing data with large language models (LLMs). Therefore, it is able to instantly support the evaluation of new domains both cost-effectively and efficiently. Besides, the new testing data is almost impossible to be covered by the training sets of any existing retrievers.

    \item \textbf{Heterogeneous:} \airbench is designed to be a heterogeneous IR benchmark including diverse tasks, domains and languages. It currently covers 2 tasks, 9 domains, and 13 languages, including a total of 69 datasets. This extensive coverage enables thorough evaluation across diverse scenarios, potentially accelerating advancements in IR technology for both established and emerging domains.
    
    \item \textbf{Dynamic:} The tasks, domains and languages covered by \airbench are planed to be augmented on regular basis. There are currently two distinct versions, 24.04 and 24.05, with more anticipated in the future. We hope \airbench is able to provide an increasingly comprehensive evaluation benchmark for community developers.
\end{enumerate}

% List all of Contributions
These features form the foundation of our proposed benchmark and directly address the limitation in existing benchmarks for information retrieval systems.
To further elucidate the impact and scope of our work, we summarize our main contributions as follows:
1) We introduce \airbench, a new information retrieval benchmark highlighted by new features: automated, heterogeneous and dynamic.
2) We demonstrate that our data generation pipeline is able to produce diverse and high-quality testing data highly consistent with human-labeled testing data, making \airbench a dependable benchmark for evaluating IR models.
3) Additionally, we develop and release software tools enabling community developers to evaluate any IR model using \airbench. To foster collaboration and progress in the field, we establish and maintain a public leaderboard\footnote{\leaderboardlink} to track and compare model performance across the community. These contributions collectively advance the field of information retrieval by providing a versatile, dynamic, and comprehensive evaluation framework.

\section{Benchmark Construction}

The entire data generation pipeline of \airbench consists of three stages: 1) Corpora preparation, 2) Candidate generation, and 3) Quality control.

\subsection{Preliminary}

\airbench focuses on the evaluation of information retrieval. The information retrieval task can be formulated as: Given a query $q$, retrieve a ranked list of $n$ most relevant documents $\mathcal{L} = [d_1, d_2, \cdots, d_n]$ from the corpus $\mathcal{D} = \{d_i\}_{i=1}^{\left| \mathcal{D} \right|}$.

To clarify the subsequent explanation, Table~\ref{tab:symbol_table} lists the symbols that appear in this section along with their corresponding meanings for reference.

\begin{table}[h]
    \centering
    \small
    \setlength{\tabcolsep}{1.5pt}
    \setlength{\extrarowheight}{2pt}
    \begin{tabular}{c|l||c|l}
        \toprule
        \textbf{Symbol} & \textbf{Meaning} & \textbf{Symbol} & \textbf{Meaning} \\
        \midrule
        \midrule
        $q$ & query & $\mathcal{Q}$ & queries set \\
        \hline
        $d$ & document & $d^+/d^-$ & \begin{tabular}[l]{@{}l@{}} positive/negative \\ document \end{tabular} \\
        \hline
        $l$ & relevance label & $n,m$ & number \\
        \hline
        $\mathcal{D}$ & documents set & $\mathcal{D}_{+}/\mathcal{D}_{-}$ & \begin{tabular}[l]{@{}l@{}} positive/negative \\ documents set \end{tabular} \\
        \hline
        $\mathcal{R}$ & relevance labels set & $\mathcal{R}_{+}/\mathcal{R}_{-}$ & \begin{tabular}[l]{@{}l@{}} positive/negative \\ relevance labels set \end{tabular} \\
        \hline
        $\mathcal{L}$ & documents list & $\mathcal{M}$ & re-ranking model \\
        \bottomrule
    \end{tabular}
    \caption{Corresponding meanings for the symbols appearing in this section.}
    \label{tab:symbol_table}
    \vspace{-10pt}
\end{table}

\subsection{Corpora Preparation}
\label{sec:corpus_preparation}

As shown in Figure~\ref{fig:generation_pipeline}, the first stage involves preparing diverse corpora. Specifically, given a task, we collect real-world datasets from diverse domains and languages, and apply distinct pre-processing strategies to the raw datasets based on the task requirements (see Appendix~\ref{appendix_sec:copora_preparation} for more details).

The corpus prepared in this stage is denoted as $\mathcal{D}_0 = \{d_i\}_{i=1}^{n_0}$, including $n_0$ documents.

\subsection{Candidate Generation}
\label{sec:candidate_generation}

The candidate data for a retrieval dataset consists of three components: corpus, queries and qrels. After preparing the corpus in the initial stage, the candidate generation stage produces the remaining two components of the dataset: queries and qrels. 

Based on the corpus, the candidate generation process is iteratively executed in a loop. As shown in Figure~\ref{fig:generation_pipeline}, the generation process can be summarized as the following steps: 1) Sample one document from the raw corpus as the positive document $d_i^{+}$. 2) Prompt LLM to generate the characters who might find the document useful. 3) Prompt LLM to generate the scenarios in which the character might find the document useful. 4) Prompt LLM to generate the query $ori\_q_i$ based on the specific character and scenario. To diversify the generated queries, we consider the following attributes when designing the prompt: query length, query type, information-based type, and expression style. 5) Prompt LLM to rewrite the generated query for multiple times to try to avoid the duplicated tokens as in the raw corpus, and finally get query $q_i$. 6) Prompt LLM to generate some hard negative documents $\{d_i^{-}(j)\}_{j=1}^{m_i}$ based on the generated query $q_i$ and the positive document $d_i^{+}$. 7) Repeat Step 1-6. Considering both simplicity and the absence of examples in a new domain, the above prompting strategies are all zero-shot. For more details, please refer to Appendix~\ref{appendix_sec:candidate_generation}.

After repeating $n$ times of the above loop, we get the queries set $\mathcal{Q}$, the positive documents set $\mathcal{D}_{+}$, the hard negative documents set $\mathcal{D}_{-}$, the corpus $\mathcal{D} = \mathcal{D}_0 \cup \mathcal{D}_{+} \cup \mathcal{D}_{-}$, the positive relevance labels set $\mathcal{R}_{+}$, and the negative relevance labels set $\mathcal{R}_{-}$.

\subsection{Quality Control}
\label{sec:quality_control}

In this stage, we design comprehensive quality control strategies to enhance the quality of the generated dataset. As shown in Figure~\ref{fig:generation_pipeline}, the quality control process can be summarized as two parts.

\textbf{Filter low-quality queries}. Since all of the queries in the candidate data are generated by LLM, there are potential low-quality queries. To improve the quality of generated queries, we utilize LLM to access the relevance between the query $q_i$ and the positive document $d_i^{+}$. If the LLM prediction is negative, indicating that $q_i$ is a low-quality query, we discard $q_i$ from $\mathcal{Q}$ and remove the relevance labels $\left\{(q_i, *, *)\right\}$ from $\mathcal{R}_+$ and $\mathcal{R}_-$. For details on how we utilize LLM to label the relevance, please refer to Appendix~\ref{appendix_sec:quality_control}.

\begin{table}[!t]
    \centering
    \resizebox{0.49\textwidth}{!}{
        \begin{tabular}{l|ll}
            \toprule
             & $l_j$ is pos & $l_j$ is neg \\
            \midrule
            Type 1: $d_j \in \mathcal{D}_+$ & - & * \\
            Type 2: $d_j \in \mathcal{D}_-$ & \begin{tabular}[l]{@{}l@{}} discard $d_j$ from $\mathcal{D}_{-}$, \\ remove $(q_i, d_j, 0)$ from $\mathcal{R}_{-}$ \end{tabular} & - \\
            Type 3: $d_j \in \mathcal{D}_0 \setminus \mathcal{D}_+ $ & \begin{tabular}[l]{@{}l@{}} add $d_j$ to $\mathcal{D}_{+}$, \\ add $(q_i, d_j, 1)$ to $\mathcal{R}_{+}$ \end{tabular} & - \\
            \bottomrule
        \end{tabular}
    }
    \caption{Specifications of different quality control strategies based on the type of document $d_j$ and the relevance label $l_j$ of $(q_i, d_j)$. Type 1 means that $d_j$ is the original positive document, Type 2 means that $d_j$ is the generated hard negative document, and Type 3 means that $(q_i, d_j)$ does not have a relevance label in the second stage. ``-'': Skip. ``*'': If the type of $d_j$ is Type 1, $l_j$ must be positive since we have filtered low-quality queries.}
    \label{tab:quality_control_class}
    \vspace{-10pt}
\end{table}

\begin{figure*}[!ht]
\centering
\includegraphics[width=1.0\textwidth]{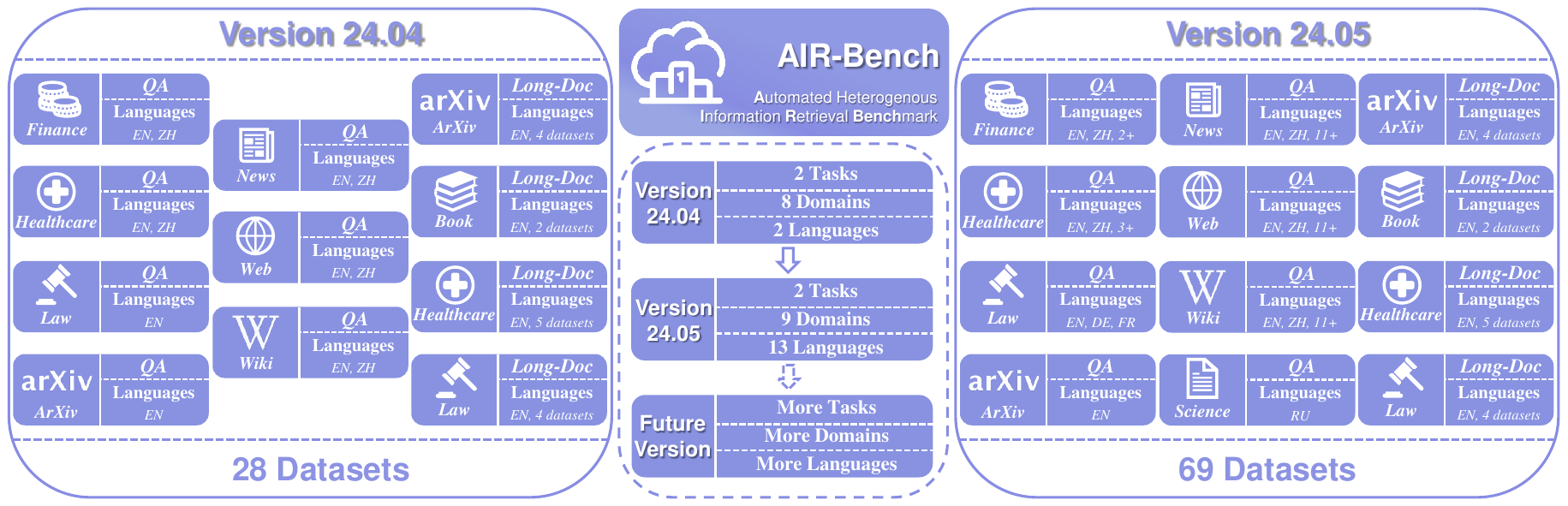}
\caption{An overview of the diverse tasks, domains, languages, and datasets in \airbench 24.04 and 24.05.}
\label{fig:air-bench_overview}
\vspace{-10pt}
\end{figure*}

\textbf{Correct the false relevance labels}. The false relevant labels comprise two types of documents: the first type includes the generated hard negative documents, and the second type consists of relevant documents that were overlooked in the corpus. Given a query $q_i$, we design a three-step pipeline to correct the false relevance labels. 1) \textit{Recall with embedding model}. Use the embedding model to search top-1000 relevant documents $\mathcal{L}_{recall} = \left[d_1, \cdots, d_{1000} \right]$ from the corpus for $q_i$. 2) \textit{Pre-label with re-ranking models}. Use multiple re-ranking models to re-rank $\mathcal{L}_{recall}$. We pre-label each document $d_j$ according to their ranking $r_j(\mathcal{M})$ in the re-ranked top-1000 relevant documents $\mathcal{L}_{rerank}(\mathcal{M})$ from the re-ranking model $\mathcal{M}$. Specifically, if $r_j(\mathcal{M})$ is higher than the predetermined threshold, the label $l_j(\mathcal{M})$ for $d_j$ from $\mathcal{M}$ is positive. If more than half of re-ranking models label $d_j$ as positive, we pre-label $d_j$ as positive, otherwise we pre-label $d_j$ as negative. After this step, each document $d_j$ in $\mathcal{L}_{recall}$ has a preliminary label $pre\_l_{j}$. 3) \textit{Label with LLM}. In this step, we also utilize LLM to access the relevance between $q_i$ and the documents $\{d_j\}_{j=1}^{m_i}$ that are pre-labeled as positive in the last step. The prediction from LLM is denoted as $l_{j}$. As shown in Table~\ref{tab:quality_control_class}, we categorize $d_j$ into three types, and take different actions by the type of $d_j$ and $l_j$. For details on how we select the embedding model and multiple re-ranking models, and set the predetermined threshold for pre-labeling, please refer to Appendix~\ref{appendix_sec:quality_control}.

After executing the above quality control process for each query, we get the new queries set $\mathcal{Q}'$, the new positive documents set $\mathcal{D}_{+}'$, the new hard negative documents set $\mathcal{D}_{-}'$, the new corpus $\mathcal{D}' = \mathcal{D}_0 \cup \mathcal{D}_{+}' \cup \mathcal{D}_{-}'$, and the new relevance labels set $\mathcal{R}' = \mathcal{R}_{+}' \cup \mathcal{R}_{-}'$, which form the final dataset.

\subsection{Design Motivations}

We elaborate the design motivations of the data generation pipeline of \airbench as follows.

\textbf{Reliance on real-world corpora}. Real-world corpora are usually diverse and available. Generating testing data based on real-world corpora not only closely aligns with real-world scenarios, but also significantly reduces the generation cost.

\textbf{Generation of characters and scenarios}. First, this step brings more transparency and interpretability on how a query is generated, compared to the naive method which directly prompts LLMs for query generation. Second, the generation of character and scenario also leads to higher diversity of queries, which contributes to the comprehensiveness of evaluation.

\textbf{Query Rewriting}. Through rewriting, queries are transformed into different forms while retaining equivalent semantics, which significantly increases the difficulty of retrieval tasks.

\textbf{Generation of hard negatives}. Similar to the introduction of query rewriting, this step increases the hardness of evaluation.

\textbf{Quality Control}. This step helps to remove low-quality queries and correct false relevance labels. Similar operations were also conducted in previous benchmark, e.g., the relevance assessment phase in MIRACL~\cite{zhang2023miracl}.

\section{The \airbench Benchmark}

\subsection{Overview}

\begin{table}[!t]
    \centering
    \small
    \setlength{\tabcolsep}{6pt}
    \begin{tabular}{l|rrrr}
        \toprule
         Task $\rightarrow$ & \multicolumn{2}{c}{\textbf{QA}} & \multicolumn{2}{c}{\textbf{Long-Doc}} \\
         \midrule
         Split $\rightarrow$ & \multicolumn{1}{c}{dev} & \multicolumn{1}{c}{test} & \multicolumn{1}{c}{dev} & \multicolumn{1}{c}{test} \\
         \# of datasets $\rightarrow$ & \multicolumn{1}{c}{54} & \multicolumn{1}{c}{53} & \multicolumn{1}{c}{4} & \multicolumn{1}{c}{11} \\
        \midrule
        \multicolumn{5}{l}{\textit{Query Type}} \\
        \midrule
        \textsc{how} & 16.4\% & 17.6\% & 17.0\% & 19.7\% \\
        \textsc{what} & 34.1\% & 30.9\% & 28.5\% & 33.1\% \\
        \textsc{when} & 4.8\% & 5.9\% & 1.1\% & 1.2\% \\
        \textsc{where} & 3.0\% & 3.2\% & 0.9\% & 0.8\% \\
        \textsc{which} & 4.7\% & 5.3\% & 4.4\% & 4.0\% \\
        \textsc{who} & 7.3\% & 7.6\% & 8.7\% & 4.0\% \\
        \textsc{why} & 3.2\% & 3.2\% & 6.4\% & 3.8\% \\
        \textsc{Yes/No} & 4.2\% & 4.1\% & 5.5\% & 6.9\% \\
        \textsc{claim} & 22.2\% & 22.1\% & 27.5\% & 26.3\% \\
        \textsc{others} & 0.1\% & 0.1\% & 0\% & 0.2\% \\
        \bottomrule
    \end{tabular}
    \caption{The type distribution of queries in each split for each task in \airbench 24.05.}
    \label{tab:query_type_diversity}
    \vspace{-10pt}
\end{table}

\textbf{LLM for Generation}. We use powerful GPT-4\footnote{gpt-4-1106-preview: \url{https://platform.openai.com/docs/models/gpt-4-turbo-and-gpt-4}}~\cite{achiam2023gpt} as the LLM through the generation pipeline. When prompting GPT-4, we set the sampling temperature to 1.0 to encourage more diversity.

\textbf{Tasks}. \airbench currently covers two retrieval tasks to meet the evaluation needs in different scenarios: 1) \textbf{QA}. This task focuses on the classic question answering scenarios~\cite{voorhees1999trec}, where the corpus consists of a large collection of documents. Following BEIR~\cite{thakur2021beir}, we utilize nDCG@10 as the main metric for the QA task. 2) \textbf{Long-Doc}. This task is closely related with today's LLM and RAG applications~\cite{lewis2020rag}, where the corpus consists of chunks from a lengthy document. Given that the proportion of positive documents precedes the ranking of positive documents in the RAG scenario, we utilize Recall@10 as the main metric for the Long-Doc task. \airbench will be extended to cover more retrieval tasks in the future.

\textbf{Datasets}. As shown in Figure~\ref{fig:air-bench_overview}, \airbench currently has two distinct versions, 24.04 and 24.05, where the latest version 24.05 consists of a total of 69 datasets, covering 9 domains\footnote{9 domains: News, Web, Wiki, Science, Finance, Healthcare, Law, ArXiv, Book.} and 13 languages\footnote{13 languages: English, Chinese, Spanish, French, German, Russian, Japanese, Korean, Arabic, Persian, Indonesian, Hindi, Bengali.} on two retrieval tasks. We hope to incorporate more domains and languages in the future version to provide an increasingly comprehensive evaluation benchmark for community developers. The specifications of all datasets in \airbench 24.05 are available in Table~\ref{tab:air-bench_datasets_2405_1}, Table~\ref{tab:air-bench_datasets_2405_2}, Table~\ref{tab:air-bench_datasets_2405_3}, Table~\ref{tab:air-bench_datasets_2405_4}, and Table~\ref{tab:air-bench_datasets_2405_5}. More details are available in Appendix~\ref{appendix_sec:air-bench_datasets}.

\textbf{Software}. We develop the \airbench software\footnote{\repolink} to facilitate the evaluation of any information retrieval methods. Besides, we maintain a Hugging Face leaderboard\footnote{\leaderboardlink} with all datasets and models. For more details, please refer to Appendix~\ref{appendix_sec:air-bench_software}.

\subsection{Diversity Analysis}

To analyze the query type diversity of \airbench, we utilize GPT-4o\footnote{
gpt-4o-2024-08-06}~\cite{achiam2023gpt}
as labeler to label the type of the generated queries. Specifically, given a query, we prompt GPT-4o to select the most suitable type for the query from the optional types. The statistics are grouped by tasks and splits in Table~\ref{tab:query_type_diversity}. Based on the results, we can make the following observations. Firstly, both the QA and Long-Doc tasks have the highest frequency of \textsc{what} queries, followed by \textsc{claim} queries as the second most common, and \textsc{how} queries as the third. Additionally, the QA task exhibits a more balanced distribution of the other query types, whereas the Long-Doc task shows a lower frequency of \textsc{when} queries and \textsc{where} queries. Lastly, a small number of queries are classified as \textsc{others}, reflecting the diverse types of queries present in \airbench to some extent. Further diversity analysis of \airbench is presented in Appendix~\ref{appendix_sec:diversity_analysis}.

\subsection{Positioning of \airbench}

We analyze the positioning of \airbench in this section to highlight extra values of \airbench over existing benchmarks. Firstly, as a diverse and continually evolving benchmark, \airbench enables comprehensive evaluation of existing retrievers while addressing the saturation issue that many popular benchmarks (e.g., MTEB~\cite{muennighoff-etal-2023-mteb} / C-MTEB~\cite{cpack}) face due to intensive in-domain fine-tuning. Furthermore, as an automated evaluation toolkit, \airbench supports ad-hoc evaluations for emerging domain-specific retrieval applications. We also provide experimental results in Section~\ref{sec:comparison_with_mteb}.

\begin{table}[!t]
    \centering
    \resizebox{0.49\textwidth}{!}{
        \begin{tabular}{l|ccc}
            \toprule
             & \textbf{\#corpus} & \textbf{\#queries} & \textbf{\#positives} \\
            \midrule
            R-MSMARCO    & 8,841,823 & 6,980 & 7,437  \\
            G-MSMARCO    & 8,872,840 & 6,319 & 31,447 \\
            \ \ \  w/o quality control & 8,878,865 & 7,429 & 7,429 \\
            \bottomrule
        \end{tabular}
    }
    \caption{Comparison of R-MSMARCO and G-MSMARCO. R-MSMARCO is the raw MS MARCO passage ranking dataset~\cite{bajaj2016ms}, and G-MSMARCO is the generated MS MARCO passage ranking dataset in \airbench. \textbf{\#corpus} represents the number of documents in the corpus, \textbf{\#queries} represents the number of queries, and \textbf{\#positives} represents the number of positive relevance labels. Since there are some generated hard negative documents in the corpus of G-MSMARCO, it is slightly larger than the corpus of R-MSMARCO.}
    \label{tab:msmarco_comparison}
    \vspace{-10pt}
\end{table}

\begin{table*}[!ht]
    \centering
    \small
    \setlength{\tabcolsep}{3pt}
    \setlength{\extrarowheight}{2pt}
    \begin{tabular}{l|c|cc|cc|cc}
        \toprule
        \multirow{3}{*}{\textbf{Model}} & \multirow{3}{*}{\textbf{Size}} & \multicolumn{2}{c|}{\textbf{R-MSMARCO}} & \multicolumn{4}{c}{\textbf{G-MSMARCO}} \\
        \cline{3-8}
         &  & \multirow{2}{*}{\textbf{nDCG@10}} & \multirow{2}{*}{\textbf{Rank}} & \multicolumn{2}{c|}{\textbf{w/ quality control}} & \multicolumn{2}{c}{\textbf{w/o quality control}} \\
         \cline{5-8}
         &  &  &  & \textbf{nDCG@10} & \textbf{Rank} & \textbf{nDCG@10} & \textbf{Rank} \\
        \midrule
        repllama-v1-7b-lora-passage~\cite{rankllama}                 & 6.74B & 48.000 & 1  & 59.625 & 1  & 33.434 & 2 \\
        e5-large-v2~\cite{wang2022text}                              & 335M  & 45.232 & 2  & 55.260 & 4  & 32.581 & 5 \\
        multilingual-e5-large~\cite{wang2024multilingual}            & 560M  & 45.119 & 3  & 54.431 & 5  & 32.099 & 6 \\
        multilingual-e5-base~\cite{wang2024multilingual}             & 278M  & 44.130 & 4  & 52.581 & 8  & 30.870 & 8 \\
        bge-large-en-v1.5~\cite{wang2024multilingual}                & 335M  & 44.122 & 5  & 55.513 & 3  & 33.119 & 4 \\
        e5-mistral-7b-instruct~\cite{wang2023improving}              & 7.11B & 43.787 & 6  & 59.015 & 2  & 36.186 & 1 \\
        e5-small-v2~\cite{wang2022text}                              & 33.4M & 43.104 & 7  & 51.456 & 10 & 30.471 & 10 \\
        e5-base-v2~\cite{wang2022text}                               & 109M  & 43.056 & 8  & 51.438 & 11 & 30.411 & 11 \\
        bge-small-en-v1.5~\cite{cpack}                               & 33.4M & 42.553 & 9  & 51.528 & 9  & 30.155 & 13 \\
        bge-base-en-v1.5~\cite{cpack}                                & 109M  & 42.388 & 10 & 54.292 & 7  & 32.067 & 7 \\
        multilingual-e5-small~\cite{wang2024multilingual}            & 118M  & 42.253 & 11 & 47.989 & 14 & 28.579 & 15 \\
        simlm-base-msmarco-finetuned~\cite{Wang2022SimLMPW}          & 110M  & 41.675 & 12 & 48.102 & 13 & 30.548 & 9 \\
        jina-embeddings-v3~\cite{sturua2024jina}                     & 572M  & 39.787 & 13 & 51.098 & 12 & 30.297 & 12  \\
        bge-m3~\cite{chen-etal-2024-m3}                              & 568M  & 39.565 & 14 & 54.404 & 6  & 33.286 & 3 \\
        contriever-msmarco~\cite{izacard2022unsupervised}            & 109M  & 36.570 & 15 & 47.127 & 15 & 29.231 & 14 \\
        msmarco-roberta-base-ance-firstp~\cite{xiong2021approximate} & 125M  & 33.637 & 16 & 42.107 & 16 & 24.798 & 16 \\
        BM25~\cite{robertson2009bm25}                                & -     & 26.211 & 17 & 34.155 & 17 & 22.582 & 17 \\
        \midrule
        \multicolumn{2}{l|}{\textbf{Spearman Rank Correlation Coefficient} (\textbf{P-value}) } & \multicolumn{2}{c|}{-} & \multicolumn{2}{c|}{0.8211 (5e-5)} & \multicolumn{2}{c}{0.6912 (2e-3)} \\
        \bottomrule
    \end{tabular}
    \caption{The consistency between the testing data generated by the pipeline of \airbench and the human-labeled testing data. We use the MS MARCO passage ranking dataset~\cite{bajaj2016ms} to evaluate the consistency. For the public link of the models appearing in the table, please refer to Table~\ref{tab:model_information}.
    }
    \label{tab:consistency_results}
    \vspace{-10pt}
\end{table*}

\section{Experiment}

In this section, we aim to address the following research questions:

\noindent \textbf{RQ1:} How well does the LLM-generated testing data in \airbench align with the human-labeled testing data? 

\noindent \textbf{RQ2:} What additional evaluation functionalities does \airbench offer compared to MTEB/BEIR?

\noindent \textbf{RQ3:} How effectively can \airbench  distinguish the capabilities of distinct IR models?

\subsection{Consistency Analysis (RQ1)}

\citet{llm_accessors} have demonstrated that LLMs like OpenAI's GPT-4 are as accurate as human labelers when generating high-quality golden labels for search system. Based on this conclusion, we attempt to examine how well the LLM-generated testing data aligns with human-labeled testing data.

\textbf{Setup}. We utilize MS MARCO passage ranking dataset~\cite{bajaj2016ms} to access the consistency between the LLM-generated testing data in \airbench and human-labeled testing data. Specifically, we use the positive passages in the raw MS MARCO dev split as the candidate positives ($d_i^+$ in Stage 2, refer to Section \ref{sec:candidate_generation}), and finally generate a new MS MARCO passage ranking dataset. The raw MS MARCO passage ranking dataset (dev split) is denoted as \textit{\underline{R-MSMARCO}}, and the new generated MS MARCO passage ranking dataset is denoted as \textit{\underline{G-MSMARCO}}. Table~\ref{tab:msmarco_comparison} shows the comparison of R-MSMARCO and G-MSMARCO.

To examine how well G-MSMARCO aligns with R-MSMARCO, we evaluate 17 IR models on R-MSMARCO and G-MSMARCO using nDCG@10, and compute the Spearman rank correlation coefficient~\cite{spearman1961proof} between their rankings on R-MSMARCO and G-MSMARCO as the consistency metric.

\textbf{Main Results}. As shown in Table~\ref{tab:consistency_results}, the Spearman rank correlation coefficient is 0.8211 with a p-value of 5e-5, indicating that the LLM-generated testing data aligns well with the human-labeled testing data. Overall, each model achieves higher nDCG@10 on G-MSMARCO than on R-MSMARCO. This can be largely attributed to more comprehensive quality control strategy of \airbench, which results in more positives for each query (see Table~\ref{tab:msmarco_comparison}).

\textbf{Ablation of Quality Control}. To demonstrate the necessity of the quality control stage in the data generation pipeline of \airbench, we also evaluate the consistency between R-MSMARCO and G-MSMARCO generated \textit{without quality control}. As shown in Table~\ref{tab:consistency_results}, the correlation coefficient shows a significant degradation (0.8211 $\rightarrow$ 0.6912). Besides, the nDCG@10 of each model on G-MSMARCO without quality control also has a huge drop, due to some low-quality queries and very limited positives (see Table~\ref{tab:msmarco_comparison}, there are 1,110 low-quality queries and only 7,429 positives). Therefore, quality control stage is necessary to ensure the data generation pipeline a reliable data generation pipeline.

\begin{figure}[!t]
\centering
\includegraphics[width=0.49\textwidth]{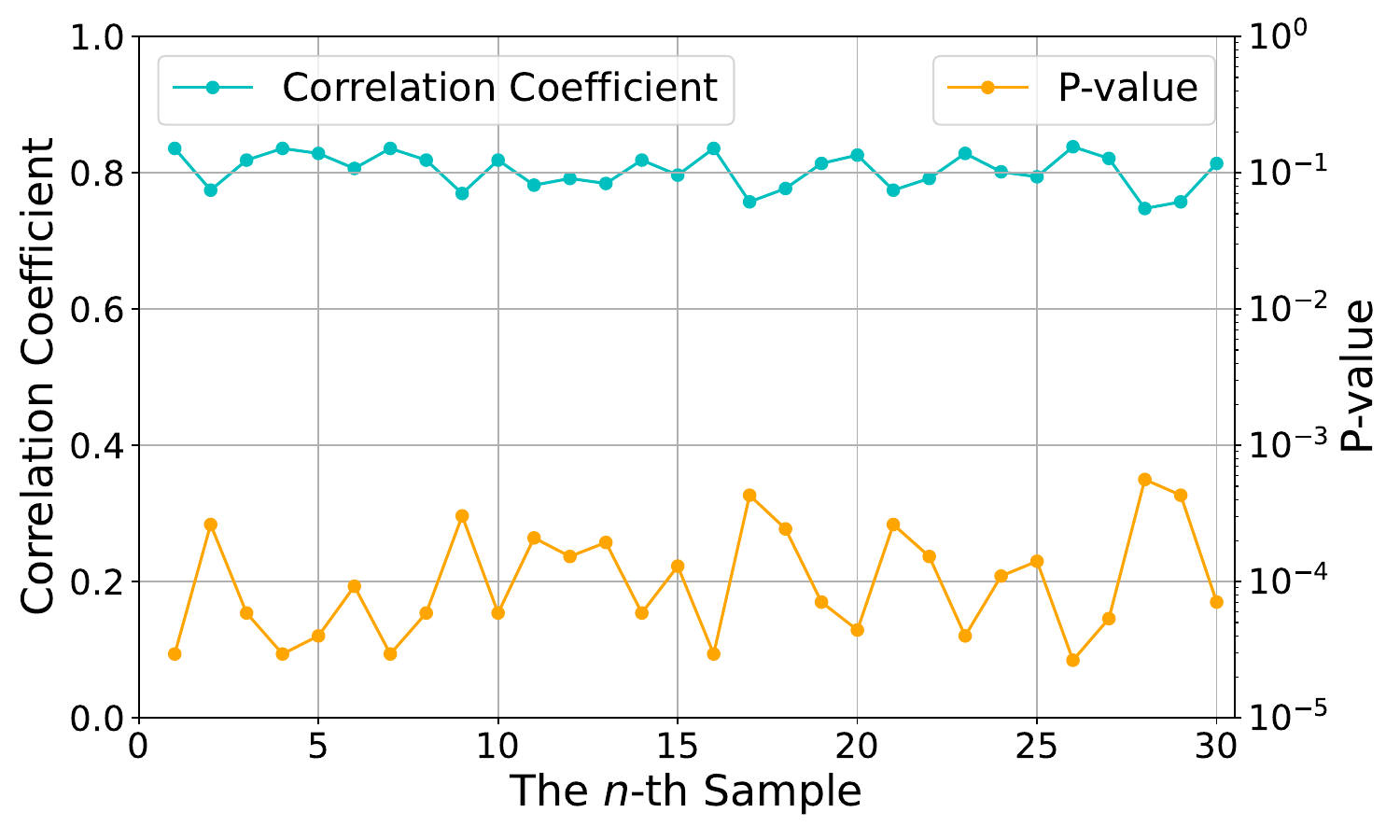}
\caption{Robustness analysis of the consistency between the LLM-generated testing data and the human-labeled testing data. The mean correlation coefficient is 0.8031 with a mean p-value of  1e-4 across 30 simulated generation processes.}
\label{fig:robustness_analysis}
\vspace{-10pt}
\end{figure}

\begin{table*}[!ht]
    \centering
    \small
    \setlength{\tabcolsep}{4pt}
    \setlength{\extrarowheight}{2pt}
    \begin{tabular}{l|c|cc|cc|cc|cc}
        \toprule
        \multirow{4}{*}{\textbf{Model}} & \multirow{4}{*}{\textbf{Size}} & \multicolumn{4}{c|}{\textbf{MTEB (English)}} & \multicolumn{4}{c}{\textbf{\airbench 24.05 (English, test)}} \\
        \cline{3-10}
         &  & \multicolumn{2}{c|}{\textbf{Overall}} & \multicolumn{2}{c|}{\textbf{Retrieval (BEIR)}} & \multicolumn{2}{c|}{\textbf{QA}} & \multicolumn{2}{c}{\textbf{Long-Doc}} \\
         &  & \multicolumn{2}{c|}{\textbf{56 datasets}} & \multicolumn{2}{c|}{\textbf{15 datasets}} & \multicolumn{2}{c|}{\textbf{7 datasets}} & \multicolumn{2}{c}{\textbf{11 datasets}} \\
        \cline{3-10}
         &  & \textbf{Avg.} & \textbf{Rank} & \textbf{nDCG@10} & \textbf{Rank} & \textbf{nDCG@10} & \textbf{Rank} & \textbf{Recall@10} & \textbf{Rank} \\
        \midrule
        \multicolumn{10}{l}{\textit{LLM-based Embedding Models}} \\
        \midrule
        NV-Embed-v2                     & 7.85B & 72.31 & 1  & 62.65 & 1  & 53.35 & 3  & 73.45 & 1 \\
        bge-en-icl (zero-shot)          & 7.11B & 71.24 & 2  & 61.67 & 2  & 53.60 & 2  & 72.62 & 3 \\
        bge-en-icl-e5data (zero-shot)   & 7.11B & 64.67 & 11 & 59.59 & 6  & 54.46 & 1  & 73.43 & 2 \\
        SFR-Embedding-2\_R              & 7.11B & 70.31 & 3  & 60.18 & 5  & 50.80 & 7  & 65.83 & 9  \\
        gte-Qwen2-7B-instruct           & 7.61B & 70.24 & 4  & 60.25 & 3  & 51.87 & 5  & 63.97 & 10 \\
        NV-Embed-v1                     & 7.85B & 69.32 & 5  & 59.36 & 7  & 50.97 & 6  & 72.08 & 4  \\
        Linq-Embed-Mistral              & 7.11B & 68.17 & 6  & 60.19 & 4  & 49.76 & 9  & 70.02 & 5  \\
        SFR-Embedding-Mistral           & 7.11B & 67.56 & 7  & 59.00 & 8  & 52.78 & 4  & 68.10 & 6  \\
        e5-mistral-7b-instruct          & 7.11B & 66.40 & 8  & 56.87 & 10 & 49.88 & 8  & 66.91 & 7  \\
        \midrule
        \multicolumn{10}{l}{\textit{Large-size Embedding Models}} \\
        \midrule
        jina-embeddings-v3              & 572M & 65.51 & 9  & 53.88 & 12 & 45.07  & 13 & 61.50 & 13 \\
        gte-large-en-v1.5               & 434M & 65.39 & 10 & 57.91 & 9  & 46.251 & 11 & 60.71 & 14 \\
        multilingual-e5-large-instruct  & 560M & 64.41 & 12 & 52.47 & 13 & 45.39  & 12 & 63.96 & 11 \\
        bge-large-en-v1.5               & 335M & 64.23 & 13 & 54.29 & 11 & 44.91  & 14 & 61.86 & 12 \\
        e5-large-v2                     & 335M & 62.20 & 14 & 50.56 & 14 & 46.253 & 10 & 66.16 & 8  \\
        \midrule
        \multicolumn{10}{l}{\textit{Lexical Method}} \\
        % \multicolumn{10}{l}{\textit{Lexical Method (+ Re-ranking Model)}} \\
        \midrule
        BM25                            &   -  &   -   & -  & 40.76 & 15 & 39.16 & 15 & 53.09 & 15 \\
        % \ \ \ + bge-reranker-v2-m3          &   -  &   -   & -  & - & - & 55.70 & - & 74.20 &  - \\
        \bottomrule
    \end{tabular}
    \caption{Comparison of the performance of 15 IR models on \airbench and MTEB/BEIR. The results on MTEB/BEIR are directly taken from the MTEB leaderboard. For detailed information of the models appearing in the table, please refer to Table~\ref{tab:model_information}. The detailed results for each dataset in \airbench are available in Appendix~\ref{appendix_sec:detailed_experiment_results}.}
    \label{tab:comparison_results}
    \vspace{-10pt}
\end{table*}

\textbf{Robustness of Consistency}. To investigate the robustness of consistency, we simulate 30 generation processes by randomly sampling 2,000 generated queries from G-MSMARCO on each occasion. After each sampling, we access the consistency between the sampled G-MSMARCO and R-MSMARCO. As illustrated in Figure~\ref{fig:robustness_analysis}, the LLM-generated testing data exhibits stable and strong consistency with the human-labeled testing data, highlighting the robustness of this consistency.

\subsection{Comparison with MTEB/BEIR (RQ2)}
\label{sec:comparison_with_mteb}

To investigate what additional evaluation functionalities \airbench can offer compared to MTEB~\cite{muennighoff-etal-2023-mteb} and BEIR~\cite{thakur2021beir}, we compare the performance of 15 IR models on \airbench and MTEB/BEIR.

\textbf{Setup}. In addition to 14 large-size and LLM-based embedding models exhibiting superior performances on MTEB/BEIR, we also evaluate the performance of lexical method BM25~\cite{robertson2009bm25}.

\textbf{Main Results}. As presented in Table~\ref{tab:comparison_results}, we can make the following observations based on the comparison results. 1) LLM-based embedding models generally outperform large-size embedding models on both \airbench and MTEB/BEIR, largely due to the superior generalization ability of LLMs. Besides, BM25 performs worse than all embedding models on both \airbench and BEIR. 2) The QA task and the Long-Doc task in \airbench exhibit a level of heterogeneity. The Spearman rank correlation coefficient between the rankings of the nine LLM-based embedding models across the two tasks is only 0.6, with a p-value of 0.0876. Moreover, as a large-size embedding model, \texttt{e5-large-v2} even outperforms some LLM-based embedding models on the Long-Doc task. 3) By comparing the results on \airbench and MTEB/BEIR, we observe that better performance on MTEB/BEIR may not indicate better performance on \airbench. For example, according to \citet{li2024making}, \texttt{bge-en-icl} utilizes more in-domain training data in MTEB/BEIR than \texttt{bge-en-icl-e5data} and achieves more superior performance on MTEB/BEIR. However, compared to \texttt{bge-en-icl-e5data}, \texttt{bge-en-icl} shows performance degradation on \airbench, including both the QA task (54.46 $\rightarrow$ 53.60) and the Long-Doc task (73.43 $\rightarrow$ 72.62). This suggests that increased in-domain training data in MTEB/BEIR may lead to over-fitting, thereby reducing the generalization ability of embedding models.
% This suggests that \airbench can offer additional evaluation functionalities.

In conclusion, as a new benchmark, \airbench can offer additional evaluation functionalities for community developers compared to MTEB/BEIR.

\subsection{Distinguishing Models (RQ3)}

\begin{table}[!t]
    \centering
    \setlength{\extrarowheight}{2pt}
    \resizebox{0.49\textwidth}{!}{
        \begin{tabular}{l|c|c}
            \toprule
            \multirow{2}{*}{\textbf{Dataset} ($\downarrow$)} & \textbf{mContriever} & \textbf{mContriever-finetuned} \\
            \cline{2-3}
             & \textbf{nDCG@10} & \textbf{nDCG@10} / \textbf{Training Data} \\
            \midrule
            finance\_en     & 39.452 & \begin{tabular}[c]{@{}c@{}} 41.281 (\textcolor{orange}{$\uparrow$ 1.829}) \\ 
                                   FiQA~\cite{maia2018fiqa} \end{tabular} \\
                                   \midrule
            healthcare\_zh  & 14.557 & \begin{tabular}[c]{@{}c@{}} 17.351 (\textcolor{orange}{$\uparrow$ 2.794}) \\ 
                                   cMedQAv2~\cite{zhang2018cmedqav2} \end{tabular} \\
                                   \midrule
            law\_de         & 5.614 & \begin{tabular}[c]{@{}c@{}} 6.687 (\textcolor{orange}{$\uparrow$ 1.073}) \\ 
                                   \citet{hoppe2021towards} \end{tabular} \\
                                   \midrule
            law\_fr         & 3.102 & \begin{tabular}[c]{@{}c@{}} 4.325 (\textcolor{orange}{$\uparrow$ 1.223}) \\ 
                                   BSARD~\cite{louis2022statutory} \end{tabular} \\
                                   \midrule
            web\_hi         & 19.067 & \begin{tabular}[c]{@{}c@{}} 30.103 (\textcolor{orange}{$\uparrow$ 11.036}) \\ 
                                   mMARCO~\cite{bonifacio2021mmarco} \end{tabular} \\
                                   \midrule
            wiki\_ar        & 38.159 & \begin{tabular}[c]{@{}c@{}} 43.470 (\textcolor{orange}{$\uparrow$ 5.311}) \\ 
                                   MIRACL~\cite{zhang2023miracl} \end{tabular} \\
            \bottomrule
        \end{tabular}
    }
    \caption{\airbench can showcase 
models' performance enhancement in specific domains. The training process takes 100 steps for cMedQAv2, and 50 steps for the other datasets.}
    \label{tab:mcontriever_adaptation}
    \vspace{-10pt}
\end{table}

To examine how effectively \airbench can distinguish the capabilities of distinct IR models, we evaluate the performance of a single model before and after fine-tuning to illustrate that \airbench can reflect the performance enhancement of IR models in specific domains.

\textbf{Setup}. We fine-tune mContriever\footnote{\url{https://huggingface.co/facebook/mcontriever-msmarco}}~\cite{izacard2021unsupervised} using domain-specific training datasets, and compare the model's performance on the corresponding datasets in \airbench before and after fine-tuning. Specifically, we fine-tune\footnote{The learning rate is $2\times 10^{-4}$, the warmup ratio is 0.1, and the weight decay is 0.01. The training process takes around a hundred steps with a total batch size of 64 on 8 A800 GPUs.} mContriever with FlagEmbedding tool\footnote{\url{https://github.com/FlagOpen/FlagEmbedding}} to enhance its domain-specific capabilities. The domain-specific training data used for fine-tuning is independent of the corresponding testing data in \airbench.

\textbf{Main Results}. Table~\ref{tab:mcontriever_adaptation} presents the detailed information about each domain-specific training dataset and compares the model's performance on the corresponding dataset in \airbench before and after fine-tuning. For example, after fine-tuning with the Hindi training data from mMARCO~\cite{bonifacio2021mmarco}, the performance of mContriever on the web\_hi dataset in \airbench improves from 19.067 to 30.103. This trend is also observed in other domains, such as finance, healthcare, law and wiki. Therefore, \airbench effectively reflects the performance enhancement of IR models in specific domains following fine-tuning with domain-specific training datasets.

We also evaluate a diverse set of IR models on \airbench to further demonstrate its capability of distinguishing different models across multiple dimensions, including model type, domain, and language. Refer to Appendix~\ref{appendix_sec:distinguishing_models} for the details.

\section{Related Work}

The related works are reviewed from two aspects: evaluation datasets for IR, and synthetic data generation for IR.

\subsection{Evaluation Datasets for IR}

Evaluation datasets are critically important for the development of IR models. 

In recent years, a series of milestone works have been introduced to the community. As the earlier contributions, MS MARCO~\cite{bajaj2016ms} includes Bing search questions paired with human-labeled relevant passages from Web documents. Natural Questions (NQ)~\cite{kwiatkowski2019natural} consists of Google search queries with human-labeled relevant Wikipedia pages. Both MS MARCO and NQ are designed for open-domain question answering tasks in English. 
% and share a similar two-stage annotation process: for each question, a few of passages are selected using an existing retrieval system as candidate data, which are then annotated by editors as relevant positives. 
Recent works like Mr.TyDi~\cite{zhang-etal-2021-mr} and MIRACL~\cite{zhang2023miracl} focus on multilingual retrieval in non-English languages. Mr.TyDi covers 11 languages and MIRACL encompasses an extended 18 languages. BEIR~\cite{thakur2021beir} and MTEB~\cite{muennighoff-etal-2023-mteb} are introduced to benchmark IR models in a general-domain zero-shot setting, including multiple existing datasets from diverse tasks and domains.

However, all of these benchmarks, which rely on pre-defined domains and human-labeled data, face limitations in addressing evaluation needs for emerging domains both cost-effectively and efficiently. Recently, several studies have explored the application of large language models for retrieval evaluation in retrieval-augmented generation (RAG) systems~\cite{es-etal-2024-ragas,saad-falcon-etal-2024-ares,salemi2024evaluating}, offering a promising solution to this challenge. Nonetheless, a comprehensive IR benchmark that addresses this limitation remains insufficiently developed.

\subsection{Synthetic Data Generation for IR}

The tasks and domains in IR applications are often diverse and dynamic, meaning that the the training and evaluation data are frequently unavailable for new tasks and domains. As a result, it becomes challenging to fine-tune and evaluate IR models in these contexts.

Several recent works~\cite{bonifacio2022inpars,dai2023promptagator,jeronymo2023inpars,khramtsova2024leveraging,thakur-etal-2024-leveraging} have focused on addressing the scarcity of domain-specific training data by prompting LLMs to generate synthetic training data. \citet{wang2023improving} and \citet{chen2024little} employ LLMs to generate synthetic task and training data. \citet{lee2024gecko} further refines the synthetic training data by using LLMs to select more relevant positives and negatives.

However, there is currently limited research addressing the scarcity of domain-specific evaluation datasets. \citet{llm_accessors} have demonstrated that powerful LLMs can generate high-quality golden labels for search system with accuracy comparable to human labelers, laying a solid foundation for our work. Our experiment results also demonstrate that the LLM-generated testing data aligns well with the human-labeled testing data. To our knowledge, \airbench is the first comprehensive IR benchmark that utilizes the LLM-generated datasets to perform evaluation.

\section{Conclusion}

In this paper, we introduce a new IR benchmark \airbench, which is highlighted for three main features: 1) Automated, 2) Heterogeneous, and 3) Dynamic. We demonstrate that the generated testing data in \airbench is highly consistent with the human-labeled testing data, which makes \airbench a dependable benchmark for evaluating IR models. Additionally, we demonstrate that \airbench can offer additional evaluation functionalities compared to MTEB/BEIR. Last but not least, we demonstrate that \airbench can effectively distinguish the capabilities of distinct IR models from multiple dimensions.

\airbench currently covers 2 tasks, 9 domains and 13 languages, including a total of 69 datasets. In the future, \airbench will be extended to cover more tasks, domains and languages to provide an increasingly comprehensive evaluation benchmark for community developers. We welcome datasets contributions to \airbench\footnote{\repolink} as well as the model submissions to our leaderboard\footnote{\leaderboardlink}.

\section*{Limitations}

While \airbench aims to be a comprehensive IR benchmark by introducing new features to address the limitations of existing benchmarks, it still has several inherent constraints: 1) Dependence on real-world corpora. The dataset generation process in \airbench begins with corpus preparation. As a result, access to real-world corpora is essential for constructing evaluation datasets. Fortunately, this requirement is typically both feasible and practical in real-world scenarios. 2) Reliance on capabilities of LLM. The quality of the generated testing data in \airbench largely depends on the LLM’s capabilities. However, This limitation can be mitigated by the rapid advancement of LLMs. 3) Potential biases from quality control models. In addition to the LLM, we incorporate several existing IR models during the quality control stage. This reliance may introduce potential biases into the final evaluation datasets. However, as these models continue to improve, the impact of such biases can be progressively reduced.

\section*{Ethics Consideration}

Since \airbench is built on testing data generated by LLM, it may inherit potential biases, toxicity, and other issues present in the LLM used during the generation process. Additionally, considering that the corpora utilized in the generation process are derived from the real-world sources, they may contain sensitive content. Therefore, the testing data in \airbench may only be used for evaluation purposes.

\section*{Acknowledgements}
This work is supported by National Science and Technology Major Project (2023ZD0121504), National Natural Science Foundation of China (No. U24A20253, 62276171), Guangdong Basic and Applied Basic Research Foundation (No. 2024A1515011938), Shenzhen Fundamental Research-General Project, China under Grant (No. JCYJ20240813141503005). We appreciate the valuable feedback from Tom Aarsen, Niklas Muennighoff, Jiajun Wang, and Linpeng Tang.

\bibliography{custom}

\begin{thebibliography}{61}
\providecommand{\natexlab}[1]{#1}

\bibitem[{Achiam et~al.(2023)Achiam, Adler, Agarwal, Ahmad, Akkaya, Aleman, Almeida, Altenschmidt, Altman, Anadkat et~al.}]{achiam2023gpt}
Josh Achiam, Steven Adler, Sandhini Agarwal, Lama Ahmad, Ilge Akkaya, Florencia~Leoni Aleman, Diogo Almeida, Janko Altenschmidt, Sam Altman, Shyamal Anadkat, et~al. 2023.
\newblock Gpt-4 technical report.
\newblock \emph{arXiv preprint arXiv:2303.08774}.

\bibitem[{Bajaj et~al.(2016)Bajaj, Campos, Craswell, Deng, Gao, Liu, Majumder, McNamara, Mitra, Nguyen et~al.}]{bajaj2016ms}
Payal Bajaj, Daniel Campos, Nick Craswell, Li~Deng, Jianfeng Gao, Xiaodong Liu, Rangan Majumder, Andrew McNamara, Bhaskar Mitra, Tri Nguyen, et~al. 2016.
\newblock Ms marco: A human generated machine reading comprehension dataset.
\newblock \emph{arXiv preprint arXiv:1611.09268}.

\bibitem[{Bonifacio et~al.(2022)Bonifacio, Abonizio, Fadaee, and Nogueira}]{bonifacio2022inpars}
Luiz Bonifacio, Hugo Abonizio, Marzieh Fadaee, and Rodrigo Nogueira. 2022.
\newblock \href {https://doi.org/10.1145/3477495.3531863} {Inpars: Unsupervised dataset generation for information retrieval}.
\newblock In \emph{Proceedings of the 45th International ACM SIGIR Conference on Research and Development in Information Retrieval}, SIGIR '22, page 2387–2392, New York, NY, USA. Association for Computing Machinery.

\bibitem[{Bonifacio et~al.(2021)Bonifacio, Campiotti, de~Alencar~Lotufo, and Nogueira}]{bonifacio2021mmarco}
Luiz~Henrique Bonifacio, Israel Campiotti, Roberto de~Alencar~Lotufo, and Rodrigo Nogueira. 2021.
\newblock mmarco: A multilingual version of ms marco passage ranking dataset. corr abs/2108.13897 (2021).
\newblock \emph{arXiv preprint arXiv:2108.13897}.

\bibitem[{Chalkidis et~al.(2023)Chalkidis, Garneau, Goanta, Katz, and S{\o}gaard}]{chalkidis-etal-2023-lexfiles}
Ilias Chalkidis, Nicolas Garneau, Catalina Goanta, Daniel Katz, and Anders S{\o}gaard. 2023.
\newblock \href {https://aclanthology.org/2023.acl-long.865} {{L}e{XF}iles and {L}egal{LAMA}: Facilitating {E}nglish multinational legal language model development}.
\newblock In \emph{Proceedings of the 61st Annual Meeting of the Association for Computational Linguistics (Volume 1: Long Papers)}, pages 15513--15535, Toronto, Canada. Association for Computational Linguistics.

\bibitem[{Chen et~al.(2024{\natexlab{a}})Chen, Wang, Yang, Zhu, Zhao, Wei, and Dou}]{chen2024little}
Haonan Chen, Liang Wang, Nan Yang, Yutao Zhu, Ziliang Zhao, Furu Wei, and Zhicheng Dou. 2024{\natexlab{a}}.
\newblock Little giants: Synthesizing high-quality embedding data at scale.
\newblock \emph{arXiv preprint arXiv:2410.18634}.

\bibitem[{Chen et~al.(2024{\natexlab{b}})Chen, Xiao, Zhang, Luo, Lian, and Liu}]{chen-etal-2024-m3}
Jianlyu Chen, Shitao Xiao, Peitian Zhang, Kun Luo, Defu Lian, and Zheng Liu. 2024{\natexlab{b}}.
\newblock \href {https://aclanthology.org/2024.findings-acl.137} {{M}3-embedding: Multi-linguality, multi-functionality, multi-granularity text embeddings through self-knowledge distillation}.
\newblock In \emph{Findings of the Association for Computational Linguistics ACL 2024}, pages 2318--2335, Bangkok, Thailand and virtual meeting. Association for Computational Linguistics.

\bibitem[{Cohan et~al.(2018)Cohan, Dernoncourt, Kim, Bui, Kim, Chang, and Goharian}]{cohan-etal-2018-discourse}
Arman Cohan, Franck Dernoncourt, Doo~Soon Kim, Trung Bui, Seokhwan Kim, Walter Chang, and Nazli Goharian. 2018.
\newblock \href {https://doi.org/10.18653/v1/N18-2097} {A discourse-aware attention model for abstractive summarization of long documents}.
\newblock In \emph{Proceedings of the 2018 Conference of the North {A}merican Chapter of the Association for Computational Linguistics: Human Language Technologies, Volume 2 (Short Papers)}, pages 615--621, New Orleans, Louisiana. Association for Computational Linguistics.

\bibitem[{Dai et~al.(2023)Dai, Zhao, Ma, Luan, Ni, Lu, Bakalov, Guu, Hall, and Chang}]{dai2023promptagator}
Zhuyun Dai, Vincent~Y Zhao, Ji~Ma, Yi~Luan, Jianmo Ni, Jing Lu, Anton Bakalov, Kelvin Guu, Keith Hall, and Ming-Wei Chang. 2023.
\newblock \href {https://openreview.net/forum?id=gmL46YMpu2J} {Promptagator: Few-shot dense retrieval from 8 examples}.
\newblock In \emph{The Eleventh International Conference on Learning Representations}.

\bibitem[{Daudert and Ahmadi(2019)}]{daudert-ahmadi-2019-cofif}
Tobias Daudert and Sina Ahmadi. 2019.
\newblock \href {https://www.aclweb.org/anthology/W19-5504} {{C}o{F}i{F}: A corpus of financial reports in {F}rench language}.
\newblock In \emph{Proceedings of the First Workshop on Financial Technology and Natural Language Processing}, pages 21--26, Macao, China.

\bibitem[{Es et~al.(2024)Es, James, Espinosa~Anke, and Schockaert}]{es-etal-2024-ragas}
Shahul Es, Jithin James, Luis Espinosa~Anke, and Steven Schockaert. 2024.
\newblock \href {https://aclanthology.org/2024.eacl-demo.16} {{RAGA}s: Automated evaluation of retrieval augmented generation}.
\newblock In \emph{Proceedings of the 18th Conference of the European Chapter of the Association for Computational Linguistics: System Demonstrations}, pages 150--158, St. Julians, Malta. Association for Computational Linguistics.

\bibitem[{Hamborg et~al.(2017)Hamborg, Meuschke, Breitinger, and Gipp}]{Hamborg2017}
Felix Hamborg, Norman Meuschke, Corinna Breitinger, and Bela Gipp. 2017.
\newblock \href {https://doi.org/10.5281/zenodo.4120316} {news-please: A generic news crawler and extractor}.
\newblock In \emph{Proceedings of the 15th International Symposium of Information Science}, pages 218--223.

\bibitem[{Henderson* et~al.(2022)Henderson*, Krass*, Zheng, Guha, Manning, Jurafsky, and Ho}]{hendersonkrass2022pileoflaw}
Peter Henderson*, Mark~S. Krass*, Lucia Zheng, Neel Guha, Christopher~D. Manning, Dan Jurafsky, and Daniel~E. Ho. 2022.
\newblock \href {https://arxiv.org/abs/2207.00220} {Pile of law: Learning responsible data filtering from the law and a 256gb open-source legal dataset}.
\newblock \emph{arXiv preprint}.

\bibitem[{Hoppe et~al.(2021)Hoppe, Pelkmann, Migenda, H{\"o}tte, and Schenck}]{hoppe2021towards}
Christoph Hoppe, David Pelkmann, Nico Migenda, Daniel H{\"o}tte, and Wolfram Schenck. 2021.
\newblock Towards intelligent legal advisors for document retrieval and question-answering in german legal documents.
\newblock In \emph{2021 IEEE Fourth International Conference on Artificial Intelligence and Knowledge Engineering (AIKE)}, pages 29--32. IEEE.

\bibitem[{Izacard et~al.(2021)Izacard, Caron, Hosseini, Riedel, Bojanowski, Joulin, and Grave}]{izacard2021unsupervised}
Gautier Izacard, Mathilde Caron, Lucas Hosseini, Sebastian Riedel, Piotr Bojanowski, Armand Joulin, and Edouard Grave. 2021.
\newblock Unsupervised dense information retrieval with contrastive learning.
\newblock \emph{arXiv preprint arXiv:2112.09118}.

\bibitem[{Izacard et~al.(2022)Izacard, Caron, Hosseini, Riedel, Bojanowski, Joulin, and Grave}]{izacard2022unsupervised}
Gautier Izacard, Mathilde Caron, Lucas Hosseini, Sebastian Riedel, Piotr Bojanowski, Armand Joulin, and Edouard Grave. 2022.
\newblock \href {https://openreview.net/forum?id=jKN1pXi7b0} {Unsupervised dense information retrieval with contrastive learning}.
\newblock \emph{Transactions on Machine Learning Research}.

\bibitem[{Jeronymo et~al.(2023)Jeronymo, Bonifacio, Abonizio, Fadaee, Lotufo, Zavrel, and Nogueira}]{jeronymo2023inpars}
Vitor Jeronymo, Luiz Bonifacio, Hugo Abonizio, Marzieh Fadaee, Roberto Lotufo, Jakub Zavrel, and Rodrigo Nogueira. 2023.
\newblock Inpars-v2: Large language models as efficient dataset generators for information retrieval.
\newblock \emph{arXiv preprint arXiv:2301.01820}.

\bibitem[{Jin et~al.(2019)Jin, Dhingra, Liu, Cohen, and Lu}]{jin2019pubmedqa}
Qiao Jin, Bhuwan Dhingra, Zhengping Liu, William Cohen, and Xinghua Lu. 2019.
\newblock Pubmedqa: A dataset for biomedical research question answering.
\newblock In \emph{Proceedings of the 2019 Conference on Empirical Methods in Natural Language Processing and the 9th International Joint Conference on Natural Language Processing (EMNLP-IJCNLP)}, pages 2567--2577.

\bibitem[{Khramtsova et~al.(2024)Khramtsova, Zhuang, Baktashmotlagh, and Zuccon}]{khramtsova2024leveraging}
Ekaterina Khramtsova, Shengyao Zhuang, Mahsa Baktashmotlagh, and Guido Zuccon. 2024.
\newblock \href {https://doi.org/10.1145/3626772.3657798} {Leveraging llms for unsupervised dense retriever ranking}.
\newblock In \emph{Proceedings of the 47th International ACM SIGIR Conference on Research and Development in Information Retrieval}, SIGIR '24, page 1307–1317, New York, NY, USA. Association for Computing Machinery.

\bibitem[{Kim et~al.(2024)Kim, Lee, Kwon, Gu, Kim, Cho, Sohn, and Choi}]{LinqAIResearch2024}
Junseong Kim, Seolhwa Lee, Jihoon Kwon, Sangmo Gu, Yejin Kim, Minkyung Cho, Jy-yong Sohn, and Chanyeol Choi. 2024.
\newblock \href {https://getlinq.com/blog/linq-embed-mistral/} {Linq-embed-mistral:elevating text retrieval with improved gpt data through task-specific control and quality refinement}.
\newblock Linq AI Research Blog.

\bibitem[{Kwiatkowski et~al.(2019)Kwiatkowski, Palomaki, Redfield, Collins, Parikh, Alberti, Epstein, Polosukhin, Devlin, Lee et~al.}]{kwiatkowski2019natural}
Tom Kwiatkowski, Jennimaria Palomaki, Olivia Redfield, Michael Collins, Ankur Parikh, Chris Alberti, Danielle Epstein, Illia Polosukhin, Jacob Devlin, Kenton Lee, et~al. 2019.
\newblock Natural questions: a benchmark for question answering research.
\newblock \emph{Transactions of the Association for Computational Linguistics}, 7:453--466.

\bibitem[{Lee et~al.(2024{\natexlab{a}})Lee, Roy, Xu, Raiman, Shoeybi, Catanzaro, and Ping}]{lee2024nvembed}
Chankyu Lee, Rajarshi Roy, Mengyao Xu, Jonathan Raiman, Mohammad Shoeybi, Bryan Catanzaro, and Wei Ping. 2024{\natexlab{a}}.
\newblock Nv-embed: Improved techniques for training llms as generalist embedding models.
\newblock \emph{arXiv:2405.17428}.

\bibitem[{Lee et~al.(2024{\natexlab{b}})Lee, Dai, Ren, Chen, Cer, Cole, Hui, Boratko, Kapadia, Ding et~al.}]{lee2024gecko}
Jinhyuk Lee, Zhuyun Dai, Xiaoqi Ren, Blair Chen, Daniel Cer, Jeremy~R Cole, Kai Hui, Michael Boratko, Rajvi Kapadia, Wen Ding, et~al. 2024{\natexlab{b}}.
\newblock Gecko: Versatile text embeddings distilled from large language models.
\newblock \emph{arXiv preprint arXiv:2403.20327}.

\bibitem[{Lewis(1997)}]{misc_reuters-21578_text_categorization_collection_137}
David Lewis. 1997.
\newblock {Reuters-21578 Text Categorization Collection}.
\newblock UCI Machine Learning Repository.
\newblock {DOI}: https://doi.org/10.24432/C52G6M.

\bibitem[{Lewis et~al.(2020)Lewis, Perez, Piktus, Petroni, Karpukhin, Goyal, K\"{u}ttler, Lewis, Yih, Rockt\"{a}schel, Riedel, and Kiela}]{lewis2020rag}
Patrick Lewis, Ethan Perez, Aleksandra Piktus, Fabio Petroni, Vladimir Karpukhin, Naman Goyal, Heinrich K\"{u}ttler, Mike Lewis, Wen-tau Yih, Tim Rockt\"{a}schel, Sebastian Riedel, and Douwe Kiela. 2020.
\newblock \href {https://proceedings.neurips.cc/paper_files/paper/2020/file/6b493230205f780e1bc26945df7481e5-Paper.pdf} {Retrieval-augmented generation for knowledge-intensive nlp tasks}.
\newblock In \emph{Advances in Neural Information Processing Systems}, volume~33, pages 9459--9474. Curran Associates, Inc.

\bibitem[{Li et~al.(2024)Li, Qin, Xiao, Chen, Luo, Shao, Lian, and Liu}]{li2024making}
Chaofan Li, MingHao Qin, Shitao Xiao, Jianlyu Chen, Kun Luo, Yingxia Shao, Defu Lian, and Zheng Liu. 2024.
\newblock Making text embedders few-shot learners.
\newblock \emph{arXiv preprint arXiv:2409.15700}.

\bibitem[{Li et~al.(2023{\natexlab{a}})Li, Wang, Wu, Zhang, Xu, Fu, Tiwari, Wan, and Wang}]{li2023huatuo26m}
Jianquan Li, Xidong Wang, Xiangbo Wu, Zhiyi Zhang, Xiaolong Xu, Jie Fu, Prayag Tiwari, Xiang Wan, and Benyou Wang. 2023{\natexlab{a}}.
\newblock \href {https://arxiv.org/abs/2305.01526} {Huatuo-26m, a large-scale chinese medical qa dataset}.
\newblock \emph{Preprint}, arXiv:2305.01526.

\bibitem[{Li et~al.(2023{\natexlab{b}})Li, Zhang, Zhang, Long, Xie, and Zhang}]{li2023towards}
Zehan Li, Xin Zhang, Yanzhao Zhang, Dingkun Long, Pengjun Xie, and Meishan Zhang. 2023{\natexlab{b}}.
\newblock Towards general text embeddings with multi-stage contrastive learning.
\newblock \emph{arXiv preprint arXiv:2308.03281}.

\bibitem[{Lin et~al.(2021)Lin, Ma, Lin, Yang, Pradeep, and Nogueira}]{lin2021pyserini}
Jimmy Lin, Xueguang Ma, Sheng-Chieh Lin, Jheng-Hong Yang, Ronak Pradeep, and Rodrigo Nogueira. 2021.
\newblock Pyserini: A python toolkit for reproducible information retrieval research with sparse and dense representations.
\newblock In \emph{Proceedings of the 44th International ACM SIGIR Conference on Research and Development in Information Retrieval}, pages 2356--2362.

\bibitem[{Liu(2022)}]{Liu_LlamaIndex_2022}
Jerry Liu. 2022.
\newblock \href {https://doi.org/10.5281/zenodo.1234} {{LlamaIndex}}.

\bibitem[{Louis and Spanakis(2022)}]{louis2022statutory}
Antoine Louis and Gerasimos Spanakis. 2022.
\newblock \href {https://doi.org/10.18653/v1/2022.acl-long.468} {A statutory article retrieval dataset in french}.
\newblock In \emph{Proceedings of the 60th Annual Meeting of the Association for Computational Linguistics}, page 6789–6803, Dublin, Ireland. Association for Computational Linguistics.

\bibitem[{Ma et~al.(2023)Ma, Wang, Yang, Wei, and Lin}]{rankllama}
Xueguang Ma, Liang Wang, Nan Yang, Furu Wei, and Jimmy Lin. 2023.
\newblock Fine-tuning llama for multi-stage text retrieval.
\newblock \emph{arXiv:2310.08319}.

\bibitem[{Maia et~al.(2018)Maia, Handschuh, Freitas, Davis, McDermott, Zarrouk, and Balahur}]{maia2018fiqa}
Macedo Maia, Siegfried Handschuh, Andr{\'e} Freitas, Brian Davis, Ross McDermott, Manel Zarrouk, and Alexandra Balahur. 2018.
\newblock Www'18 open challenge: financial opinion mining and question answering.
\newblock In \emph{Companion proceedings of the the web conference 2018}, pages 1941--1942.

\bibitem[{Muennighoff et~al.(2023)Muennighoff, Tazi, Magne, and Reimers}]{muennighoff-etal-2023-mteb}
Niklas Muennighoff, Nouamane Tazi, Loic Magne, and Nils Reimers. 2023.
\newblock \href {https://aclanthology.org/2023.eacl-main.148} {{MTEB}: Massive text embedding benchmark}.
\newblock In \emph{Proceedings of the 17th Conference of the European Chapter of the Association for Computational Linguistics}, pages 2014--2037, Dubrovnik, Croatia. Association for Computational Linguistics.

\bibitem[{NetEase~Youdao(2023)}]{youdao_bcembedding_2023}
Inc. NetEase~Youdao. 2023.
\newblock Bcembedding: Bilingual and crosslingual embedding for rag.
\newblock \url{https://github.com/netease-youdao/BCEmbedding}.

\bibitem[{Niklaus et~al.(2023)Niklaus, Matoshi, Stürmer, Chalkidis, and Ho}]{niklaus2023multilegalpile}
Joel Niklaus, Veton Matoshi, Matthias Stürmer, Ilias Chalkidis, and Daniel~E. Ho. 2023.
\newblock \href {https://arxiv.org/abs/2306.02069} {Multilegalpile: A 689gb multilingual legal corpus}.
\newblock \emph{Preprint}, arXiv:2306.02069.

\bibitem[{Raffel et~al.(2020)Raffel, Shazeer, Roberts, Lee, Narang, Matena, Zhou, Li, and Liu}]{mc4}
Colin Raffel, Noam Shazeer, Adam Roberts, Katherine Lee, Sharan Narang, Michael Matena, Yanqi Zhou, Wei Li, and Peter~J. Liu. 2020.
\newblock Exploring the limits of transfer learning with a unified text-to-text transformer.
\newblock \emph{J. Mach. Learn. Res.}, 21(1).

\bibitem[{Reimers and Gurevych(2019)}]{reimers-2019-sentence-bert}
Nils Reimers and Iryna Gurevych. 2019.
\newblock \href {https://arxiv.org/abs/1908.10084} {Sentence-bert: Sentence embeddings using siamese bert-networks}.
\newblock In \emph{Proceedings of the 2019 Conference on Empirical Methods in Natural Language Processing}. Association for Computational Linguistics.

\bibitem[{Robertson and Zaragoza(2009)}]{robertson2009bm25}
Stephen Robertson and Hugo Zaragoza. 2009.
\newblock \href {https://doi.org/10.1561/1500000019} {The probabilistic relevance framework: Bm25 and beyond}.
\newblock \emph{Foundations and Trends® in Information Retrieval}, 3(4):333--389.

\bibitem[{Saad-Falcon et~al.(2024)Saad-Falcon, Khattab, Potts, and Zaharia}]{saad-falcon-etal-2024-ares}
Jon Saad-Falcon, Omar Khattab, Christopher Potts, and Matei Zaharia. 2024.
\newblock \href {https://doi.org/10.18653/v1/2024.naacl-long.20} {{ARES}: An automated evaluation framework for retrieval-augmented generation systems}.
\newblock In \emph{Proceedings of the 2024 Conference of the North American Chapter of the Association for Computational Linguistics: Human Language Technologies (Volume 1: Long Papers)}, pages 338--354, Mexico City, Mexico. Association for Computational Linguistics.

\bibitem[{Salemi and Zamani(2024)}]{salemi2024evaluating}
Alireza Salemi and Hamed Zamani. 2024.
\newblock \href {https://doi.org/10.1145/3626772.3657957} {Evaluating retrieval quality in retrieval-augmented generation}.
\newblock In \emph{Proceedings of the 47th International ACM SIGIR Conference on Research and Development in Information Retrieval}, SIGIR '24, page 2395–2400, New York, NY, USA. Association for Computing Machinery.

\bibitem[{Scialom et~al.(2020)Scialom, Dray, Lamprier, Piwowarski, and Staiano}]{scialom-etal-2020-mlsum}
Thomas Scialom, Paul-Alexis Dray, Sylvain Lamprier, Benjamin Piwowarski, and Jacopo Staiano. 2020.
\newblock \href {https://doi.org/10.18653/v1/2020.emnlp-main.647} {{MLSUM}: The multilingual summarization corpus}.
\newblock In \emph{Proceedings of the 2020 Conference on Empirical Methods in Natural Language Processing (EMNLP)}, pages 8051--8067, Online. Association for Computational Linguistics.

\bibitem[{Spearman(1961)}]{spearman1961proof}
Charles Spearman. 1961.
\newblock The proof and measurement of association between two things.

\bibitem[{Sturua et~al.(2024)Sturua, Mohr, Akram, G{\"u}nther, Wang, Krimmel, Wang, Mastrapas, Koukounas, Wang et~al.}]{sturua2024jina}
Saba Sturua, Isabelle Mohr, Mohammad~Kalim Akram, Michael G{\"u}nther, Bo~Wang, Markus Krimmel, Feng Wang, Georgios Mastrapas, Andreas Koukounas, Nan Wang, et~al. 2024.
\newblock jina-embeddings-v3: Multilingual embeddings with task lora.
\newblock \emph{arXiv preprint arXiv:2409.10173}.

\bibitem[{Thakur et~al.(2024)Thakur, Ni, Hernandez~Abrego, Wieting, Lin, and Cer}]{thakur-etal-2024-leveraging}
Nandan Thakur, Jianmo Ni, Gustavo Hernandez~Abrego, John Wieting, Jimmy Lin, and Daniel Cer. 2024.
\newblock \href {https://doi.org/10.18653/v1/2024.naacl-long.426} {Leveraging {LLM}s for synthesizing training data across many languages in multilingual dense retrieval}.
\newblock In \emph{Proceedings of the 2024 Conference of the North American Chapter of the Association for Computational Linguistics: Human Language Technologies (Volume 1: Long Papers)}, pages 7699--7724, Mexico City, Mexico. Association for Computational Linguistics.

\bibitem[{Thakur et~al.(2021)Thakur, Reimers, R{\"u}ckl{\'e}, Srivastava, and Gurevych}]{thakur2021beir}
Nandan Thakur, Nils Reimers, Andreas R{\"u}ckl{\'e}, Abhishek Srivastava, and Iryna Gurevych. 2021.
\newblock \href {https://openreview.net/forum?id=wCu6T5xFjeJ} {{BEIR}: A heterogeneous benchmark for zero-shot evaluation of information retrieval models}.
\newblock In \emph{Thirty-fifth Conference on Neural Information Processing Systems Datasets and Benchmarks Track (Round 2)}.

\bibitem[{Thomas et~al.(2024)Thomas, Spielman, Craswell, and Mitra}]{llm_accessors}
Paul Thomas, Seth Spielman, Nick Craswell, and Bhaskar Mitra. 2024.
\newblock \href {https://doi.org/10.1145/3626772.3657707} {Large language models can accurately predict searcher preferences}.
\newblock In \emph{Proceedings of the 47th International ACM SIGIR Conference on Research and Development in Information Retrieval}, SIGIR '24, page 1930–1940, New York, NY, USA. Association for Computing Machinery.

\bibitem[{Villena(2019)}]{fabian_villena_2019_3463379}
Fabián Villena. 2019.
\newblock \href {https://doi.org/10.5281/zenodo.3463379} {Multilingual medical corpora}.

\bibitem[{Voorhees et~al.(1999)}]{voorhees1999trec}
Ellen~M Voorhees et~al. 1999.
\newblock The trec-8 question answering track report.
\newblock In \emph{Trec}, volume~99, pages 77--82.

\bibitem[{Wang et~al.(2022{\natexlab{a}})Wang, Yang, Huang, Jiao, Yang, Jiang, Majumder, and Wei}]{Wang2022SimLMPW}
Liang Wang, Nan Yang, Xiaolong Huang, Binxing Jiao, Linjun Yang, Daxin Jiang, Rangan Majumder, and Furu Wei. 2022{\natexlab{a}}.
\newblock Simlm: Pre-training with representation bottleneck for dense passage retrieval.
\newblock \emph{ArXiv}, abs/2207.02578.

\bibitem[{Wang et~al.(2022{\natexlab{b}})Wang, Yang, Huang, Jiao, Yang, Jiang, Majumder, and Wei}]{wang2022text}
Liang Wang, Nan Yang, Xiaolong Huang, Binxing Jiao, Linjun Yang, Daxin Jiang, Rangan Majumder, and Furu Wei. 2022{\natexlab{b}}.
\newblock Text embeddings by weakly-supervised contrastive pre-training.
\newblock \emph{arXiv preprint arXiv:2212.03533}.

\bibitem[{Wang et~al.(2023)Wang, Yang, Huang, Yang, Majumder, and Wei}]{wang2023improving}
Liang Wang, Nan Yang, Xiaolong Huang, Linjun Yang, Rangan Majumder, and Furu Wei. 2023.
\newblock Improving text embeddings with large language models.
\newblock \emph{arXiv preprint arXiv:2401.00368}.

\bibitem[{Wang et~al.(2024)Wang, Yang, Huang, Yang, Majumder, and Wei}]{wang2024multilingual}
Liang Wang, Nan Yang, Xiaolong Huang, Linjun Yang, Rangan Majumder, and Furu Wei. 2024.
\newblock Multilingual e5 text embeddings: A technical report.
\newblock \emph{arXiv preprint arXiv:2402.05672}.

\bibitem[{Wolf et~al.(2020)Wolf, Debut, Sanh, Chaumond, Delangue, Moi, Cistac, Rault, Louf, Funtowicz, Davison, Shleifer, von Platen, Ma, Jernite, Plu, Xu, Scao, Gugger, Drame, Lhoest, and Rush}]{wolf-etal-2020-transformers}
Thomas Wolf, Lysandre Debut, Victor Sanh, Julien Chaumond, Clement Delangue, Anthony Moi, Pierric Cistac, Tim Rault, Rémi Louf, Morgan Funtowicz, Joe Davison, Sam Shleifer, Patrick von Platen, Clara Ma, Yacine Jernite, Julien Plu, Canwen Xu, Teven~Le Scao, Sylvain Gugger, Mariama Drame, Quentin Lhoest, and Alexander~M. Rush. 2020.
\newblock \href {https://www.aclweb.org/anthology/2020.emnlp-demos.6} {Transformers: State-of-the-art natural language processing}.
\newblock In \emph{Proceedings of the 2020 Conference on Empirical Methods in Natural Language Processing: System Demonstrations}, pages 38--45, Online. Association for Computational Linguistics.

\bibitem[{Xiao et~al.(2024)Xiao, Liu, Zhang, Muennighoff, Lian, and Nie}]{cpack}
Shitao Xiao, Zheng Liu, Peitian Zhang, Niklas Muennighoff, Defu Lian, and Jian-Yun Nie. 2024.
\newblock \href {https://doi.org/10.1145/3626772.3657878} {C-pack: Packed resources for general chinese embeddings}.
\newblock In \emph{Proceedings of the 47th International ACM SIGIR Conference on Research and Development in Information Retrieval}, SIGIR '24, page 641–649, New York, NY, USA. Association for Computing Machinery.

\bibitem[{Xiong et~al.(2021)Xiong, Xiong, Li, Tang, Liu, Bennett, Ahmed, and Overwijk}]{xiong2021approximate}
Lee Xiong, Chenyan Xiong, Ye~Li, Kwok-Fung Tang, Jialin Liu, Paul~N. Bennett, Junaid Ahmed, and Arnold Overwijk. 2021.
\newblock \href {https://openreview.net/forum?id=zeFrfgyZln} {Approximate nearest neighbor negative contrastive learning for dense text retrieval}.
\newblock In \emph{International Conference on Learning Representations}.

\bibitem[{Zhang et~al.(2018)Zhang, Zhang, Wang, Guo, and Liu}]{zhang2018cmedqav2}
Sheng Zhang, Xin Zhang, Hui Wang, Lixiang Guo, and Shanshan Liu. 2018.
\newblock \href {https://doi.org/10.1109/ACCESS.2018.2883637} {Multi-scale attentive interaction networks for chinese medical question answer selection}.
\newblock \emph{IEEE Access}, 6:74061--74071.

\bibitem[{Zhang et~al.(2024)Zhang, Zhang, Long, Xie, Dai, Tang, Lin, Yang, Xie, Huang, Zhang, Li, and Zhang}]{zhang2024mgte}
Xin Zhang, Yanzhao Zhang, Dingkun Long, Wen Xie, Ziqi Dai, Jialong Tang, Huan Lin, Baosong Yang, Pengjun Xie, Fei Huang, Meishan Zhang, Wenjie Li, and Min Zhang. 2024.
\newblock mgte: Generalized long-context text representation and reranking models for multilingual text retrieval.

\bibitem[{Zhang et~al.(2021)Zhang, Ma, Shi, and Lin}]{zhang-etal-2021-mr}
Xinyu Zhang, Xueguang Ma, Peng Shi, and Jimmy Lin. 2021.
\newblock \href {https://aclanthology.org/2021.mrl-1.12} {Mr. {T}y{D}i: A multi-lingual benchmark for dense retrieval}.
\newblock In \emph{Proceedings of the 1st Workshop on Multilingual Representation Learning}, pages 127--137, Punta Cana, Dominican Republic. Association for Computational Linguistics.

\bibitem[{Zhang et~al.(2023)Zhang, Thakur, Ogundepo, Kamalloo, Alfonso-Hermelo, Li, Liu, Rezagholizadeh, and Lin}]{zhang2023miracl}
Xinyu Zhang, Nandan Thakur, Odunayo Ogundepo, Ehsan Kamalloo, David Alfonso-Hermelo, Xiaoguang Li, Qun Liu, Mehdi Rezagholizadeh, and Jimmy Lin. 2023.
\newblock Miracl: A multilingual retrieval dataset covering 18 diverse languages.
\newblock \emph{Transactions of the Association for Computational Linguistics}, 11:1114--1131.

\bibitem[{Zhuang et~al.(2024)Zhuang, Qin, Hui, Wu, Yan, Wang, and Bendersky}]{zhuang-etal-2024-beyond}
Honglei Zhuang, Zhen Qin, Kai Hui, Junru Wu, Le~Yan, Xuanhui Wang, and Michael Bendersky. 2024.
\newblock \href {https://doi.org/10.18653/v1/2024.naacl-short.31} {Beyond yes and no: Improving zero-shot {LLM} rankers via scoring fine-grained relevance labels}.
\newblock In \emph{Proceedings of the 2024 Conference of the North American Chapter of the Association for Computational Linguistics: Human Language Technologies (Volume 2: Short Papers)}, pages 358--370, Mexico City, Mexico. Association for Computational Linguistics.

\end{thebibliography}

\newpage

\appendix

\section*{Overview of Appendix}

\begin{itemize}
    \item Appendix~\ref{appendix_sec:benchmark_construction}: Details on Benchmark Construction.
    \item Appendix~\ref{appendix_sec:air_bench_datasets_all}: \airbench Datasets.
    \item Appendix~\ref{appendix_sec:air-bench_software}: \airbench Software.
    \item Appendix~\ref{appendix_sec:air_bench_examples}: \airbench Data Examples.
    \item Appendix~\ref{appendix_sec:evaluation_details}: Evaluation Details.
    \item Appendix~\ref{appendix_sec:more_experiment_results}: More Experiment Results.
\end{itemize}

\section{Details on Benchmark Construction}
\label{appendix_sec:benchmark_construction}

In this section, we provide more details on the construction of datasets in \airbench.

\subsection{Corpora Preparation}
\label{appendix_sec:copora_preparation}

\airbench currently covers two different tasks: QA and Long-Doc. For QA task, we directly use the real-world dataset as the corpus, such as Wikipedia, mC4\cite{mc4}, CC-News\cite{Hamborg2017}, etc. We filter out text that is either too short or too long and make a straightforward attempt to remove any information that names or uniquely identifies individuals, as well as any offensive content. For Long-Doc task, we first select one long document for each dataset, such as book, ArXiv paper, legal document, etc., and remove table of contents and references. Then we use the node parser\footnote{SimpleNodeParser: \url{https://github.com/run-llama/llama_index}} tool from LlamaIndex\cite{Liu_LlamaIndex_2022} to split the long document into fixed-size chunks\footnote{\texttt{chunk\_size=200, chunk\_overlap=50}} as the corpus. All corpora used in \airbench are available in Appendix~\ref{appendix_sec:air-bench_datasets}.

\subsection{Candidate Generation}
\label{appendix_sec:candidate_generation}

\subsubsection{Query Generation}

To diversify the generated queries, we consider the following attributes when designing the prompt.

\textbf{Query Length}. This refers to the length of the query. We consider four different categories based on word count: \textit{less than 5 words}, \textit{less than 10 words}, \textit{10 to 20 words}, and \textit{at least 20 words}. The ratio of the number of queries in these categories is 1:4:2:1.

\textbf{Query Type}. This refers to the type of the query. We consider three different types: \textit{question}, \textit{problem}, and \textit{claim}. Based on our observation, the ``problem'' type is usually more difficult than the ``question'' type. The ratio of the number of queries in these three types is 3:1:1. For Long-Doc task, considering that the chunks in the corpus are derived from the same long document, the topics of these chunks are highly related. Therefore, we only utilized two types for Long-Doc task: question and claim. For the ``claim'' type, we observe that when the claim is too short, it will become too ambiguous to be a high-quality query. Therefore, the query length for the ``claim'' type is only sampled from ``between 10 and 20 words'' and ``at least 20 words''.

\textbf{Information-based Type}. This refers to the type of the information used when formulating queries. We consider two different types: \textit{queries based on the overall information in the document}, and \textit{queries based on the partial information beyond the main topic of the document}. The ratio of the number of queries in these two types is 1:1.

\textbf{Expression Style}. This refers to the style of query formulation. The three attributes mentioned above are used in Step 4. In Step 5, we consider different types of expression styles, allowing the LLM to rewrite the queries using various styles, thereby enhancing the diversity of query formulations. There are seven different styles in total: \textit{concise}, \textit{casual}, \textit{informal}, \textit{formal}, \textit{professional}, \textit{complicated}, and \textit{academic}. During the rewriting process, the sampling probabilities for these styles are in the ratio of 5:3:3:1:1:1:1.

\subsubsection{Hard Negative Generation}

To improve the difficulty of the generated datasets, we prompt LLM to generate 3-7 hard negative documents based on the rewritten query and the original positive document. For Long-Doc task, considering the chunks are extracted from the same long document and some of them have been hard enough, we do not generate additional hard negatives. The statistics of the number of hard negatives in each dataset are available in Appendix~\ref{appendix_sec:air-bench_datasets}.

\subsection{Quality Control}
\label{appendix_sec:quality_control}

We present more details on how we use LLM as labeler to label the relevance, select the embedding model and multiple re-ranking models, and set the predetermined threshold for pre-labeling.

\textbf{Use LLM as labeler}. \citet{llm_accessors} demonstrated that LLMs like OpenAI's GPT-4 are as accurate as human labelers when generating high-quality golden labels for search system. \citet{zhuang-etal-2024-beyond} showed that incorporating fine-grained relevance labels into the prompt for LLM rerankers significantly improves their performance on zero-shot reranking. In our paper, we use GPT-4 as labeler with a 4-level relevance generation strategy. The prompt we used is shown in Table~\ref{tab:label_prompt}.

\begin{table}[!h]
    \centering
    \small
    \resizebox{0.49\textwidth}{!}{
    \begin{tabular}{l}
    \toprule
For the following query and document, judge whether the \\ document is relevant to the query. \\
\\
Query:\\
```\\
\{query\}\\
'''\\
\\
Document:\\
```\\
\{doc\}\\
'''\\
\\
Your output must be one of the following:\\
- 0: The document is not relevant to the query.\\
- 1: The document is superficially relevant but actually not\\ \ \ \ \ \ \ \ relevant to the query.\\
- 2: The document is somewhat relevant to the query.\\
- 3: The document is relevant to the query.\\
\\
Do not explain your answer in the output. Your output must\\ be a single number.\\
    \bottomrule
    \end{tabular}
    }
    \caption{Prompt used for LLM to label the relevance. \{query\} and \{doc\} are placeholders of query and document, respectively.}
    \label{tab:label_prompt}
\end{table}

\textbf{Embedding Model}. Considering that the corpora in \airbench are multilingual, we use bge-m3\footnote{\url{https://huggingface.co/BAAI/bge-m3}} as the embedding model to recall the top-1000 relevant documents.

\textbf{Multiple Re-ranking Models}. For the datasets in English and Chinese, we use the following three re-ranking models: bge-reranker-large\footnote{\url{https://huggingface.co/BAAI/bge-reranker-large}}, bce-reranker-base\_v1\footnote{\url{https://huggingface.co/maidalun1020/bce-reranker-base_v1}}, mmarco-mMiniLMv2-L12-H384-v1\footnote{\url{https://huggingface.co/nreimers/mmarco-mMiniLMv2-L12-H384-v1}}. For the datasets in the other languages, we use the following three re-ranking models: bge-reranker-v2-m3\footnote{\url{https://huggingface.co/BAAI/bge-reranker-v2-m3}}, mmarco-mMiniLMv2-L12-H384-v1, bge-reranker-v2-gemma\footnote{\url{https://huggingface.co/BAAI/bge-reranker-v2-gemma}}.

\textbf{Predetermined Threshold}. For the hard negative documents, we set the threshold to 20. For the other documents, we set the threshold to 10.

\subsection{Queries Split}

After the quality control stage, we split the generated queries into different sets. For QA task, we split the queries in each dataset into dev set and test set in a 1:4 ratio. For Long-Doc task, we select one dataset as the dev set for each domain, and remain other datasets as the test set. Refer to Appendix~\ref{appendix_sec:air-bench_datasets} for more details.

\section{\airbench Datasets}
\label{appendix_sec:air_bench_datasets_all}

\subsection{Specifications}
\label{appendix_sec:air-bench_datasets}

The available versions of \airbench are listed in Table~\ref{tab:available_versions}.

\begin{table}[!h]
    \centering
    \resizebox{0.49\textwidth}{!}{
        \begin{tabular}{c|c|c|c|c|c}
            \toprule
            \textbf{Version} & \textbf{Release Date} & \textbf{\#domains} & \textbf{\#languages} & \textbf{\#datasets} & \textbf{Statistics} \\
            \midrule
            24.04 & May 21, 2024 & 8 & 2 & 28 & Table~\ref{tab:air-bench_datasets_2404} \\
            24.05 & Oct 17, 2024 & 9 & 13 & 69 & Table~\ref{tab:air-bench_datasets_2405_1}-\ref{tab:air-bench_datasets_2405_5} \\
            \bottomrule
        \end{tabular}
    }
    \caption{Available versions of \airbench.}
    \label{tab:available_versions}
\end{table}

For each dataset, we use the same format as BEIR, i.e. corpus, queries and qrels\footnote{\textit{qrels} are the relevance labels for queries. The relevance label is 1 for the positive document, and 0 for the negative document.}, which are all available in the Hugging Face Hub\footnote{\hforglink} of \airbench. To avoid the possible data leakage, we keep the qrels in test splits private. For the qrels in dev splits, we make them public to enable the developers to perform evaluation by themselves.

As the initial version, \airbench 24.04 only covered 2 languages, English and Chinese. Additionally, each dataset in \airbench 24.04 only contains the test set, which means that the developers could not know the evaluation results until they submit their model's search results to the leaderboard. As for the latest version \airbench 24.05, we have covered 13 languages, and included dev set and test set. The golden labels of dev set are made public, and the golden labels of test set remain private. Furthermore, the corpus size of some datasets in \airbench 24.04 is too large (such as 6.7M for wiki\_en dataset and 2.4M for finance\_zh dataset in QA task), which makes the download of datasets and the evaluation of models relatively inefficient. Therefore, in \airbench 24.05, we trimmed the large corpora to maintain a corpus size of around 1M for each dataset.

For the detailed statistics of all datasets in \airbench 24.04 and 24.05, please refer to Table~\ref{tab:air-bench_datasets_2404} and Table~\ref{tab:air-bench_datasets_2405_1}-\ref{tab:air-bench_datasets_2405_5}, respectively. Note that we use the tokenizer\footnote{\url{https://github.com/openai/tiktoken}} of OpenAI's GPT-4o\footnote{\url{https://openai.com/index/hello-gpt-4o/}} to count the token number for every language.

\subsection{Licenses}

In Table~\ref{tab:air-bench_datasets_2404}-\ref{tab:air-bench_datasets_2405_5}, we also list the licenses of the source corpora used for the dataset generation in \airbench. All generated testing data in \airbench is licensed under CC BY-NC-SA-4.0\footnote{\url{https://creativecommons.org/licenses/by-nc-sa/4.0}}. The testing data in \airbench may only be used for evaluation purposes.

\subsection{Additional Diversity Analysis}
\label{appendix_sec:diversity_analysis}

We provide more analysis of diversity to better characterize \airbench.

\begin{table}[!t]
    \centering
    \small
    \setlength{\tabcolsep}{6pt}
    \begin{tabular}{l|rrrr}
        \toprule
          & \multicolumn{2}{c}{\textbf{QA}} & \multicolumn{2}{c}{\textbf{Long-Doc}} \\
         \midrule
         split $\rightarrow$ & \multicolumn{1}{c}{dev} & \multicolumn{1}{c}{test} & \multicolumn{1}{c}{dev} & \multicolumn{1}{c}{test} \\
         \# of datasets $\rightarrow$ & \multicolumn{1}{c}{54} & \multicolumn{1}{c}{53} & \multicolumn{1}{c}{4} & \multicolumn{1}{c}{11} \\
        \midrule
        \multicolumn{5}{l}{\textit{Query Style}} \\
        \midrule
        \textsc{formal} & 31.3\% & 35.3\% & 17.7\% & 17.9\% \\
        \textsc{informal} & 44.3\% & 44.7\% & 28.3\% & 27.4\% \\
        \textsc{professional} & 7.0\% & 6.8\% & 8.2\% & 10.3\% \\
        \textsc{casual} & 0.8\% & 0.7\% & 0.5\% & 0.6\% \\
        \textsc{complicated} & 0.5\% & 0.3\% & 0.7\% & 1.4\% \\
        \textsc{concise} & 8.3\% & 5.4\% & 12.3\% & 12.5\% \\
        \textsc{academic} & 7.8\% & 6.8\% & 32.2\% & 29.8\% \\
        \textsc{others} & < 0.1\% & < 0.1\% & 0.1\% & 0.1\% \\
        \bottomrule
    \end{tabular}
    \caption{The style distribution of queries in each split for each task in \airbench 24.05.}
    \label{tab:query_style_diversity}
    \vspace{-10pt}
\end{table}

\begin{figure*}[!t]
\vspace{-1cm}
\centering
\includegraphics[width=1.0\textwidth]{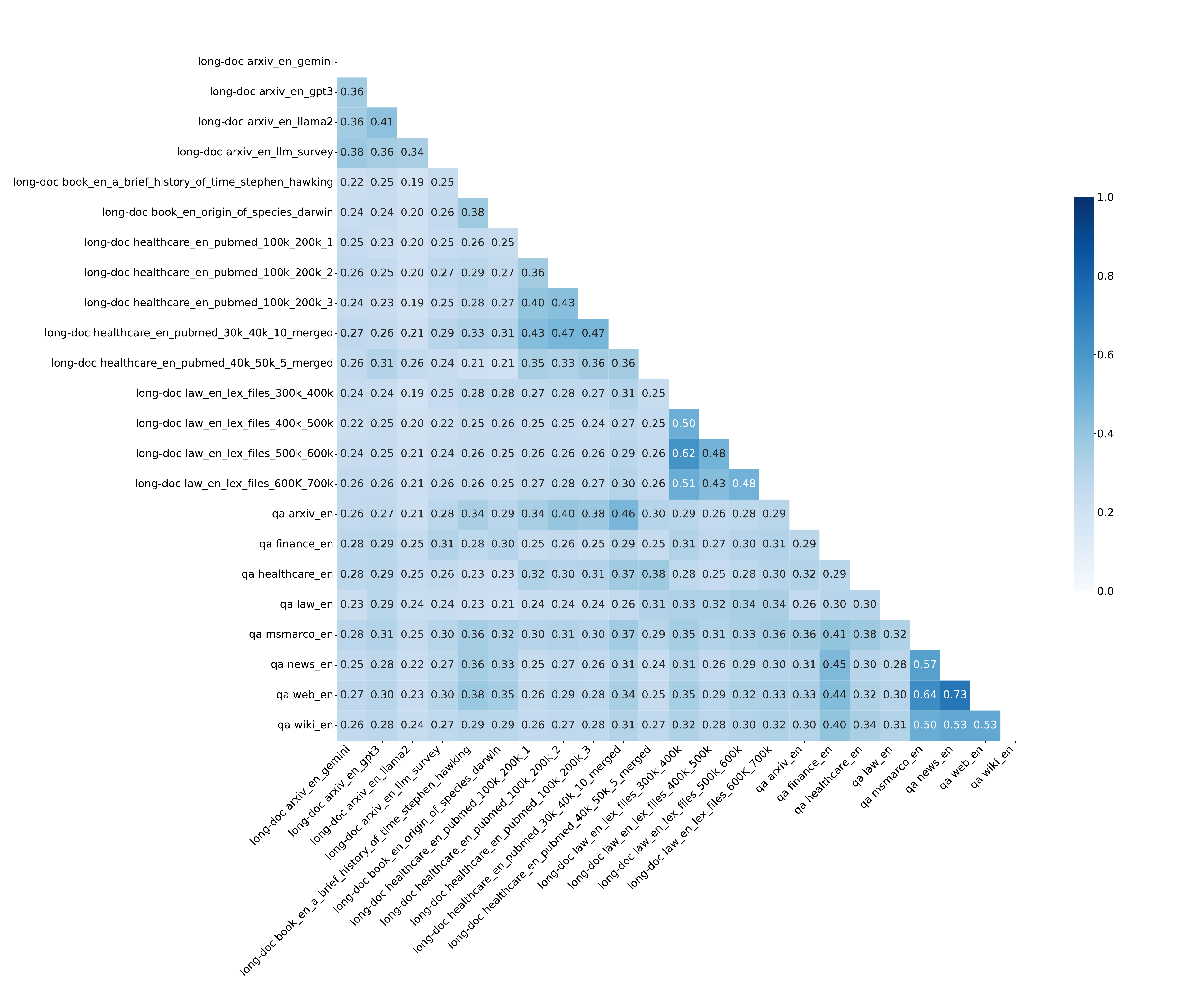}
\vspace{-10pt}
\caption{Pairwise weighted Jaccard similarity scores between \airbench English datasets. We use the tokenizer of GPT-4o to tokenize the corpus of each dataset.}
\label{fig:corpus_similarity}
\vspace{-10pt}
\end{figure*}

\subsubsection{Query Diversity}

We also analyze the \textbf{style diversity} of the generated queries in \airbench. We still utilize GPT-4o\footnote{
gpt-4o-2024-08-06: 

\url{https://platform.openai.com/docs/models/gpt-4o}}
as labeler to label the style of the queries in \airbench. The optional query styles include: \textsc{formal}, \textsc{informal}, \textsc{professional}, \textsc{casual}, \textsc{complicated}, \textsc{concise}, \textsc{academic}, and \textsc{others}. The statistics are grouped by tasks and splits in Table~\ref{tab:query_style_diversity}. 

We can make the following observations according to the results. First of all, since the optional styles given to GPT-4o are not mutually exclusive, the ratio of the number of different styles is not consistent with the ratio we set in the generation stage (Step 5 of the Candidate Generation stage). Secondly, the QA task tends to have more \textsc{informal} queries, and Long-Doc task tends to have more \textsc{academic} queries, which may be due to the fact that the long documents in the Long-Doc task are more academic related, such as ArXiv papers, books, etc. Finally, \textsc{professional} queries and \textsc{complicated} queries account for a certain portion, which means that some queries in \airbench are probably challenging for IR models.

\subsubsection{Corpus Diversity}

Following the work of BEIR~\cite{thakur2021beir}, we compute the pairwise weighted Jaccard similaity scores between the datasets in \airbench. Considering that there are 69 datasets in total, we only present the results of datasets in English here. As shown in Figure~\ref{fig:corpus_similarity}, we can observe that the corpora from different domains have a low weighted Jaccard similarity word overlap, indicating that \airbench is a challenging benchmark where the IR methods must generalize well to diverse out-of-distribution domains.

\section{\airbench Software}
\label{appendix_sec:air-bench_software}

The \airbench software\footnote{\repolink} makes it convenient for the evaluation of any information retrieval methods. With the provided Python framework, in order to evaluate a retrieval method, users only need to implement a \texttt{Retriever} that takes the queries and the corpus as the inputs, and returns the top-$k$ relevant documents for each query as the outputs. If the users want to evaluate the performance of retrieval-then-reranking method, they only need to additionally implement a \texttt{Reranker}, which takes the queries, the corpus, and the top-$k$ search results from \texttt{Retriever} as the inputs, and returns the re-ranked top-$k'$ ($k' \leq k$) relevant documents as the outputs.

We also maintain a Hugging Face leaderboard\footnote{\leaderboardlink} with all datasets and models. To make the leaderboard more readable, we classify the submissions into three categories: 1) \textbf{Retrieval Only}. It means that this submission only uses a specific retrieval method to generate the top-$k$ search results. 2) \textbf{Reranking Only}. It means that this submission uses BM25 as the retrieval method and then uses a specific reranking method to re-rank the search results from BM25 to generate the re-ranked top-$k$ search results. 3) \textbf{Retrieval+Reranking}. It means that this submission first uses a specific retrieval method to generate the top-$k$ search results, and then uses a specific reranking method to re-rank to get the final search results. It should be noted that our leaderboard only maintain the evaluation results for the test splits, and the evaluations results for the dev splits will be available on the MTEB leaderboard\footnote{\url{https://huggingface.co/spaces/mteb/leaderboard}}.

To facilitate the evaluation of existing IR models, we also develop the evaluation scripts based on two mainstream architectures: HuggingFace Transformers\footnote{\url{https://github.com/huggingface/transformers}}~\cite{wolf-etal-2020-transformers} and Sentence Transformers\footnote{\url{https://github.com/UKPLab/sentence-transformers}}~\cite{reimers-2019-sentence-bert}. These scripts are all available in our repository\footnote{\repolink}.

\section{\airbench Data Examples}
\label{appendix_sec:air_bench_examples}

We list some examples of the generated testing data in Table~\ref{tab:generated_examples_1}-\ref{tab:generated_examples_3}.

\section{Evaluation Details}
\label{appendix_sec:evaluation_details}

\subsection{Models}

For detailed information of the models appearing in this paper, please refer to Table~\ref{tab:model_information}. For the BM25 method, we employ the implementation from Pyserini\footnote{\url{https://github.com/castorini/pyserini}}~\cite{lin2021pyserini}. For the evaluation of BM25-based re-ranking models, we evaluate the performance by re-ranking the top-100 search results from BM25 with the re-ranking models.

The models used in this paper are all publicly available (see Table~\ref{tab:model_information} for the public link). We confirm that we did not violate the license of any model used in our paper.

\subsection{Parameters}

When performing evaluation, we set the max length of both query and passage to 512 tokens. If the embedding models need task specific instruction, such as \texttt{e5-mistral-7b-instruct}~\cite{wang2023improving}, \texttt{SFR-Embedding-Mistral}, etc., we use the same instruction for all datasets: ``Given a question, retrieve passages that answer the question'', which is denoted as Instr-1. Considering that the queries in \airbench include both questions and claims, we also evaluate the performance of \texttt{e5-mistral-7b-instruct} with a more reasonable but more complex instruction: ``Given a question or claim, retrieve passages that answer the question or support the claim'', which is denoted as Instr-2. However, as shown in Table~\ref{tab:eval_para_comparison}, the performance of e5-mistral-7b-instruct using Instr-2 is slightly worse than that using Instr-1, which may indicate that current models are not yet able to adapt well to more complex instruction.

For the total computational budget, we did not perform detailed statistics. However, based on our estimates, all evaluations in this paper required approximately 2000 GPU hours using 24 A800 (80GB) GPUs.

\begin{table*}[!ht]
    \centering
    \small
    \begin{tabular}{l|ccc}
        \toprule
         & \textbf{QA (English, test)} & \textbf{QA (Multilingual, test)} & \textbf{Long-Doc (English, test)} \\
        \# of datasets $\rightarrow$ & 7 datasets & 53 datasets & 11 datasets \\
        \midrule
        e5-mistral-7b-instruct (Instr-1) & 49.880 & 48.077 & 66.908 \\
        e5-mistral-7b-instruct (Instr-2) & 49.252 & 47.772 & 66.766 \\
        \bottomrule
    \end{tabular}
    \caption{Comparison of performances when using different evaluation parameters on \airbench. The metric for QA task is nDCG@10, and the metric for Long-Doc task is Recall@10.}
    \label{tab:eval_para_comparison}
\end{table*}

\section{More Experiment Results}
\label{appendix_sec:more_experiment_results}

\subsection{Distinguishing Models}
\label{appendix_sec:distinguishing_models}

We evaluate a diverse set of IR models on \airbench to demonstrate its capability of distinguishing different models from multiple dimensions: model type, domain, language.

\begin{figure*}[!ht]
    \centering
    \begin{subfigure}[t]{0.49\textwidth}
        \centering
        \includegraphics[width=\textwidth]{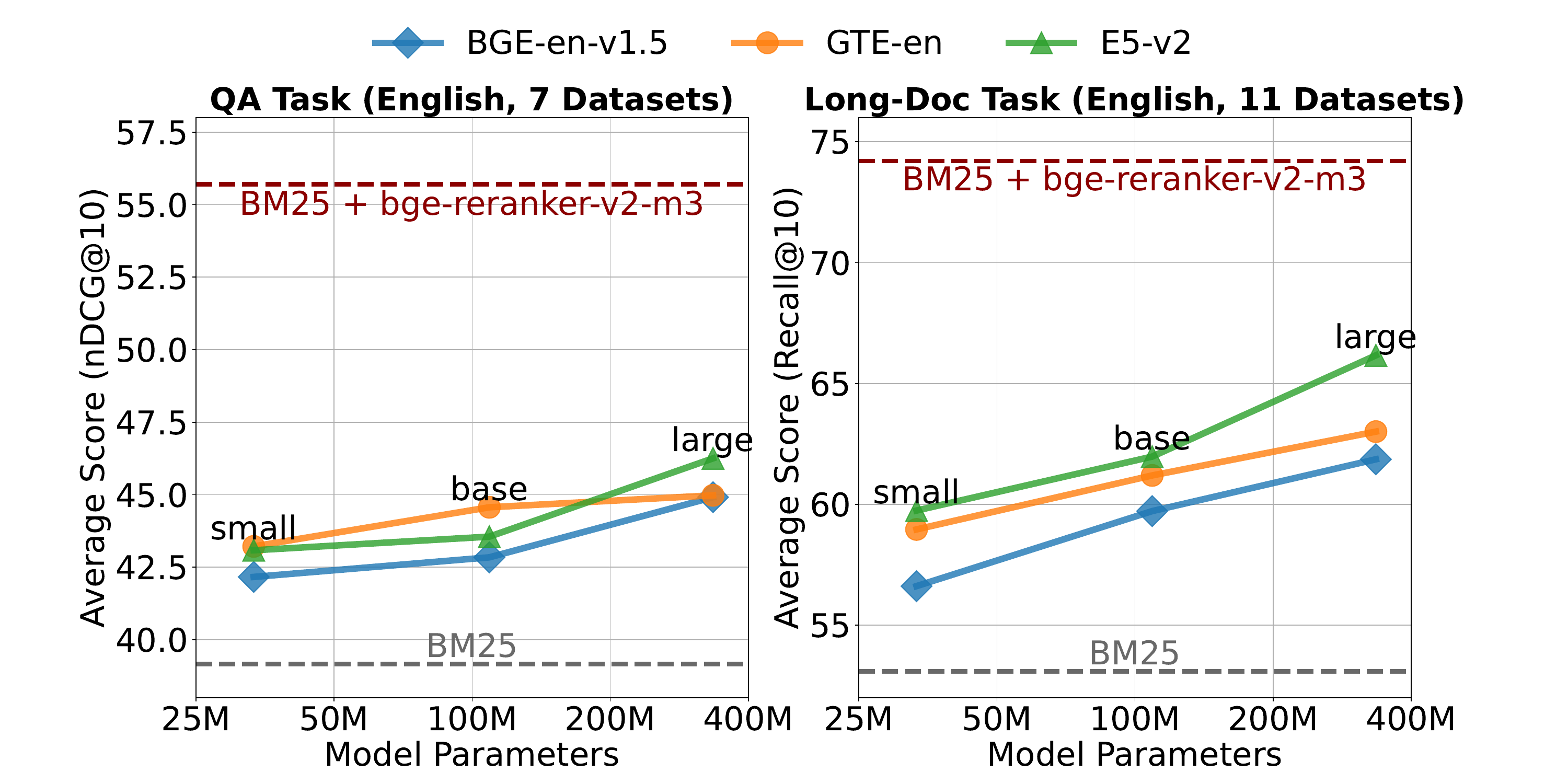}
        \caption{\textbf{Model dimension} comparison results (\textbf{English}).}
        \label{fig:model_dimension_en}
    \end{subfigure}
    \hfill
    \begin{subfigure}[t]{0.49\textwidth}
        \centering
        \includegraphics[width=\textwidth]{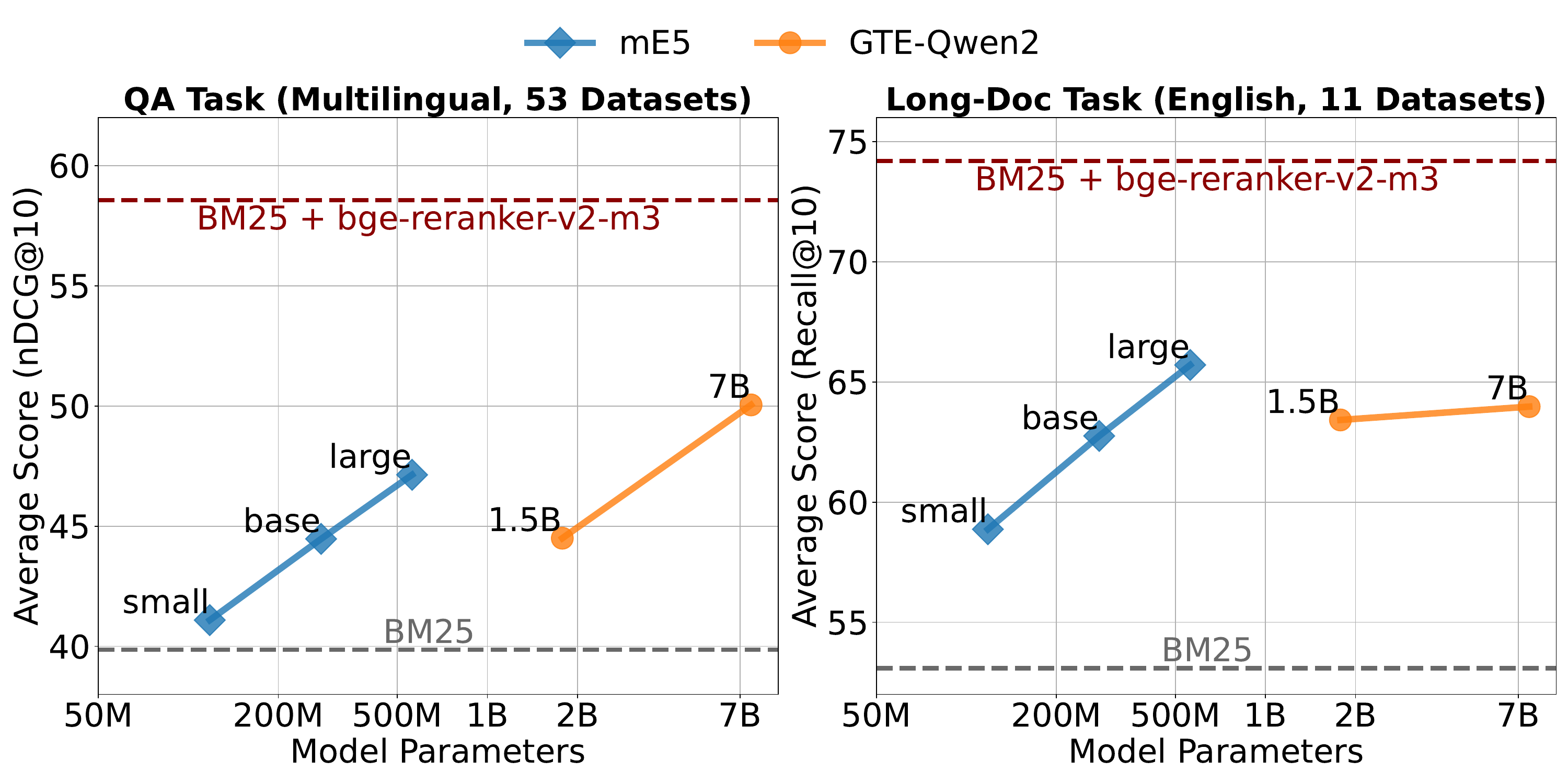}
        \caption{\textbf{Model dimension} comparison results (\textbf{Multilingual}).}
        \label{fig:model_dimension_multilingual}
    \end{subfigure}

    \vspace{5pt}
    
    \begin{subfigure}[t]{0.49\textwidth}
        \centering
        \includegraphics[width=\textwidth]{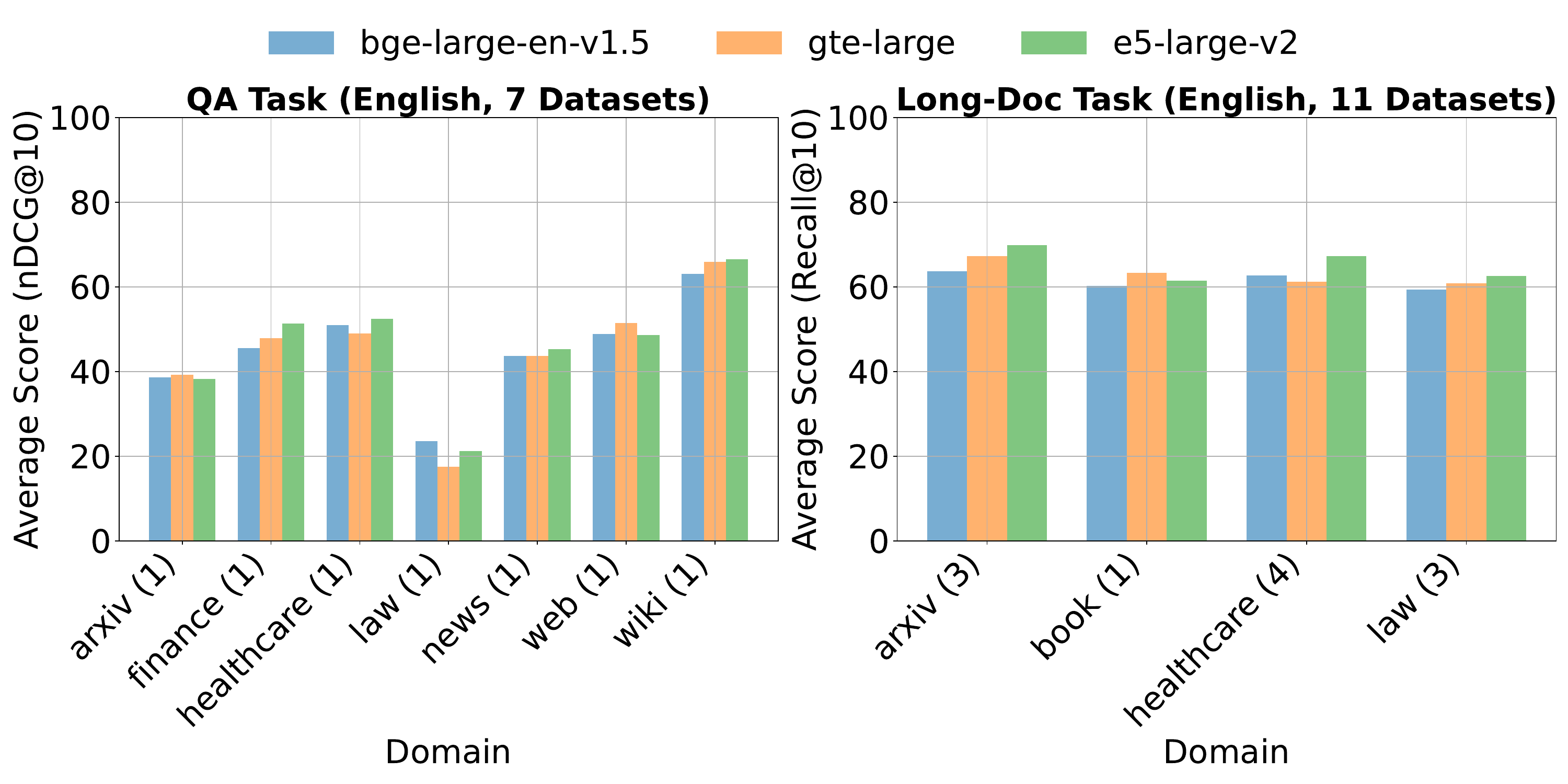}
        \caption{\textbf{Domain dimension} comparison results (\textbf{English}, \textit{large-size} embedding models).}
        \label{fig:domain_dimension_en}
    \end{subfigure}
    \hfill
    \begin{subfigure}[t]{0.49\textwidth}
        \centering
        \includegraphics[width=\textwidth]{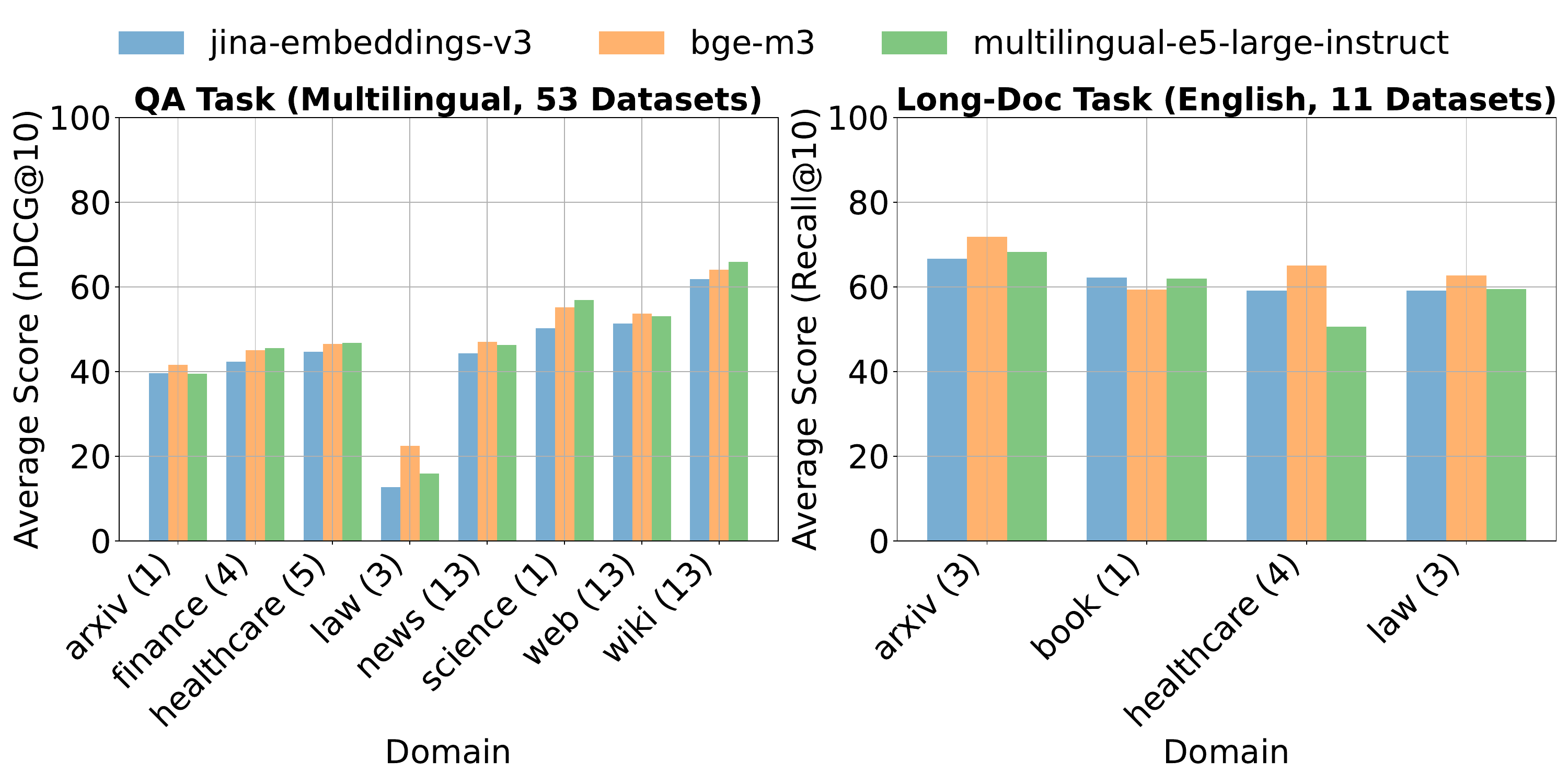}
        \caption{\textbf{Domain dimension} comparison results (\textbf{Multilingual}, \textit{large-size} embedding models).}
        \label{fig:domain_dimension_multilingual}
    \end{subfigure}

    \vspace{5pt}

    \begin{subfigure}[t]{0.98\textwidth}
        \centering
        \includegraphics[width=\linewidth]{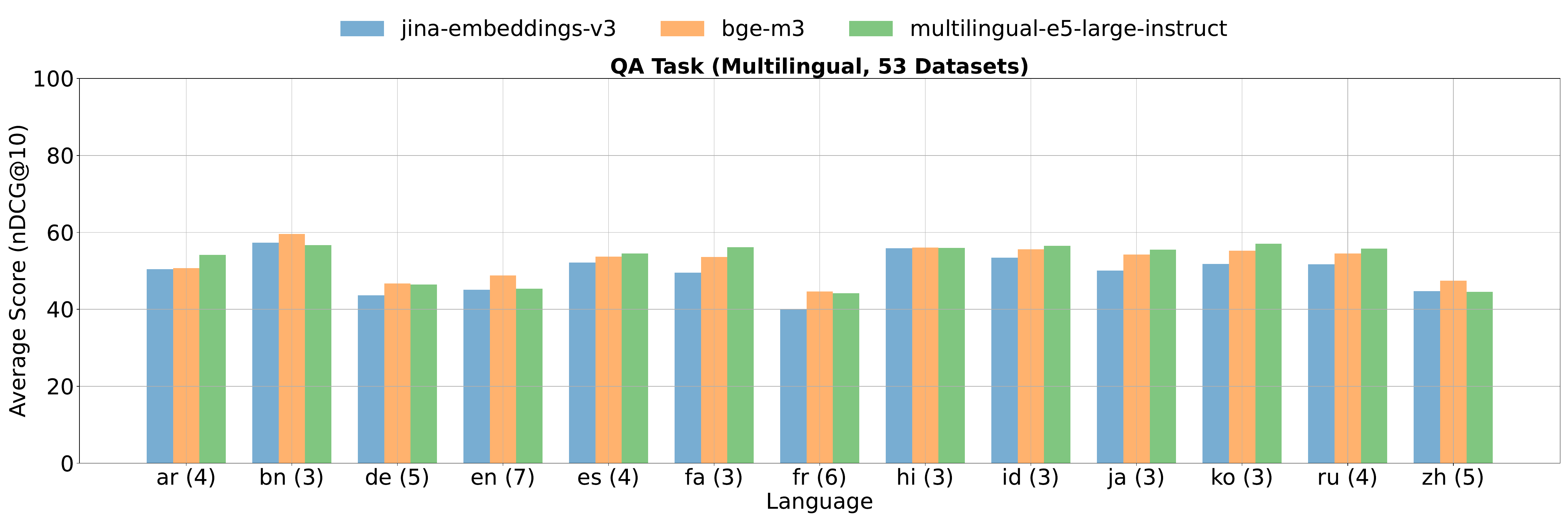}
        \caption{\textbf{Language dimension} comparison results (\textbf{Multilingual}, \textit{large-size} embedding models).}
        \label{fig:language_dimension}
    \end{subfigure}
    
    \vspace{-10pt}
    \caption{\airbench can distinguish models in different dimensions, including model dimension, domain dimension, and language dimension. For detailed information of the models appearing in this figure, please refer to Table~\ref{tab:model_information}. The detailed metric value and additional results on other model size are all available in Appendix~\ref{appendix_sec:detailed_experiment_results}.}
    \label{fig:distinguishing_models}
    \vspace{-10pt}
\end{figure*}

\textbf{Model Type}. As shown in Figure~\ref{fig:model_dimension_en} and Figure~\ref{fig:model_dimension_multilingual}, we can observe the following three points on both QA task and Long-Doc task, regardless of whether the datasets are only in English or multilingual: 1) BM25 performs worse than all embedding models. 2) BM25 + bge-reranker-v2-m3 achieves more excellent performance than all of the embedding models. 3) The performance of embedding models from the same series scales with model size.
% For the English datasets, we select three popular series, including BGE-en-v1.5-\{small/base/large\}, E5-v2-\{small/base/large\} and GTE-en-\{small/base/large\}, and for the multilingual datasets, we select two popular series, including mE5-\{small/base/large\} and GTE-Qwen2-\{1.5B/7B\}. 

\textbf{Domain}. We evaluate three kinds of embedding models with the same model size (\textit{large-size}), and compare their performances in each domain on \airbench. As shown in Figure~\ref{fig:domain_dimension_en} and Figure~\ref{fig:domain_dimension_multilingual}, regardless of whether the task is QA or Long-Doc and whether the datasets are only in English or multilingual, no model is able to achieve the best performance on all domains.

\textbf{Language}. We evaluate three kinds of embedding models with the same model size (\textit{large-size}) on the multilingual datasets of \airbench, and compare their performance on the datasets of each language. As shown in Figure~\ref{fig:language_dimension}, we also observe that no model is able to achieve the best performance on all languages.

\begin{figure*}[!ht]
    \centering
    \begin{subfigure}[t]{0.49\textwidth}
        \centering
        \includegraphics[width=\textwidth]{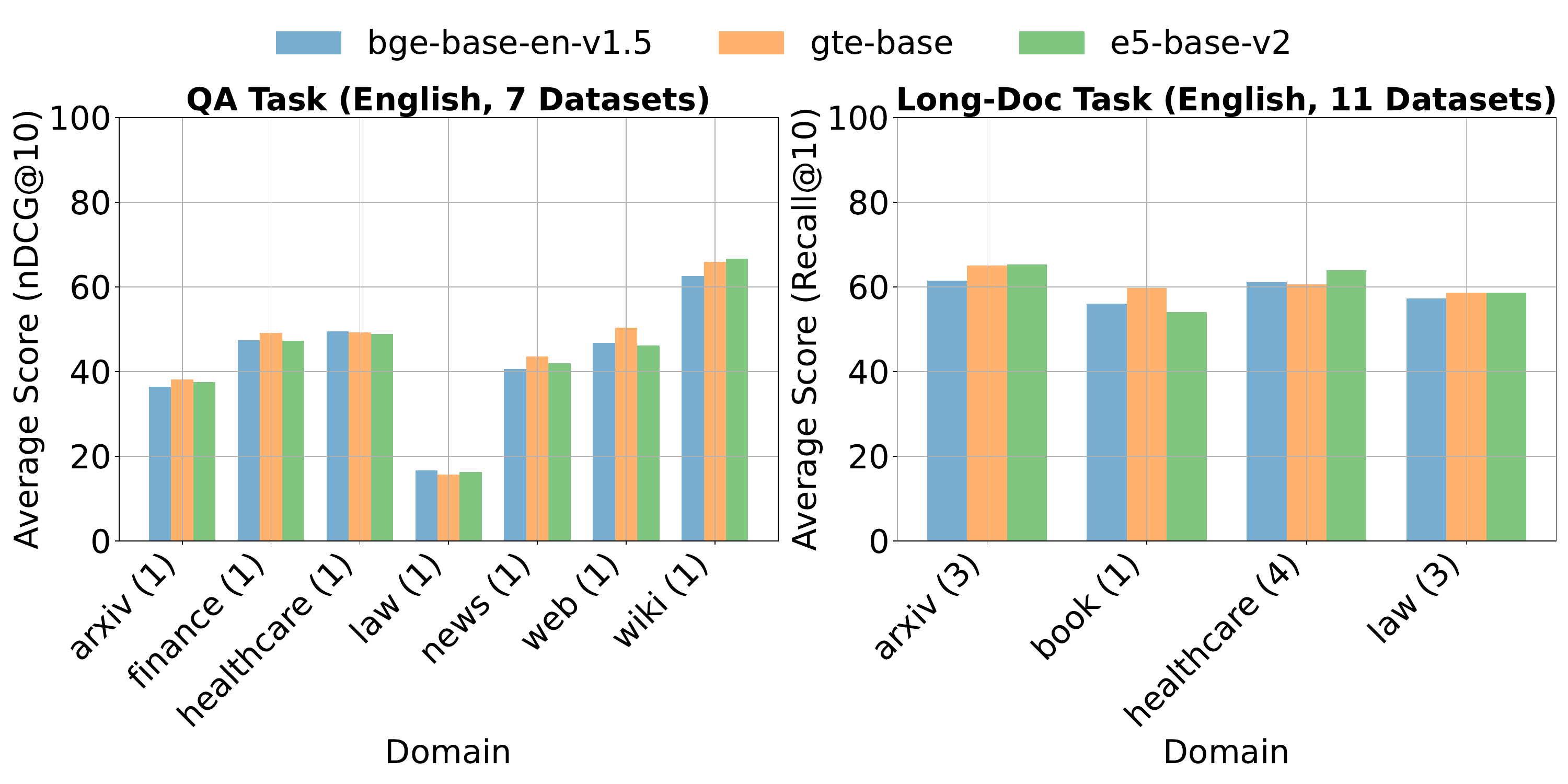}
        \caption{\textbf{Domain dimension} comparison results (\textbf{English}, \textit{base-size} embedding models).}
        \label{fig:domain_dimension_en_base}
    \end{subfigure}
    \hfill
    \begin{subfigure}[t]{0.49\textwidth}
        \centering
        \includegraphics[width=\textwidth]{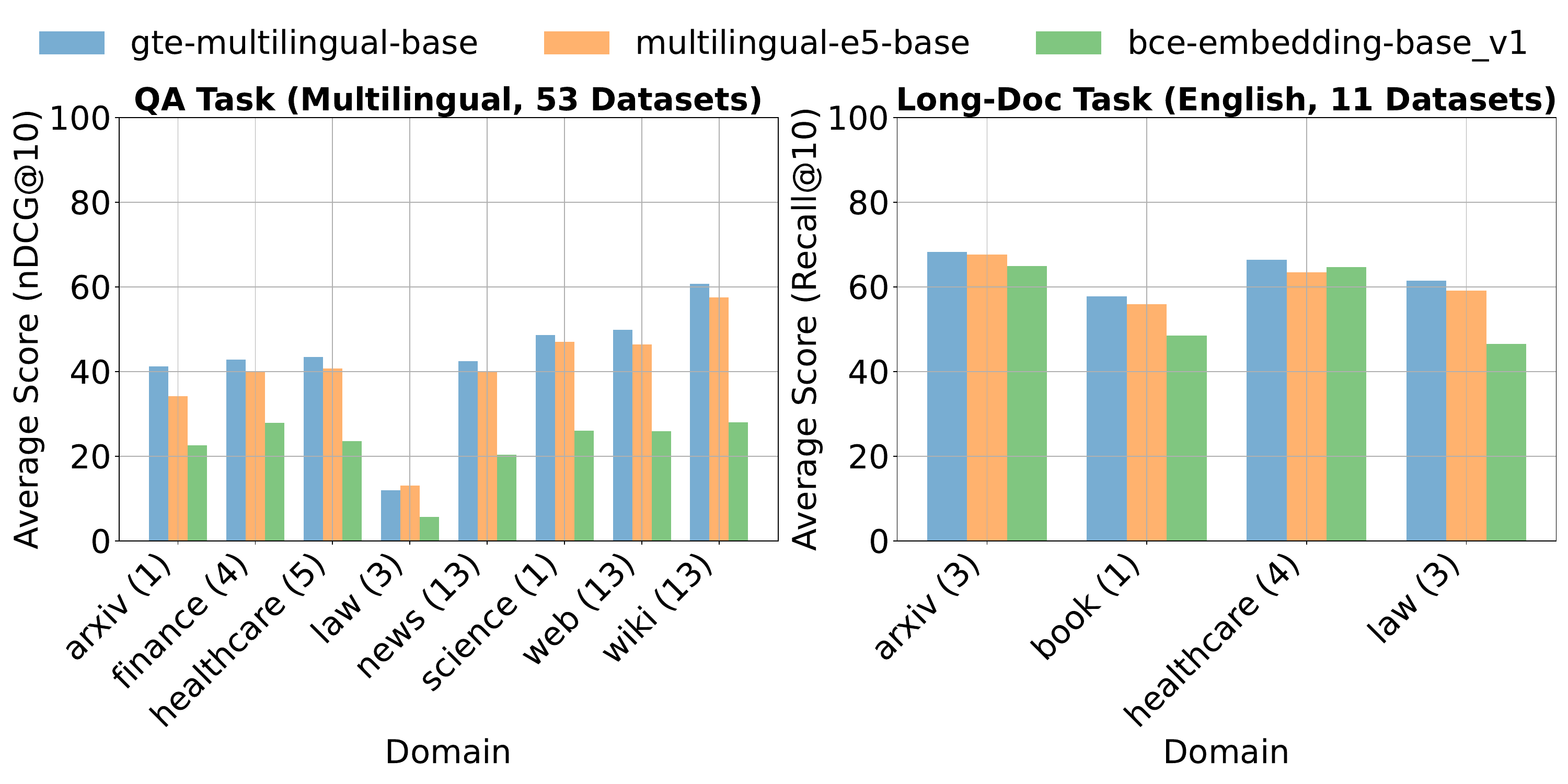}
        \caption{\textbf{Domain dimension} comparison results (\textbf{Multilingual}, \textit{base-size} embedding models).}
        \label{fig:domain_dimension_multilingual_base}
    \end{subfigure}

    \vspace{5pt}

    \begin{subfigure}[t]{0.49\textwidth}
        \centering
        \includegraphics[width=\textwidth]{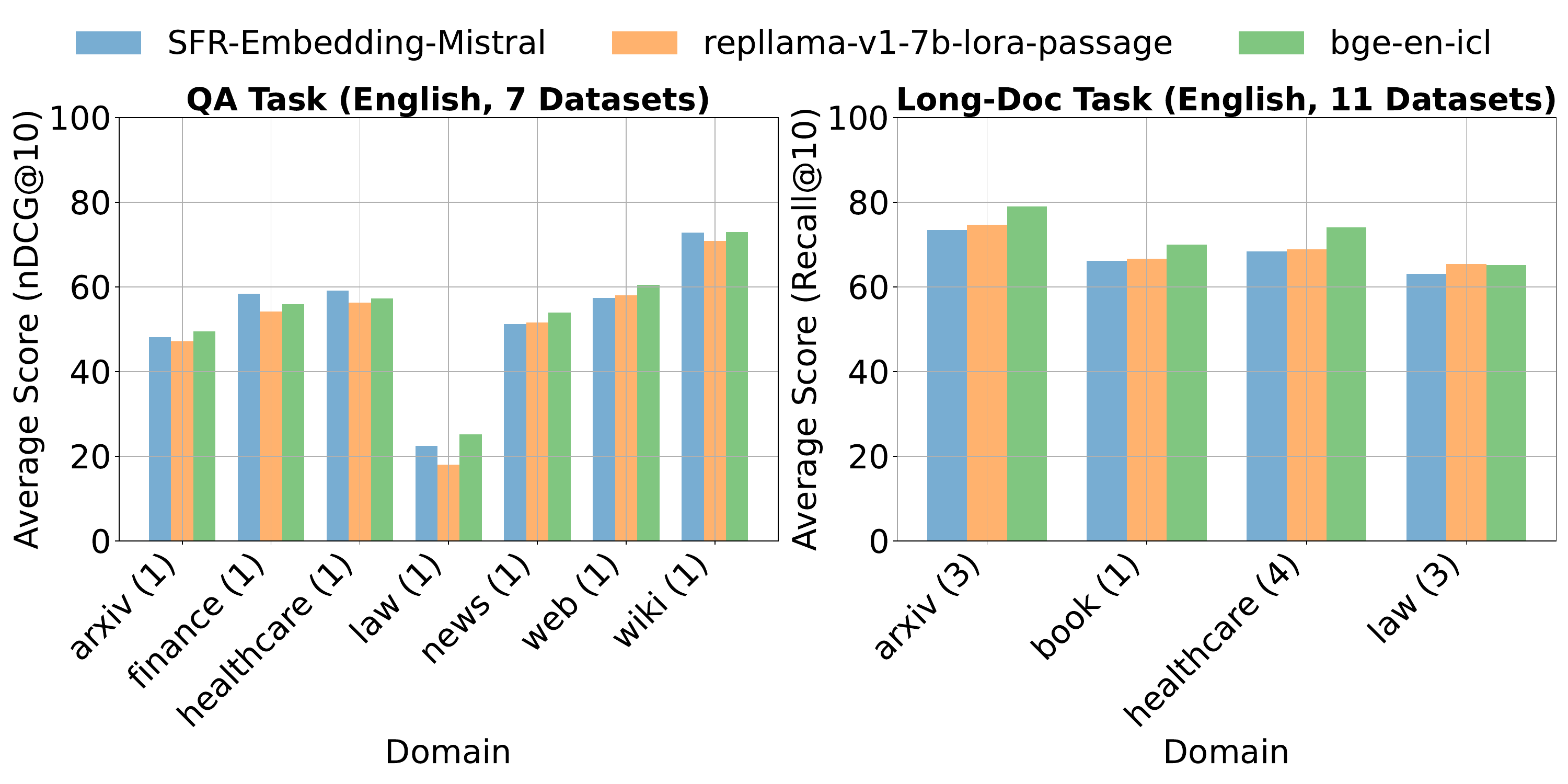}
        \caption{\textbf{Domain dimension} comparison results (\textbf{English}, \textit{LLM-based} embedding models).}
        \label{fig:domain_dimension_en_llm}
    \end{subfigure}
    \hfill
    \begin{subfigure}[t]{0.49\textwidth}
        \centering
        \includegraphics[width=\textwidth]{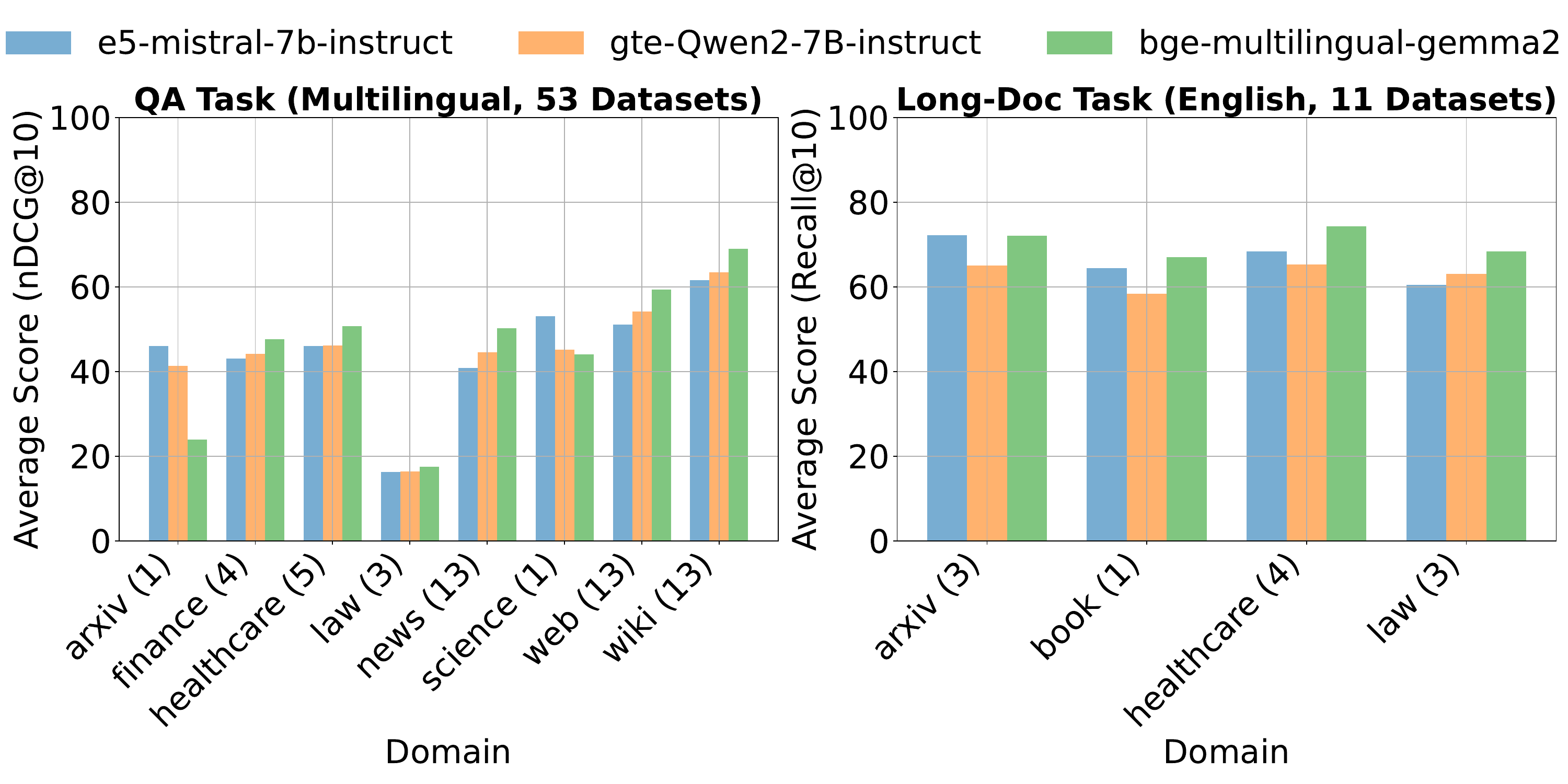}
        \caption{\textbf{Domain dimension} comparison results (\textbf{Multilingual}, \textit{LLM-based} embedding models).}
        \label{fig:domain_dimension_multilingual_llm}
    \end{subfigure}

    \vspace{5pt}

    \begin{subfigure}[t]{0.98\textwidth}
        \centering
        \includegraphics[width=\linewidth]{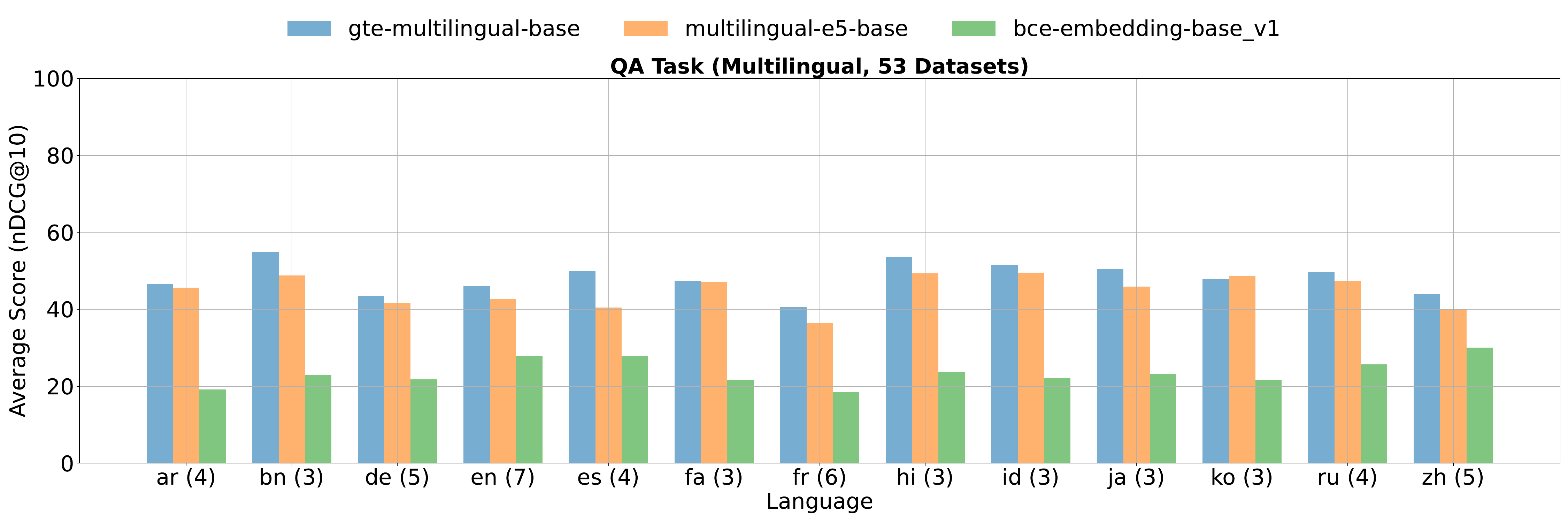}
        \caption{\textbf{Language dimension} comparison results (\textbf{Multilingual}, \textit{base-size} embedding models).}
        \label{fig:language_dimension_base}
    \end{subfigure}

    \vspace{5pt}

    \begin{subfigure}[t]{0.98\textwidth}
        \centering
        \includegraphics[width=\linewidth]{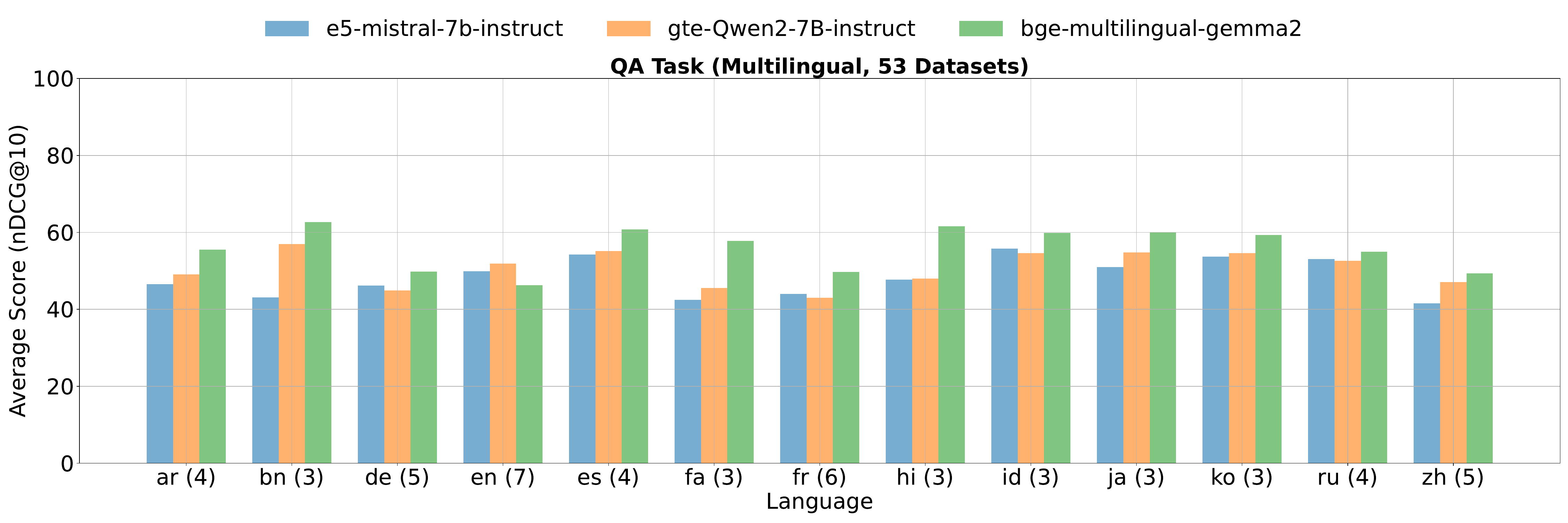}
        \caption{\textbf{Language dimension} comparison results in \textbf{multilingual} datasets (\textit{LLM-based} embedding models).}
        \label{fig:language_dimension_llm}
    \end{subfigure}
    
    \vspace{-10pt}
    \caption{Additional results indicating that \airbench can distinguish models in different dimensions. For detailed information of the models appearing in this figure, please refer to Table~\ref{tab:model_information}.}
    \label{fig:distinguishing_models_additional}
    \vspace{-10pt}
\end{figure*}

Apart from the results of large-size embedding models in Figure~\ref{fig:distinguishing_models}, we also perform investigation with base-size embedding models and LLM-based embedding models. The additional results are shown in Figure~\ref{fig:distinguishing_models_additional}.

\subsection{Detailed Evaluation Results}
\label{appendix_sec:detailed_experiment_results}

In this section, we provide the detailed evaluation results of each model on \airbench 24.05. Table~\ref{tab:detailed_en_results} presents the detailed evaluation results of English IR models on \airbench 24.05. Table~\ref{tab:detailed_multilingual_results} presents the detailed evaluation results of multilingual IR models on \airbench. For detailed information of the models appearing in these tables, please refer to Table~\ref{tab:model_information}.

\newpage

\begin{table*}[!h]
    \centering
    \small
    \resizebox{0.98\textwidth}{!}{
    \begin{tabular}{>{\raggedright\arraybackslash}p{0.9\textwidth}}
    \toprule
    \textbf{Domain:} news; \textbf{Language:} English  \\
    \textbf{Original Positive:} \\
    ``It’s hard to think of a part of the world that hasn’t been touched by robotic advances this year.\textbackslash n In 2016, strides were taken in the areas of robotic home delivery, cooking, tough terrain navigation and even attempts to conquer the beautiful game of football.\textbackslash n Here are the top five robots of the year.\textbackslash n While we’re not quite at the singularity yet, more sophisticated automation is an inevitability of the future.\textbackslash n The strides in Artificial Intelligence (AI) over the past decade have been huge, so expect to see a lot more in this area in the coming years.\textbackslash n We just hope the tech guys making super AI fit it with an “off” switch so it can be unplugged when it wants to, you know, take over the world and destroy everything.'' \\
    \midrule
    \textbf{Character:} Robotics Engineer \\
    \midrule
    \textbf{Scenario:} Preparing a presentation on the yearly advancements in robotics technology. \\
    \midrule
    \textbf{Original Query:} Which industries implemented robotic home delivery? \\
    \textbf{Rewritten Query:} In which sectors has the implementation of autonomous delivery robots for residential services been observed? \\
    \midrule
    \textbf{Hard Negative 1:} \\
    ``Autonomous technologies have been expanding rapidly across various industries, with drones making headway in aerial inspections and surveillance. Companies are investing in autonomous flight for package delivery, but primarily in commercial settings. The convenience and efficiency improvements in logistics are undeniable, but residential use isn't widespread yet.'' \\
    \textbf{Hard Negative 2:} \\
    ``Residential sectors are increasingly relying on technology, with smart homes integrating systems for automated cleaning, energy management and advanced security. These innovations in domestic tech have redefined the way we live, promising a future where household chores are managed seamlessly through digital interfaces and remote controls.'' \\
    \textbf{Hard Negative 3:} \\
    ``Recent developments in the robotics industry have witnessed significant progress in various sectors, such as industrial manufacturing, precision agriculture, and automated warehousing solutions. These robots have revolutionized production efficiency, crop management, and inventory control, enhancing economic output.'' \\
    \textbf{Hard Negative 4:} \\
    ``In recent years, residential areas have seen an uptick in smart home innovations that include automated climate control, security systems with facial recognition, and voice-activated appliances. The integration of AI in household management has significantly enhanced the convenience and efficiency of daily living.'' \\
    \textbf{Hard Negative 5:} \\
    ``Experts predict an expansion in the use of unmanned vehicles for military logistics and combat support missions. The autonomous systems being developed are designed for supply transport, surveillance, and even tactical offense, set to revolutionize battlefield strategies in the near future.'' \\
    \bottomrule
    \end{tabular}
    }
    \caption{Random sampled examples for the generated testing data. Domain: news, Language: English.}
    \label{tab:generated_examples_1}
\end{table*}

\begin{table*}[!h]
    \centering
    \small
    \resizebox{0.98\textwidth}{!}{
    \begin{tabular}{>{\raggedright\arraybackslash}p{0.9\textwidth}}
    \toprule
    \textbf{Domain:} healthcare; \textbf{Language:} English  \\
    \textbf{Original Positive:} \\
    ``Only two patients, 5 and 12 years old, with primary gastric NHL were found. Upper gastroduodenal endoscopy detected an ulcer in the lesser curvature of the body of the stomach, in both cases. Endoscopy revealed a moderate chronic gastritis in the antrum of both patients that was H. pylori associated in one of them who also suffered from chronic gastritis. Biopsy specimens demonstrated infiltration by Burkitt lymphoma (BL). The two patients received chemotherapy for 6 months. Additionally, one of the two patients received a triple therapy regimen with bismuth, amoxicillin, and metronidazole for H. pylori. Fifteen and six years later they are in complete remission, free of symptoms.'' \\
    \midrule
    \textbf{Character:} College student \\
    \midrule
    \textbf{Scenario:} Creating a presentation on the clinical manifestations and treatment outcomes of primary gastric non-Hodgkin's lymphoma in pediatric patients. \\
    \midrule
    \textbf{Original Query:} How long did the pediatric patients receive chemotherapy for primary gastric NHL? \\
    \textbf{Rewritten Query:} How long were the kids treated with chemo for their stomach lymphoma? \\
    \midrule
    \textbf{Hard Negative 1:} \\
    ``In a recent clinical review, five pediatric cases of gastrointestinal complaints were assessed. The patients, ranging in age from 3 to 14 years, presented with various symptoms including abdominal pain, vomiting, and weight loss. In-depth medical evaluations, including blood tests, abdominal ultrasonography, and, for three patients, an upper gastroduodenal endoscopy, were conducted. The endoscopic examination in these three patients showed mild inflammation in the stomach lining and superficial gastric erosions in the antrum and the lesser curvature. None of the patients had a history of gastric malignancies, and there were no indications of Non-Hodgkin Lymphoma (NHL) or any other types of cancer. Helicobacter pylori infection was not detected in any of the cases. The patients' symptoms were managed with dietary modifications and antacid medications. Symptom relief was noted in all cases, and follow-up visits over the course of six months revealed significant improvement and no further gastrointestinal issues. The clinical team concluded that the symptoms were likely due to functional dyspepsia and emphasized the importance of considering less severe diagnoses when pediatric patients present with gastrointestinal symptoms.'' \\
    \textbf{Hard Negative 2:} \\
    ``Two young individuals, aged 6 and 11, presented with abdominal discomfort and were subsequently screened for gastrointestinal disorders. Initial evaluation through pediatric upper gastrointestinal series indicated irregularities in the stomach lining, prompting further investigation. Comprehensive upper gastrointestinal endoscopies were performed, illuminating significant gastroesophageal reflux disease (GERD) in both patients, characterized by distinctive erosions in the esophagus and transient lower esophageal sphincter relaxations. GERD was particularly pronounced along the greater curvature of the stomach. Their evaluations also included biopsies of the gastric tissue, which fortunately ruled out malignancy, including lymphomas and other gastric cancers. To manage the GERD, both patients were placed on a rigorous treatment regimen including lifestyle modifications and proton-pump inhibitors (PPIs). Each was monitored regularly via follow-up endoscopies which demonstrated gradual improvements in esophageal tissue integrity. Concurrently, both were tested for H. pylori, with one testing positive. The H. pylori-positive patient underwent an eradication protocol with a combination therapy of clarithromycin, amoxicillin, and a PPI, resulting in successful elimination of the infection. Years later, through diligent management and follow-up, both individuals have achieved excellent control over their symptoms and maintain a good quality of life.'' \\
    \textbf{Hard Negative 3:} \\
    ``Numerous pediatric cases have been reviewed to understand the duration and efficacy of chemotherapy in treating various forms of juvenile cancer. One study outlines the treatment plan for a pair of siblings, aged 7 and 14, diagnosed with acute lymphoblastic leukemia (ALL). The treatment protocol involved a comprehensive induction regimen followed by a consolidation phase. During the induction phase, which lasted for about a month, the patients were administered a combination of vincristine, prednisone, asparaginase, and an anthracycline. The consolidation phase incorporated methotrexate and 6-mercaptopurine and extended over several months. Intrathecal chemotherapy was included to prevent CNS disease. Maintenance therapy was subsequently initiated, which is scheduled to continue for a period of three years, with regular follow-ups to monitor remission status. It was observed that the older child had to face additional challenges due to the emergence of several therapy-related side effects. Despite the intensive treatment, both patients are currently responding positively with substantial remission observed in follow-up examinations. The study emphasizes the importance of a tailored approach to pediatric chemotherapy, taking into account not only the type of cancer but also individual patient factors and potential long-term outcomes.'' \\
    \bottomrule
    \end{tabular}
    }
    \caption{Random sampled examples for the generated testing data. Domain: healthcare, Language: English.}
    \label{tab:generated_examples_2}
\end{table*}

\begin{table*}[!h]
    \centering
    \small
    \resizebox{0.98\textwidth}{!}{
    \begin{tabular}{>{\raggedright\arraybackslash}p{0.9\textwidth}}
    \toprule
    \textbf{Domain:} wiki; \textbf{Language:} English  \\
    \textbf{Original Positive:} \\
    ``Caffeine/ergotamine (trade name Cafergot) is the proprietary name of a medication consisting of ergotamine tartrate and caffeine.  This combination is used for the treatment of headaches, such as migraine headache.\textbackslash n\textbackslash n Use\textbackslash n\textbackslash n Correct timing of use is important. Cafergot is an abortive headache treatment, which prevents the development of the headache, rather than a treatment for an established headache. The medication should be administered at the first sign of headache.\textbackslash n\textbackslash n There exist some limitations as to the maximum number of tablets that can be taken per day per week. Different sources of drug information may carry different information, and patients are encouraged to ask their pharmacist or prescriber about such details.\textbackslash n\textbackslash n Cafergot is currently available as a generic drug (ergotamine tartrate/caffeine)\textbackslash n\textbackslash n Mechanism of action\textbackslash n\textbackslash n According to a topic review on UpToDate, \"ergotamine and dihydroergotamine (DHE 45) bind to 5HT 1b/d receptors, just as triptans do.\" This along with binding to other serotonergic and dopaminergic receptors is their presumed mechanism of action in treating migraine.\textbackslash n\textbackslash n Adverse effects\textbackslash n\textbackslash n Because the vasoconstrictive effects of ergotamine and caffeine are not selective for the brain, adverse effects due to systemic vasoconstriction can occur. Cold feet or hands, angina pectoris, myocardial infarction, or dizziness are some examples. \textbackslash n\textbackslash n It has also been shown to be associated with mitral valve stenosis.\textbackslash n\textbackslash n References \textbackslash n\textbackslash n Antimigraine drugs\textbackslash n Combination drugs'' \\
    \midrule
    \textbf{Character:} Pharmacist \\
    \midrule
    \textbf{Scenario:} Advising a patient on the proper usage of Cafergot, including timing and dosage limits. \\
    \midrule
    \textbf{Original Query:} What is the optimal timing for administering Cafergot to treat migraine headaches? \\
    \textbf{Rewritten Query:} At which temporal juncture is it considered most optimal to commence administration of Cafergot for the alleviation of cephalalgic discomfort characteristic of a migraine? \\
    \midrule
    \textbf{Hard Negative 1:} \\
    ``The importance of adherence to a prescribed treatment regimen cannot be overstated, especially when managing chronic conditions such as hypertension and diabetes. Medications for these diseases, while different in function and timing from migraine treatments like Cafergot, require consistent and timely dosing to maintain health and prevent complications. For example, antihypertensive drugs must be taken daily to effectively control blood pressure and reduce the risk of heart attack and stroke. Similarly, diabetic patients must monitor their blood sugar levels regularly and administer insulin or oral hypoglycemic agents as directed to avoid hyperglycemic or hypoglycemic episodes. Although the precise timing may differ from abortive headache therapies, the principle of timing in medication administration is universally critical. Patients are advised to follow the specific instructions provided by their healthcare provider or pharmacist to achieve the best outcomes from their medication regimen. Furthermore, lifestyle modifications, such as diet and exercise, also play a vital role in the management of these conditions and should be initiated in conjunction with pharmacotherapy for an integrated approach to treatment.'' \\
    \textbf{Hard Negative 2:} \\
    ``Caffeine and its Role in Pain Relief: An Overview\textbackslash n\textbackslash n Caffeine, a central nervous system stimulant, has been widely recognized for its ability to increase alertness and alleviate fatigue. Commonly found in various beverages such as coffee, tea, and energy drinks, caffeine is also included in certain pain relief medications. Its application in pain management is based on its pharmacological properties that enhance the efficacy of other analgesic compounds.\textbackslash n\textbackslash n Although not a primary treatment for migraine pain, caffeine is sometimes combined with analgesics like acetaminophen or aspirin to increase their effectiveness. The precise timing for administration of such combination therapies is generally flexible and tailored to individual patient needs. Unlike migraine-specific treatments, these over-the-counter remedies aim to reduce the severity of pain after onset of symptoms.\textbackslash n\textbackslash n Research into caffeine's role in pain relief extends beyond headaches to muscle soreness and other types of pain. While it possesses some anti-inflammatory properties, the exact mechanism through which caffeine exerts its effect on pain pathways is still being investigated. However, it is thought to involve adenosine receptor antagonism.\textbackslash n\textbackslash n Knowing the right amount of caffeine consumption for pain relief is crucial since excessive intake can cause side effects such as jitteriness, insomnia, and an increased heart rate. As with any medication or supplement, users should consult healthcare professionals to determine the appropriate dosage for their condition.'' \\
    \bottomrule
    \end{tabular}
    }
    \caption{Random sampled examples for the generated testing data. Domain: wiki, Language: English.}
    \label{tab:generated_examples_3}
\end{table*}

\begin{table*}[!h]
    \centering
    \small
    \resizebox{\textwidth}{!}{
        \begin{tabular}{l|c|c}
        \toprule
        \textbf{Model} & \textbf{Size} & \textbf{Model Link} \\
        \midrule
        \multicolumn{3}{l}{\textit{Lexical Method}} \\
        \midrule
        BM25~\cite{robertson2009bm25} & - & \href{https://github.com/castorini/pyserini}{\texttt{https://github.com/castorini/pyserini}} \\
        \midrule
        \multicolumn{3}{l}{\textit{English Embedding Models}} \\
        \midrule
        bge-small-en-v1.5~\cite{cpack} & 33.4M & \href{https://huggingface.co/BAAI/bge-small-en-v1.5}{\texttt{https://huggingface.co/BAAI/bge-small-en-v1.5}} \\
        bge-base-en-v1.5~\cite{cpack} & 109M & \href{https://huggingface.co/BAAI/bge-base-en-v1.5}{\texttt{https://huggingface.co/BAAI/bge-base-en-v1.5}} \\
        bge-large-en-v1.5~\cite{cpack} & 335M & \href{https://huggingface.co/BAAI/bge-large-en-v1.5}{\texttt{https://huggingface.co/BAAI/bge-large-en-v1.5}} \\
        bge-en-icl~\cite{li2024making} & 7.11B & \href{https://huggingface.co/BAAI/bge-en-icl}{\texttt{https://huggingface.co/BAAI/bge-en-icl}} \\
        bge-en-icl-e5data~\cite{li2024making} & 7.11B & \href{https://huggingface.co/BAAI/bge-en-icl-e5-data}{\texttt{https://huggingface.co/BAAI/bge-en-icl-e5data}} \\
        e5-small-v2~\cite{wang2022text} & 33.4M & \href{https://huggingface.co/intfloat/e5-small-v2}{\texttt{https://huggingface.co/intfloat/e5-small-v2}} \\
        e5-base-v2~\cite{wang2022text} & 109M & \href{https://huggingface.co/intfloat/e5-base-v2}{\texttt{https://huggingface.co/intfloat/e5-base-v2}} \\
        e5-large-v2~\cite{wang2022text} & 335M & \href{https://huggingface.co/intfloat/e5-large-v2}{\texttt{https://huggingface.co/intfloat/e5-large-v2}} \\
        gte-small~\cite{li2023towards} & 33.4M & \href{https://huggingface.co/thenlper/gte-small}{\texttt{https://huggingface.co/thenlper/gte-small}} \\
        gte-base~\cite{li2023towards} & 109M & \href{https://huggingface.co/thenlper/gte-base}{\texttt{https://huggingface.co/thenlper/gte-base}} \\
        gte-large~\cite{li2023towards} & 335M & \href{https://huggingface.co/thenlper/gte-large}{\texttt{https://huggingface.co/thenlper/gte-large}} \\
        gte-large-en-v1.5~\cite{li2023towards} & 434M & \href{https://huggingface.co/Alibaba-NLP/gte-large-en-v1.5}{\texttt{https://huggingface.co/Alibaba-NLP/gte-large-en-v1.5}} \\
        repllama-v1-7b-lora-passage~\cite{rankllama} & 6.74B & \href{https://huggingface.co/castorini/repllama-v1-7b-lora-passage}{\texttt{https://huggingface.co/castorini/repllama-v1-7b-lora-passage}} \\
        SFR-Embedding-Mistral & 7.11B & \href{https://huggingface.co/Salesforce/SFR-Embedding-Mistral}{\texttt{https://huggingface.co/Salesforce/SFR-Embedding-Mistral}} \\
        SFR-Embedding-2\_R & 7.11B & \href{https://huggingface.co/Salesforce/SFR-Embedding-2_R}{\texttt{https://huggingface.co/Salesforce/SFR-Embedding-2\_R}} \\
        NV-Embed-v1~\cite{lee2024nvembed} & 7.85B & \href{https://huggingface.co/nvidia/NV-Embed-v1}{\texttt{https://huggingface.co/nvidia/NV-Embed-v1}} \\
        NV-Embed-v2~\cite{lee2024nvembed} & 7.85B & \href{https://huggingface.co/nvidia/NV-Embed-v2}{\texttt{https://huggingface.co/nvidia/NV-Embed-v2}} \\
        Linq-Embed-Mistral~\cite{LinqAIResearch2024} & 7.11B & \href{https://huggingface.co/Linq-AI-Research/Linq-Embed-Mistral}{\texttt{https://huggingface.co/Linq-AI-Research/Linq-Embed-Mistral}} \\
        simlm-base-msmarco-finetuned~\cite{Wang2022SimLMPW} & 110M & \href{https://huggingface.co/intfloat/simlm-base-msmarco-finetuned}{\texttt{https://huggingface.co/intfloat/simlm-base-msmarco-finetuned}} \\
        msmarco-roberta-base-ance-firstp~\cite{xiong2021approximate} & 125M & \href{https://huggingface.co/sentence-transformers/msmarco-roberta-base-ance-firstp}{\texttt{https://huggingface.co/sentence-transformers/msmarco-roberta-base-ance-firstp}} \\
        contriever-msmarco~\cite{izacard2022unsupervised} & 109M & \href{https://huggingface.co/facebook/contriever-msmarco}{\texttt{https://huggingface.co/facebook/contriever-msmarco}} \\
        \midrule
        \multicolumn{3}{l}{\textit{Multilingual Embedding Models}} \\
        \midrule
        bge-m3~\cite{chen-etal-2024-m3} & 568M & \href{https://huggingface.co/BAAI/bge-m3}{\texttt{https://huggingface.co/BAAI/bge-m3}} \\
        bge-multilingual-gemma2~\cite{li2024making} & 9.24B & \href{https://huggingface.co/BAAI/bge-multilingual-gemma2}{\texttt{https://huggingface.co/BAAI/bge-multilingual-gemma2}} \\
        jina-embeddings-v3~\cite{sturua2024jina} & 572M & \href{https://huggingface.co/jinaai/jina-embeddings-v3}{\texttt{https://huggingface.co/jinaai/jina-embeddings-v3}} \\
        e5-mistral-7b-instruct~\cite{wang2023improving} & 7.11B & \href{https://huggingface.co/intfloat/e5-mistral-7b-instruct}{\texttt{https://huggingface.co/intfloat/e5-mistral-7b-instruct}} \\
        multilingual-e5-small~\cite{wang2024multilingual} & 118M & \href{https://huggingface.co/intfloat/multilingual-e5-small}{\texttt{https://huggingface.co/intfloat/multilingual-e5-small}} \\
        multilingual-e5-base~\cite{wang2024multilingual} & 278M & \href{https://huggingface.co/intfloat/multilingual-e5-base}{\texttt{https://huggingface.co/intfloat/multilingual-e5-base}} \\
        multilingual-e5-large~\cite{wang2024multilingual} & 560M & \href{https://huggingface.co/intfloat/multilingual-e5-large}{\texttt{https://huggingface.co/intfloat/multilingual-e5-large}} \\
        multilingual-e5-large-instruct~\cite{wang2024multilingual} & 560M & \href{https://huggingface.co/intfloat/multilingual-e5-large-instruct}{\texttt{https://huggingface.co/intfloat/multilingual-e5-large-instruct}} \\
        gte-multilingual-base~\cite{zhang2024mgte} & 305M & \href{https://huggingface.co/Alibaba-NLP/gte-multilingual-base}{\texttt{https://huggingface.co/Alibaba-NLP/gte-multilingual-base}} \\
        bce-embedding-base\_v1~\cite{youdao_bcembedding_2023} & 278M & \href{https://huggingface.co/maidalun1020/bce-embedding-base_v1}{\texttt{https://huggingface.co/maidalun1020/bce-embedding-base\_v1}} \\
        gte-Qwen2-1.5B-instruct~\cite{li2023towards} & 1.78B & \href{https://huggingface.co/Alibaba-NLP/gte-Qwen2-1.5B-instruct}{\texttt{https://huggingface.co/Alibaba-NLP/gte-Qwen2-1.5B-instruct}} \\
        gte-Qwen2-7B-instruct~\cite{li2023towards} & 7.61B & \href{https://huggingface.co/Alibaba-NLP/gte-Qwen2-7B-instruct}{\texttt{https://huggingface.co/Alibaba-NLP/gte-Qwen2-7B-instruct}} \\
        \midrule
        \multicolumn{3}{l}{\textit{Re-ranking Models}} \\
        \midrule
        bge-reranker-large~\cite{cpack} & 560M & \href{https://huggingface.co/BAAI/bge-reranker-large}{\texttt{https://huggingface.co/BAAI/bge-reranker-large}} \\
        bge-reranker-v2-m3 & 568M & \href{https://huggingface.co/BAAI/bge-reranker-v2-m3}{\texttt{https://huggingface.co/BAAI/bge-reranker-v2-m3}} \\
        bge-reranker-v2-gemma & 2.51B & \href{https://huggingface.co/BAAI/bge-reranker-v2-gemma}{\texttt{https://huggingface.co/BAAI/bge-reranker-v2-gemma}} \\
        bce-reranker-base\_v1~\cite{youdao_bcembedding_2023} & 278M & \href{https://huggingface.co/maidalun1020/bce-reranker-base_v1}{\texttt{https://huggingface.co/maidalun1020/bce-reranker-base\_v1}} \\
        mmarco-mMiniLMv2-L12-H384-v1 & 118M & \href{https://huggingface.co/nreimers/mmarco-mMiniLMv2-L12-H384-v1}{\texttt{https://huggingface.co/nreimers/mmarco-mMiniLMv2-L12-H384-v1}} \\
        \bottomrule
    \end{tabular}
    }
    \vspace{-10pt}
    \caption{Detailed information on all of the models appearing in our paper.}
    \label{tab:model_information}
\end{table*}

\newpage

% ====================== 24.04 ==========================
\begin{sidewaystable*}[t]
    \centering
    \setlength{\extrarowheight}{4pt}
    \resizebox{\textwidth}{!}{
        \begin{tabular}{|c|c|c|c|c|c|c|c|c|c|c|c|c|c|}
        \toprule
            \multirow{2}{*}{\textbf{Task}} & \multirow{2}{*}{\textbf{Domain}} & \multicolumn{2}{c|}{\textbf{Language}} & \multirow{2}{*}{\textbf{Dataset Name}} & \multicolumn{2}{c|}{\textbf{Source of Corpus}} & \multirow{2}{*}{\textbf{\#corpus}} & \multirow{2}{*}{\begin{tabular}[c]{@{}c@{}} \textbf{Avg \#token} \\ \textbf{of corpus} \end{tabular}} & \multirow{2}{*}{\textbf{Split}} & \multirow{2}{*}{\begin{tabular}[c]{@{}c@{}} \textbf{\# of} \\ \textbf{queries} \end{tabular}} & \multirow{2}{*}{\begin{tabular}[c]{@{}c@{}} \textbf{Avg \#token} \\ \textbf{of queries} \end{tabular}} & \multirow{2}{*}{\begin{tabular}[c]{@{}c@{}} \textbf{\# of} \\ \textbf{positives} \end{tabular}} & \multirow{2}{*}{\begin{tabular}[c]{@{}c@{}} \textbf{\# of hard} \\ \textbf{negatives} \end{tabular}} \\
            \cline{3-4} \cline{6-7}
             &  & \textbf{Name} & \textbf{ISO Code} &  & \textbf{Link \& Citation} & \textbf{License} &  &  &  &  &  &  & \\
            \midrule
            \midrule
            % qa arxiv en
            \multirow{13}{*}{qa} & arxiv & English & en & default & \href{https://github.com/armancohan/long-summarization}{long-summarization}~\cite{cohan-etal-2018-discourse} & Apache 2.0 & 222,877 & 334 & 
            test  & 1,731 & 19 & 5,340 & 6,288 \\
            \cline{2-14}
            % qa finance en
             & \multirow{2}{*}{finance} & English & en & default & \href{https://huggingface.co/datasets/reuters21578}{Reuters-21578}~\cite{misc_reuters-21578_text_categorization_collection_137} & CC BY 4.0 & 26,226 & 202 & 
            test  & 1,585 & 17 & 3,357 & 5,595 \\
            \cline{3-14}
            % qa finance zh
             &  & Chinese & zh & default & \href{https://huggingface.co/datasets/Duxiaoman-DI/FinCorpus}{Duxiaoman-DI/FinCorpus} & Apache 2.0 & 2,398,095 & 1,616 & 
            test  & 1,805 & 29 & 7,836 & 7,211 \\
            \cline{2-14}
            % qa healthcare en
             & \multirow{2}{*}{healthcare} & English & en & default & \href{https://huggingface.co/datasets/qiaojin/PubMedQA}{PubMedQA}~\cite{jin2019pubmedqa} & MIT & 847,395 & 103 & 
            test  & 1,707 & 19 & 5,052 & 7,023 \\
            \cline{3-14}
            % qa healthcare zh
             &  & Chinese & zh & default & \href{https://huggingface.co/datasets/FreedomIntelligence/huatuo_encyclopedia_qa}{Huatuo-26M}~\cite{li2023huatuo26m} & Apache 2.0 & 360,218 & 751 & 
            test  & 1,874 & 31 & 10,029 & 7,336 \\
            \cline{2-14}
            % qa law en
             & law & English & en & default & \href{https://huggingface.co/datasets/pile-of-law/pile-of-law}{Pile of Law}~\cite{hendersonkrass2022pileoflaw} & CC BY-NC-SA 4.0 & 141,678 & 1,509 & 
            test  & 1,801 & 19 & 5,372 & 6,574 \\
            \cline{2-14}
            % qa news en
             & \multirow{2}{*}{news} & English & en & default & \href{https://huggingface.co/datasets/cc_news}{CC-News}~\cite{Hamborg2017} & Unknown & 574,417 & 531 & 
            test  & 1,614 & 16 & 5,798 & 6,784 \\
            \cline{3-14}
            % qa news zh
             &  & Chinese & zh & default & \href{https://huggingface.co/datasets/intfloat/multilingual_cc_news}{intfloat/multilingual\_cc\_news} & Unknown & 935,162 & 1,263 & 
            test  & 1,697 & 31 & 7,381 & 6,618 \\
            \cline{2-14}
            % qa web en
             & \multirow{2}{*}{web} & English & en & default & \href{https://huggingface.co/datasets/allenai/c4}{mC4}~\cite{mc4} & ODC-BY & 2,459,587 & 840 & 
            test  & 1,707 & 16 & 5,543 & 7,439 \\
            \cline{3-14}
            % qa web zh
             &  & Chinese & zh & default & \href{https://huggingface.co/datasets/allenai/c4}{mC4}~\cite{mc4} & ODC-BY & 956,699 & 1,208 & 
            test  & 1,683 & 29 & 6,250 & 6,721 \\
            \cline{2-14}
            % qa wiki en
             & \multirow{2}{*}{wiki} & English & en & default & \href{https://huggingface.co/datasets/NeuML/wikipedia-20240101}{Wikipedia 20240101} & CC BY-SA 3.0, GFDL & 6,738,498 & 667 & 
            test  & 1,727 & 17 & 4,260 & 7,882 \\
            \cline{3-14}
            % qa wiki zh
             &  & Chinese & zh & default & \href{https://huggingface.co/datasets/wikipedia}{Wikipedia 20240401} & CC BY-SA 3.0, GFDL & 1,161,226 & 557 & 
            test  & 1,679 & 30 & 4,745 & 6,963 \\
            \cline{2-14}
            % qa msmarco en
             & web (msmarco) & English & en & default & \href{https://huggingface.co/datasets/intfloat/simlm-msmarco}{MS MARCO}~\cite{bajaj2016ms} & MIT & 8,872,840 & 81 & 
            test  & 6,319 & 16 & 31,447 & 26,828 \\
            \hline
            % long-doc arxiv en
            \multirow{15}{*}{long-doc} & \multirow{4}{*}{arxiv} & \multirow{4}{*}{English} & \multirow{4}{*}{en} & gemini & \href{https://arxiv.org/pdf/2312.11805.pdf}{Paper of Gemini} & CC BY 4.0 & 276 & 136 & 
            test  & 249 & 18 & 249 & 0 \\
            \cline{5-14}
             &  &  &  & gpt3 & \href{https://arxiv.org/pdf/2005.14165.pdf}{Paper of GPT-3} & \begin{tabular}[c]{@{}c@{}} arXiv.org perpetual, \\ non-exclusive license 1.0 \end{tabular} & 515 & 137 & 
            test  & 337 & 16 & 496 & 0 \\
            \cline{5-14}
             &  &  &  & llama2 & \href{https://arxiv.org/pdf/2307.09288.pdf}{Paper of Llama 2} & \begin{tabular}[c]{@{}c@{}} arXiv.org perpetual, \\ non-exclusive license 1.0 \end{tabular} & 566 & 136 & 
            test  & 326 & 18 & 635 & 0 \\
            \cline{5-14}
             &  &  &  & llm-survey & \href{https://arxiv.org/pdf/2303.18223.pdf}{Survey of LLM} & \begin{tabular}[c]{@{}c@{}} arXiv.org perpetual, \\ non-exclusive license 1.0 \end{tabular} & 1,144 & 135 & 
            test  & 357 & 17 & 924 & 0 \\
            \cline{2-14}
            % long-doc book en
             & \multirow{2}{*}{book} & \multirow{2}{*}{English} & \multirow{2}{*}{en} & a-brief-history-of-time\_stephen-hawking & \href{https://www.docdroid.net/GCLN82v/stephen-hawking-a-brief-history-of-time-pdf}{\textit{A Brief History of Time}} & Unknown & 778 & 127 & 
            test  & 370 & 16 & 876 & 0 \\
            \cline{5-14}
             &  &  &  & origin-of-species\_darwin & \href{https://www.vliz.be/docs/Zeecijfers/Origin_of_Species.pdf}{\textit{On the Origin of Species}} & Unknown & 1,758 & 126 & 
            test  & 366 & 16 & 1,145 & 0 \\
            \cline{2-14}
            % long-doc healthcare en
             & \multirow{5}{*}{healthcare} & \multirow{5}{*}{English} & \multirow{5}{*}{en} & pubmed\_100K-200K\_1 & \href{https://github.com/armancohan/long-summarization}{long-summarization}~\cite{cohan-etal-2018-discourse} & Apache 2.0 & 899 & 133 & 
            test  & 372 & 20 & 1,008 & 0 \\
            \cline{5-14}
             &  &  &  & pubmed\_100K-200K\_2 & \href{https://github.com/armancohan/long-summarization}{long-summarization}~\cite{cohan-etal-2018-discourse} & Apache 2.0 & 872 & 136 & 
            test  & 355 & 18 & 980 & 0 \\
            \cline{5-14}
             &  &  &  & pubmed\_100K-200K\_3 & \href{https://github.com/armancohan/long-summarization}{long-summarization}~\cite{cohan-etal-2018-discourse} & Apache 2.0 & 873 & 133 & 
            test  & 357 & 19 & 978 & 0 \\
            \cline{5-14}
             &  &  &  & pubmed\_30K-40K\_10-merged & \href{https://github.com/armancohan/long-summarization}{long-summarization}~\cite{cohan-etal-2018-discourse} & Apache 2.0 & 2,154 & 133 & 
            test  & 368 & 18 & 1,485 & 0 \\
            \cline{5-14}
             &  &  &  & pubmed\_40K-50K\_5-merged & \href{https://github.com/armancohan/long-summarization}{long-summarization}~\cite{cohan-etal-2018-discourse} & Apache 2.0 & 1,731 & 136 & 
            test  & 336 & 21 & 1,046 & 0 \\
            \cline{2-14}
            % long-doc law en
             & \multirow{4}{*}{law} & \multirow{4}{*}{English} & \multirow{4}{*}{en} & lex\_files\_300K-400K & \href{https://huggingface.co/datasets/lexlms/lex_files}{LexFiles}~\cite{chalkidis-etal-2023-lexfiles} & CC BY-NC-SA 4.0 & 2,797 & 137 & 
            test  & 339 & 15 & 1,307 & 0 \\
            \cline{5-14}
             &  &  &  & lex\_files\_400K-500K & \href{https://huggingface.co/datasets/lexlms/lex_files}{LexFiles}~\cite{chalkidis-etal-2023-lexfiles} & CC BY-NC-SA 4.0 & 3,320 & 137 & 
            test  & 333 & 17 & 1,427 & 0 \\
            \cline{5-14}
             &  &  &  & lex\_files\_500K-600K & \href{https://huggingface.co/datasets/lexlms/lex_files}{LexFiles}~\cite{chalkidis-etal-2023-lexfiles} & CC BY-NC-SA 4.0 & 4,087 & 136 &
            test  & 346 & 17 & 1,324 & 0 \\
            \cline{5-14}
             &  &  &  & lex\_files\_600K-700K & \href{https://huggingface.co/datasets/lexlms/lex_files}{LexFiles}~\cite{chalkidis-etal-2023-lexfiles} & CC BY-NC-SA 4.0 & 5,049 & 138 & 
            test  & 338 & 18 & 1,442 & 0 \\
            
            % \cline{2-14}
            % % \cline{3-14}
            %  & \multirow{2}{*}{} & \multirow{2}{*}{} & \multirow{2}{*}{} & \multirow{2}{*}{default} & \multirow{2}{*}{\href{}{}~\cite{}} & \multirow{2}{*}{} & \multirow{2}{*}{} & 
            % dev  &  &  &  &  \\
            % \cline{10-14}
            %  &  &  &  &  &  &  &  &  &
            % test &  &  &  &  \\
        \bottomrule
        \end{tabular}
    }
    \vspace{-5pt}
    \caption{Statistics of all datasets in \airbench 24.04.}
    \label{tab:air-bench_datasets_2404}
\end{sidewaystable*}

% ====================== 24.05 Part 1 ==========================
\begin{sidewaystable*}[t]
    \centering
    \setlength{\extrarowheight}{2pt}
    \resizebox{0.9\textwidth}{!}{
        \begin{tabular}{|c|c|c|c|c|c|c|c|c|c|c|c|c|c|}
        \toprule
            \multirow{2}{*}{\textbf{Task}} & \multirow{2}{*}{\textbf{Domain}} & \multicolumn{2}{c|}{\textbf{Language}} & \multirow{2}{*}{\textbf{Dataset Name}} & \multicolumn{2}{c|}{\textbf{Source of Corpus}} & \multirow{2}{*}{\textbf{\#corpus}} & \multirow{2}{*}{\begin{tabular}[c]{@{}c@{}} \textbf{Avg \#token} \\ \textbf{of corpus} \end{tabular}} & \multirow{2}{*}{\textbf{Split}} & \multirow{2}{*}{\begin{tabular}[c]{@{}c@{}} \textbf{\# of} \\ \textbf{queries} \end{tabular}} & \multirow{2}{*}{\begin{tabular}[c]{@{}c@{}} \textbf{Avg \#token} \\ \textbf{of queries} \end{tabular}} & \multirow{2}{*}{\begin{tabular}[c]{@{}c@{}} \textbf{\# of} \\ \textbf{positives} \end{tabular}} & \multirow{2}{*}{\begin{tabular}[c]{@{}c@{}} \textbf{\# of hard} \\ \textbf{negatives} \end{tabular}} \\
            \cline{3-4} \cline{6-7}
             &  & \textbf{Name} & \textbf{ISO Code} &  & \textbf{Link \& Citation} & \textbf{License} &  &  &  &  &  &  & \\
            \midrule
            \midrule
            
            % qa arxiv en
            \multirow{29}{*}{qa} & \multirow{2}{*}{arxiv} & \multirow{2}{*}{English} & \multirow{2}{*}{en} & \multirow{2}{*}{default} & \multirow{2}{*}{\href{https://github.com/armancohan/long-summarization}{long-summarization}~\cite{cohan-etal-2018-discourse}} & \multirow{2}{*}{Apache 2.0} & \multirow{2}{*}{222,877} & \multirow{2}{*}{334} & 
            dev  & 346 & 19 & 1,091 & 1,230 \\
            \cline{10-14}
             &  &  &  &  &  &  &  &  &
            test & 1,385 & 19 & 4,249 & 5,058 \\
            \cline{2-14}
            % qa finance en
             & \multirow{8}{*}{finance} & \multirow{2}{*}{English} & \multirow{2}{*}{en} & \multirow{2}{*}{default} & \multirow{2}{*}{\href{https://huggingface.co/datasets/reuters21578}{Reuters-21578}~\cite{misc_reuters-21578_text_categorization_collection_137}} & \multirow{2}{*}{CC BY 4.0} & \multirow{2}{*}{26,226} & \multirow{2}{*}{202} & 
            dev  & 317 & 17 & 627 & 1,122 \\
            \cline{10-14}
             &  &  &  &  &  &  &  &  &
            test & 1,268 & 17 & 2,730 & 4,473 \\
            \cline{3-14}
            % qa finance ar
             &  & \multirow{2}{*}{Arabic} & \multirow{2}{*}{ar} & \multirow{2}{*}{default} & \multirow{2}{*}{\href{https://huggingface.co/datasets/asas-ai/financial_news}{asas-ai/financial\_news}} & \multirow{2}{*}{Apache 2.0} & \multirow{2}{*}{11,235} & \multirow{2}{*}{397} & 
            dev  & 293 & 49 & 635 & 727 \\
            \cline{10-14}
             &  &  &  &  &  &  &  &  &
            test & 1,175 & 46 & 2,796 & 2,959 \\
            \cline{3-14}
            % qa finance fr
             &  & \multirow{2}{*}{French} & \multirow{2}{*}{fr} & \multirow{2}{*}{default} & \multirow{2}{*}{\href{https://huggingface.co/datasets/FrancophonIA/CoFiF}{CoFiF}~\cite{daudert-ahmadi-2019-cofif}} & \multirow{2}{*}{CC BY-NC 4.0} & \multirow{2}{*}{1,006,801} & \multirow{2}{*}{92} & 
            dev  & 310 & 21 & 1,841 & 1,071 \\
            \cline{10-14}
             &  &  &  &  &  &  &  &  &
            test & 1,243 & 20 & 7,206 & 4,362 \\
            \cline{3-14}
            % qa finance zh
             &  & \multirow{2}{*}{Chinese} & \multirow{2}{*}{zh} & \multirow{2}{*}{default} & \multirow{2}{*}{\href{https://huggingface.co/datasets/Duxiaoman-DI/FinCorpus}{Duxiaoman-DI/FinCorpus}} & \multirow{2}{*}{Apache 2.0} & \multirow{2}{*}{1,014,974} & \multirow{2}{*}{1,613} & 
            dev  & 361 & 29 & 1,516 & 1,471 \\
            \cline{10-14}
             &  &  &  &  &  &  &  &  &
            test & 1,444 & 29 & 6,320 & 5,740 \\
            \cline{2-14}
            % qa healthcare en
             & \multirow{10}{*}{healthcare} & \multirow{2}{*}{English} & \multirow{2}{*}{en} & \multirow{2}{*}{default} & \multirow{2}{*}{\href{https://huggingface.co/datasets/qiaojin/PubMedQA}{PubMedQA}~\cite{jin2019pubmedqa}} & \multirow{2}{*}{MIT} & \multirow{2}{*}{847,395} & \multirow{2}{*}{103} & 
            dev  & 341 & 20 & 1,008 & 1,382 \\
            \cline{10-14}
             &  &  &  &  &  &  &  &  &
            test & 1,366 & 19 & 4,044 & 5,641 \\
            \cline{3-14}
            % qa healthcare de
             &  & \multirow{2}{*}{German} & \multirow{2}{*}{de} & \multirow{2}{*}{default} & \multirow{2}{*}{\href{https://huggingface.co/datasets/GEM/mlsum}{MLSUM}~\cite{scialom-etal-2020-mlsum}} & \multirow{2}{*}{MIT} & \multirow{2}{*}{27,934} & \multirow{2}{*}{909} & 
            dev  & 360 & 21 & 1,102 & 1,137 \\
            \cline{10-14}
             &  &  &  &  &  &  &  &  &
            test & 1,441 & 20 & 4,667 & 4,306 \\
            \cline{3-14}
            % qa healthcare es
             &  & \multirow{2}{*}{Spanish} & \multirow{2}{*}{es} & \multirow{2}{*}{default} & \multirow{2}{*}{\href{https://zenodo.org/records/3463379}{Multilingual Medical Corpora}~\cite{fabian_villena_2019_3463379}} & \multirow{2}{*}{Unknown} & \multirow{2}{*}{1,006,093} & \multirow{2}{*}{60} & 
            dev  & 300 & 21 & 1,210 & 930 \\
            \cline{10-14}
             &  &  &  &  &  &  &  &  &
            test & 1,201 & 22 & 4,695 & 3,809 \\
            \cline{3-14}
            % qa healthcare fr
             &  & \multirow{2}{*}{French} & \multirow{2}{*}{fr} & \multirow{2}{*}{default} & \multirow{2}{*}{\href{https://zenodo.org/records/3463379}{Multilingual Medical Corpora}~\cite{fabian_villena_2019_3463379}} & \multirow{2}{*}{Unknown} & \multirow{2}{*}{972,938} & \multirow{2}{*}{202} & 
            dev  & 331 & 23 & 1,885 & 1,261 \\
            \cline{10-14}
             &  &  &  &  &  &  &  &  &
            test & 1,326 & 24 & 7,460 & 5,119 \\
            \cline{3-14}
            % qa healthcare zh
             &  & \multirow{2}{*}{Chinese} & \multirow{2}{*}{zh} & \multirow{2}{*}{default} & \multirow{2}{*}{\href{https://huggingface.co/datasets/FreedomIntelligence/huatuo_encyclopedia_qa}{Huatuo-26M}~\cite{li2023huatuo26m}} & \multirow{2}{*}{Apache 2.0} & \multirow{2}{*}{360,218} & \multirow{2}{*}{751} & 
            dev  & 374 & 31 & 2,030 & 1,490 \\
            \cline{10-14}
             &  &  &  &  &  &  &  &  &
            test & 1,500 & 31 & 7,999 & 5,846 \\
            \cline{2-14}
            % qa law en
             & \multirow{6}{*}{law} & \multirow{2}{*}{English} & \multirow{2}{*}{en} & \multirow{2}{*}{default} & \multirow{2}{*}{\href{https://huggingface.co/datasets/pile-of-law/pile-of-law}{Pile of Law}~\cite{hendersonkrass2022pileoflaw}} & \multirow{2}{*}{CC BY-NC-SA 4.0} & \multirow{2}{*}{141,678} & \multirow{2}{*}{1,509} & 
            dev  & 360 & 20 & 1,080 & 1,341 \\
            \cline{10-14}
             &  &  &  &  &  &  &  &  &
            test & 1,441 & 19 & 4,292 & 5,233 \\
            \cline{3-14}
            % qa law de
             &  & \multirow{2}{*}{German} & \multirow{2}{*}{de} & \multirow{2}{*}{default} & \multirow{2}{*}{\href{https://huggingface.co/datasets/joelniklaus/Multi_Legal_Pile}{MultiLegalPile}~\cite{niklaus2023multilegalpile}}& \multirow{2}{*}{CC BY-NC-SA 4.0} & \multirow{2}{*}{752,913} & \multirow{2}{*}{3,361} & 
            dev  & 345 & 24 & 1,373 & 1,099 \\
            \cline{10-14}
             &  &  &  &  &  &  &  &  &
            test & 1,382 & 25 & 5,500 & 4,622 \\
            \cline{3-14}
            % qa law fr
             &  & \multirow{2}{*}{French} & \multirow{2}{*}{fr} & \multirow{2}{*}{default} & \multirow{2}{*}{\href{https://huggingface.co/datasets/joelniklaus/Multi_Legal_Pile}{MultiLegalPile}~\cite{niklaus2023multilegalpile}}& \multirow{2}{*}{CC BY-NC-SA 4.0} & \multirow{2}{*}{649,017} & \multirow{2}{*}{2,540} & 
            dev  & 348 & 23 & 1,371 & 1,260 \\
            \cline{10-14}
             &  &  &  &  &  &  &  &  &
            test & 1,394 & 22 & 5,535 & 4,968 \\
            \cline{2-14}
            % qa science ru
             & \multirow{2}{*}{science} & \multirow{2}{*}{Russian} & \multirow{2}{*}{ru} & \multirow{2}{*}{default} & \multirow{2}{*}{\href{https://huggingface.co/datasets/mlsa-iai-msu-lab/ru_sci_bench}{mlsa-iai-msu-lab/ru\_sci\_bench}} & \multirow{2}{*}{MIT} & \multirow{2}{*}{200,532} & \multirow{2}{*}{347} & 
            dev  & 345 & 34 & 1,577 & 1,160 \\
            \cline{10-14}
             &  &  &  &  &  &  &  &  &
            test & 1,382 & 33 & 6,018 & 4,655 \\
            \cline{2-14}
            % qa msmarco en
             & web (msmarco) & English & en & default & \href{https://huggingface.co/datasets/intfloat/simlm-msmarco}{MS MARCO}~\cite{bajaj2016ms} & MIT & 8,872,840 & 81 & 
            dev  & 6,319 & 16 & 31,447 & 26,828 \\
        \bottomrule
        \end{tabular}
    }
    \vspace{-5pt}
    \caption{Statistics of all datasets in \airbench 24.05 (Part 1).}
    \label{tab:air-bench_datasets_2405_1}
\end{sidewaystable*}

% ====================== 24.05 Part 2 ==========================
\begin{sidewaystable*}[t]
    \centering
    \setlength{\extrarowheight}{2pt}
    \resizebox{0.9\textwidth}{!}{
        \begin{tabular}{|c|c|c|c|c|c|c|c|c|c|c|c|c|c|}
        \toprule
            \multirow{2}{*}{\textbf{Task}} & \multirow{2}{*}{\textbf{Domain}} & \multicolumn{2}{c|}{\textbf{Language}} & \multirow{2}{*}{\textbf{Dataset Name}} & \multicolumn{2}{c|}{\textbf{Source of Corpus}} & \multirow{2}{*}{\textbf{\#corpus}} & \multirow{2}{*}{\begin{tabular}[c]{@{}c@{}} \textbf{Avg \#token} \\ \textbf{of corpus} \end{tabular}} & \multirow{2}{*}{\textbf{Split}} & \multirow{2}{*}{\begin{tabular}[c]{@{}c@{}} \textbf{\# of} \\ \textbf{queries} \end{tabular}} & \multirow{2}{*}{\begin{tabular}[c]{@{}c@{}} \textbf{Avg \#token} \\ \textbf{of queries} \end{tabular}} & \multirow{2}{*}{\begin{tabular}[c]{@{}c@{}} \textbf{\# of} \\ \textbf{positives} \end{tabular}} & \multirow{2}{*}{\begin{tabular}[c]{@{}c@{}} \textbf{\# of hard} \\ \textbf{negatives} \end{tabular}} \\
            \cline{3-4} \cline{6-7}
             &  & \textbf{Name} & \textbf{ISO Code} &  & \textbf{Link \& Citation} & \textbf{License} &  &  &  &  &  &  & \\
            \midrule
            \midrule
            
            % qa news en
            \multirow{26}{*}{qa} & \multirow{26}{*}{news} & \multirow{2}{*}{English} & \multirow{2}{*}{en} & \multirow{2}{*}{default} & \multirow{2}{*}{\href{https://huggingface.co/datasets/cc_news}{CC-News}~\cite{Hamborg2017}} & \multirow{2}{*}{Unknown} & \multirow{2}{*}{574,417} & \multirow{2}{*}{531} & 
            dev  & 322 & 16 & 1,206 & 1,375 \\
            \cline{10-14}
             &  &  &  &  &  &  &  &  &
            test & 1,292 & 16 & 4,592 & 5,409 \\
            \cline{3-14}
            % qa news ar
             &  & \multirow{2}{*}{Arabic} & \multirow{2}{*}{ar} & \multirow{2}{*}{default} & \multirow{2}{*}{\href{https://huggingface.co/datasets/intfloat/multilingual_cc_news}{intfloat/multilingual\_cc\_news}} & \multirow{2}{*}{Unknown} & \multirow{2}{*}{1,006,308} & \multirow{2}{*}{992} & 
            dev  & 349 & 42 & 1,810 & 1,307 \\
            \cline{10-14}
             &  &  &  &  &  &  &  &  &
            test & 1,396 & 43 & 7,169 & 5,266 \\
            \cline{3-14}
            % qa news bn
             &  & \multirow{2}{*}{Bengali} & \multirow{2}{*}{bn} & \multirow{2}{*}{default} & \multirow{2}{*}{\href{https://huggingface.co/datasets/intfloat/multilingual_cc_news}{intfloat/multilingual\_cc\_news}} & \multirow{2}{*}{Unknown} & \multirow{2}{*}{20,681} & \multirow{2}{*}{912} & 
            dev  & 289 & 84 & 562 & 741 \\
            \cline{10-14}
             &  &  &  &  &  &  &  &  &
            test & 1,159 & 78 & 2,269 & 3,013 \\
            \cline{3-14}
            % qa news de
             &  & \multirow{2}{*}{German} & \multirow{2}{*}{de} & \multirow{2}{*}{default} & \multirow{2}{*}{\href{https://huggingface.co/datasets/intfloat/multilingual_cc_news}{intfloat/multilingual\_cc\_news}} & \multirow{2}{*}{Unknown} & \multirow{2}{*}{1,006,876} & \multirow{2}{*}{659} & 
            dev  & 336 & 23 & 1,448 & 1,234 \\
            \cline{10-14}
             &  &  &  &  &  &  &  &  &
            test & 1,348 & 23 & 5,990 & 5,176 \\
            \cline{3-14}
            % qa news es
             &  & \multirow{2}{*}{Spanish} & \multirow{2}{*}{es} & \multirow{2}{*}{default} & \multirow{2}{*}{\href{https://huggingface.co/datasets/intfloat/multilingual_cc_news}{intfloat/multilingual\_cc\_news}} & \multirow{2}{*}{Unknown} & \multirow{2}{*}{1,007,104} & \multirow{2}{*}{615} & 
            dev  & 337 & 23 & 1,541 & 1,240 \\
            \cline{10-14}
             &  &  &  &  &  &  &  &  &
            test & 1,351 & 23 & 6,246 & 5,257 \\
            \cline{3-14}
            % qa news fa
             &  & \multirow{2}{*}{Persian} & \multirow{2}{*}{fa} & \multirow{2}{*}{default} & \multirow{2}{*}{\href{https://huggingface.co/datasets/intfloat/multilingual_cc_news}{intfloat/multilingual\_cc\_news}} & \multirow{2}{*}{Unknown} & \multirow{2}{*}{1,002,797} & \multirow{2}{*}{1,351} & 
            dev  & 346 & 50 & 1,952 & 1,341 \\
            \cline{10-14}
             &  &  &  &  &  &  &  &  &
            test & 1,386 & 48 & 7,885 & 5,353 \\
            \cline{3-14}
            % qa news fr
             &  & \multirow{2}{*}{French} & \multirow{2}{*}{fr} & \multirow{2}{*}{default} & \multirow{2}{*}{\href{https://huggingface.co/datasets/intfloat/multilingual_cc_news}{intfloat/multilingual\_cc\_news}} & \multirow{2}{*}{Unknown} & \multirow{2}{*}{1,007,592} & \multirow{2}{*}{641} & 
            dev  & 345 & 23 & 1,548 & 1,424 \\
            \cline{10-14}
             &  &  &  &  &  &  &  &  &
            test & 1,383 & 22 & 6,224 & 5,594 \\
            \cline{3-14}
            % qa news hi
             &  & \multirow{2}{*}{Hindi} & \multirow{2}{*}{hi} & \multirow{2}{*}{default} & \multirow{2}{*}{\href{https://huggingface.co/datasets/intfloat/multilingual_cc_news}{intfloat/multilingual\_cc\_news}} & \multirow{2}{*}{Unknown} & \multirow{2}{*}{1,006,218} & \multirow{2}{*}{1,465} & 
            dev  & 349 & 66 & 1,716 & 1,264 \\
            \cline{10-14}
             &  &  &  &  &  &  &  &  &
            test & 1,398 & 67 & 7,039 & 5,162 \\
            \cline{3-14}
            % qa news id
             &  & \multirow{2}{*}{Indonesian} & \multirow{2}{*}{id} & \multirow{2}{*}{default} & \multirow{2}{*}{\href{https://huggingface.co/datasets/intfloat/multilingual_cc_news}{intfloat/multilingual\_cc\_news}} & \multirow{2}{*}{Unknown} & \multirow{2}{*}{1,007,724} & \multirow{2}{*}{548} & 
            dev  & 338 & 24 & 1,799 & 1,397 \\
            \cline{10-14}
             &  &  &  &  &  &  &  &  &
            test & 1,356 & 24 & 7,485 & 5,618 \\
            \cline{3-14}
            % qa news ja
             &  & \multirow{2}{*}{Japanese} & \multirow{2}{*}{ja} & \multirow{2}{*}{default} & \multirow{2}{*}{\href{https://huggingface.co/datasets/intfloat/multilingual_cc_news}{intfloat/multilingual\_cc\_news}} & \multirow{2}{*}{Unknown} & \multirow{2}{*}{834,364} & \multirow{2}{*}{1,559} & 
            dev  & 344 & 35 & 1,817 & 1,334 \\
            \cline{10-14}
             &  &  &  &  &  &  &  &  &
            test & 1,378 & 36 & 6,590 & 5,197 \\
            \cline{3-14}
            % qa news ko
             &  & \multirow{2}{*}{Korean} & \multirow{2}{*}{ko} & \multirow{2}{*}{default} & \multirow{2}{*}{\href{https://huggingface.co/datasets/intfloat/multilingual_cc_news}{intfloat/multilingual\_cc\_news}} & \multirow{2}{*}{Unknown} & \multirow{2}{*}{1,006,798} & \multirow{2}{*}{1,072} & 
            dev  & 361 & 34 & 1,967 & 1,413 \\
            \cline{10-14}
             &  &  &  &  &  &  &  &  &
            test & 1,447 & 36 & 7,908 & 5,665 \\
            \cline{3-14}
            % qa news ru
             &  & \multirow{2}{*}{Russian} & \multirow{2}{*}{ru} & \multirow{2}{*}{default} & \multirow{2}{*}{\href{https://huggingface.co/datasets/intfloat/multilingual_cc_news}{intfloat/multilingual\_cc\_news}} & \multirow{2}{*}{Unknown} & \multirow{2}{*}{1,004,550} & \multirow{2}{*}{776} & 
            dev  & 337 & 34 & 1,676 & 1,301 \\
            \cline{10-14}
             &  &  &  &  &  &  &  &  &
            test & 1,352 & 33 & 6,689 & 5,158 \\
            \cline{3-14}
            % qa news zh
             &  & \multirow{2}{*}{Chinese} & \multirow{2}{*}{zh} & \multirow{2}{*}{default} & \multirow{2}{*}{\href{https://huggingface.co/datasets/intfloat/multilingual_cc_news}{intfloat/multilingual\_cc\_news}} & \multirow{2}{*}{Unknown} & \multirow{2}{*}{935,162} & \multirow{2}{*}{1,263} & 
            dev  & 339 & 32 & 1,477 & 1,354 \\
            \cline{10-14}
             &  &  &  &  &  &  &  &  &
            test & 1,358 & 30 & 5,904 & 5,264 \\
        \bottomrule
        \end{tabular}
    }
    \vspace{-5pt}
    \caption{Statistics of all datasets in \airbench 24.05 (Part 2).}
    \label{tab:air-bench_datasets_2405_2}
\end{sidewaystable*}

% ====================== 24.05 Part 3 ==========================
\begin{sidewaystable*}[t]
    \centering
    \setlength{\extrarowheight}{2pt}
    \resizebox{0.9\textwidth}{!}{
        \begin{tabular}{|c|c|c|c|c|c|c|c|c|c|c|c|c|c|}
        \toprule
            \multirow{2}{*}{\textbf{Task}} & \multirow{2}{*}{\textbf{Domain}} & \multicolumn{2}{c|}{\textbf{Language}} & \multirow{2}{*}{\textbf{Dataset Name}} & \multicolumn{2}{c|}{\textbf{Source of Corpus}} & \multirow{2}{*}{\textbf{\#corpus}} & \multirow{2}{*}{\begin{tabular}[c]{@{}c@{}} \textbf{Avg \#token} \\ \textbf{of corpus} \end{tabular}} & \multirow{2}{*}{\textbf{Split}} & \multirow{2}{*}{\begin{tabular}[c]{@{}c@{}} \textbf{\# of} \\ \textbf{queries} \end{tabular}} & \multirow{2}{*}{\begin{tabular}[c]{@{}c@{}} \textbf{Avg \#token} \\ \textbf{of queries} \end{tabular}} & \multirow{2}{*}{\begin{tabular}[c]{@{}c@{}} \textbf{\# of} \\ \textbf{positives} \end{tabular}} & \multirow{2}{*}{\begin{tabular}[c]{@{}c@{}} \textbf{\# of hard} \\ \textbf{negatives} \end{tabular}} \\
            \cline{3-4} \cline{6-7}
             &  & \textbf{Name} & \textbf{ISO Code} &  & \textbf{Link \& Citation} & \textbf{License} &  &  &  &  &  &  & \\
            \midrule
            \midrule
            
            % qa web en
            \multirow{26}{*}{qa} & \multirow{26}{*}{web} & \multirow{2}{*}{English} & \multirow{2}{*}{en} & \multirow{2}{*}{default} & \multirow{2}{*}{\href{https://huggingface.co/datasets/allenai/c4}{mC4}~\cite{mc4}} & \multirow{2}{*}{ODC-BY} & \multirow{2}{*}{1,012,910} & \multirow{2}{*}{838} & 
            dev  & 341 & 16 & 1,087 & 1,511 \\
            \cline{10-14}
             &  &  &  &  &  &  &  &  &
            test & 1,366 & 16 & 4,456 & 5,928 \\
            \cline{3-14}
            % qa web ar
             &  & \multirow{2}{*}{Arabic} & \multirow{2}{*}{ar} & \multirow{2}{*}{default} & \multirow{2}{*}{\href{https://huggingface.co/datasets/allenai/c4}{mC4}~\cite{mc4}} & \multirow{2}{*}{ODC-BY} & \multirow{2}{*}{165,902} & \multirow{2}{*}{1,686} & 
            dev  & 334 & 42 & 1,133 & 1,119 \\
            \cline{10-14}
             &  &  &  &  &  &  &  &  &
            test & 1,338 & 42 & 4,782 & 4,717 \\
            \cline{3-14}
            % qa web bn
             &  & \multirow{2}{*}{Bengali} & \multirow{2}{*}{bn} & \multirow{2}{*}{default} & \multirow{2}{*}{\href{https://huggingface.co/datasets/allenai/c4}{mC4}~\cite{mc4}} & \multirow{2}{*}{ODC-BY} & \multirow{2}{*}{45,375} & \multirow{2}{*}{2,161} & 
            dev  & 362 & 73 & 1,142 & 1,073 \\
            \cline{10-14}
             &  &  &  &  &  &  &  &  &
            test & 1,451 & 77 & 4,759 & 4,449 \\
            \cline{3-14}
            % qa web de
             &  & \multirow{2}{*}{German} & \multirow{2}{*}{de} & \multirow{2}{*}{default} & \multirow{2}{*}{\href{https://huggingface.co/datasets/allenai/c4}{mC4}~\cite{mc4}} & \multirow{2}{*}{ODC-BY} & \multirow{2}{*}{441,182} & \multirow{2}{*}{1,025} & 
            dev  & 357 & 20 & 1,320 & 1,345 \\
            \cline{10-14}
             &  &  &  &  &  &  &  &  &
            test & 1,432 & 20 & 5,539 & 5,253 \\
            \cline{3-14}
            % qa web es
             &  & \multirow{2}{*}{Spanish} & \multirow{2}{*}{es} & \multirow{2}{*}{default} & \multirow{2}{*}{\href{https://huggingface.co/datasets/allenai/c4}{mC4}~\cite{mc4}} & \multirow{2}{*}{ODC-BY} & \multirow{2}{*}{403,020} & \multirow{2}{*}{912} & 
            dev  & 341 & 23 & 1,317 & 1,281 \\
            \cline{10-14}
             &  &  &  &  &  &  &  &  &
            test & 1,368 & 24 & 5,317 & 5,302 \\
            \cline{3-14}
            % qa web fa
             &  & \multirow{2}{*}{Persian} & \multirow{2}{*}{fa} & \multirow{2}{*}{default} & \multirow{2}{*}{\href{https://huggingface.co/datasets/allenai/c4}{mC4}~\cite{mc4}} & \multirow{2}{*}{ODC-BY} & \multirow{2}{*}{181,463} & \multirow{2}{*}{2,114} & 
            dev  & 338 & 49 & 1,389 & 1,160 \\
            \cline{10-14}
             &  &  &  &  &  &  &  &  &
            test & 1,354 & 47 & 5,532 & 4,839 \\
            \cline{3-14}
            % qa web fr
             &  & \multirow{2}{*}{French} & \multirow{2}{*}{fr} & \multirow{2}{*}{default} & \multirow{2}{*}{\href{https://huggingface.co/datasets/allenai/c4}{mC4}~\cite{mc4}} & \multirow{2}{*}{ODC-BY} & \multirow{2}{*}{387,210} & \multirow{2}{*}{1,076} & 
            dev  & 364 & 20 & 1,444 & 1,451 \\
            \cline{10-14}
             &  &  &  &  &  &  &  &  &
            test & 1,457 & 20 & 5,572 & 5,552 \\
            \cline{3-14}
            % qa web hi
             &  & \multirow{2}{*}{Hindi} & \multirow{2}{*}{hi} & \multirow{2}{*}{default} & \multirow{2}{*}{\href{https://huggingface.co/datasets/allenai/c4}{mC4}~\cite{mc4}} & \multirow{2}{*}{ODC-BY} & \multirow{2}{*}{50,501} & \multirow{2}{*}{2,396} & 
            dev  & 355 & 68 & 1,180 & 1,180 \\
            \cline{10-14}
             &  &  &  &  &  &  &  &  &
            test & 1,423 & 64 & 4,706 & 4,481 \\
            \cline{3-14}
            % qa web id
             &  & \multirow{2}{*}{Indonesian} & \multirow{2}{*}{id} & \multirow{2}{*}{default} & \multirow{2}{*}{\href{https://huggingface.co/datasets/allenai/c4}{mC4}~\cite{mc4}} & \multirow{2}{*}{ODC-BY} & \multirow{2}{*}{245,878} & \multirow{2}{*}{1,059} & 
            dev  & 339 & 23 & 1,395 & 1,295 \\
            \cline{10-14}
             &  &  &  &  &  &  &  &  &
            test & 1,356 & 23 & 5,605 & 5,202 \\
            \cline{3-14}
            % qa web ja
             &  & \multirow{2}{*}{Japanese} & \multirow{2}{*}{ja} & \multirow{2}{*}{default} & \multirow{2}{*}{\href{https://huggingface.co/datasets/allenai/c4}{mC4}~\cite{mc4}} & \multirow{2}{*}{ODC-BY} & \multirow{2}{*}{547,419} & \multirow{2}{*}{1,026} & 
            dev  & 323 & 35 & 1,106 & 1,253 \\
            \cline{10-14}
             &  &  &  &  &  &  &  &  &
            test & 1,293 & 36 & 4,473 & 4,976 \\
            \cline{3-14}
            % qa web ko
             &  & \multirow{2}{*}{Korean} & \multirow{2}{*}{ko} & \multirow{2}{*}{default} & \multirow{2}{*}{\href{https://huggingface.co/datasets/allenai/c4}{mC4}~\cite{mc4}} & \multirow{2}{*}{ODC-BY} & \multirow{2}{*}{250,605} & \multirow{2}{*}{1,137} & 
            dev  & 327 & 34 & 1,156 & 1,083 \\
            \cline{10-14}
             &  &  &  &  &  &  &  &  &
            test & 1,309 & 36 & 4,239 & 4,457 \\
            \cline{3-14}
            % qa web ru
             &  & \multirow{2}{*}{Russian} & \multirow{2}{*}{ru} & \multirow{2}{*}{default} & \multirow{2}{*}{\href{https://huggingface.co/datasets/allenai/c4}{mC4}~\cite{mc4}} & \multirow{2}{*}{ODC-BY} & \multirow{2}{*}{490,581} & \multirow{2}{*}{1,266} & 
            dev  & 324 & 32 & 1,330 & 1,277 \\
            \cline{10-14}
             &  &  &  &  &  &  &  &  &
            test & 1,297 & 33 & 5,096 & 5,152 \\
            \cline{3-14}
            % qa web zh
             &  & \multirow{2}{*}{Chinese} & \multirow{2}{*}{zh} & \multirow{2}{*}{default} & \multirow{2}{*}{\href{https://huggingface.co/datasets/allenai/c4}{mC4}~\cite{mc4}} & \multirow{2}{*}{ODC-BY} & \multirow{2}{*}{956,699} & \multirow{2}{*}{1,208} & 
            dev  & 336 & 30 & 1,230 & 1,366 \\
            \cline{10-14}
             &  &  &  &  &  &  &  &  &
            test & 1,347 & 29 & 5,020 & 5,355 \\
        \bottomrule
        \end{tabular}
    }
    \vspace{-5pt}
    \caption{Statistics of all datasets in \airbench 24.05 (Part 3).}
    \label{tab:air-bench_datasets_2405_3}
\end{sidewaystable*}

% ====================== 24.05 Part 4 ==========================
\begin{sidewaystable*}[t]
    \centering
    \setlength{\extrarowheight}{2pt}
    \resizebox{0.9\textwidth}{!}{
        \begin{tabular}{|c|c|c|c|c|c|c|c|c|c|c|c|c|c|}
        \toprule
            \multirow{2}{*}{\textbf{Task}} & \multirow{2}{*}{\textbf{Domain}} & \multicolumn{2}{c|}{\textbf{Language}} & \multirow{2}{*}{\textbf{Dataset Name}} & \multicolumn{2}{c|}{\textbf{Source of Corpus}} & \multirow{2}{*}{\textbf{\#corpus}} & \multirow{2}{*}{\begin{tabular}[c]{@{}c@{}} \textbf{Avg \#token} \\ \textbf{of corpus} \end{tabular}} & \multirow{2}{*}{\textbf{Split}} & \multirow{2}{*}{\begin{tabular}[c]{@{}c@{}} \textbf{\# of} \\ \textbf{queries} \end{tabular}} & \multirow{2}{*}{\begin{tabular}[c]{@{}c@{}} \textbf{Avg \#token} \\ \textbf{of queries} \end{tabular}} & \multirow{2}{*}{\begin{tabular}[c]{@{}c@{}} \textbf{\# of} \\ \textbf{positives} \end{tabular}} & \multirow{2}{*}{\begin{tabular}[c]{@{}c@{}} \textbf{\# of hard} \\ \textbf{negatives} \end{tabular}} \\
            \cline{3-4} \cline{6-7}
             &  & \textbf{Name} & \textbf{ISO Code} &  & \textbf{Link \& Citation} & \textbf{License} &  &  &  &  &  &  & \\
            \midrule
            \midrule
            
            % qa wiki en
            \multirow{26}{*}{qa} & \multirow{26}{*}{wiki} & \multirow{2}{*}{English} & \multirow{2}{*}{en} & \multirow{2}{*}{default} & \multirow{2}{*}{\href{https://huggingface.co/datasets/NeuML/wikipedia-20240101}{Wikipedia 20240101}} & \multirow{2}{*}{CC BY-SA 3.0, GFDL} & \multirow{2}{*}{1,012,092} & \multirow{2}{*}{665} & 
            dev  & 345 & 18 & 863 & 1,576 \\
            \cline{10-14}
             &  &  &  &  &  &  &  &  &
            test & 1,382 & 17 & 3,397 & 6,306 \\
            \cline{3-14}
            % qa wiki ar
             &  & \multirow{2}{*}{Arabic} & \multirow{2}{*}{ar} & \multirow{2}{*}{default} & \multirow{2}{*}{\href{https://huggingface.co/datasets/wikipedia}{Wikipedia 20240401}} & \multirow{2}{*}{CC BY-SA 3.0, GFDL} & \multirow{2}{*}{1,008,232} & \multirow{2}{*}{787} & 
            dev  & 338 & 40 & 1,112 & 1,438 \\
            \cline{10-14}
             &  &  &  &  &  &  &  &  &
            test & 1,355 & 38 & 4,467 & 5,778 \\
            \cline{3-14}
            % qa wiki bn
             &  & \multirow{2}{*}{Bengali} & \multirow{2}{*}{bn} & \multirow{2}{*}{default} & \multirow{2}{*}{\href{https://huggingface.co/datasets/wikipedia}{Wikipedia 20240401}} & \multirow{2}{*}{CC BY-SA 3.0, GFDL} & \multirow{2}{*}{152,064} & \multirow{2}{*}{2,129} & 
            dev  & 364 & 69 & 1,016 & 1,542 \\
            \cline{10-14}
             &  &  &  &  &  &  &  &  &
            test & 1,456 & 71 & 4,203 & 5,841 \\
            \cline{3-14}
            % qa wiki de
             &  & \multirow{2}{*}{German} & \multirow{2}{*}{de} & \multirow{2}{*}{default} & \multirow{2}{*}{\href{https://huggingface.co/datasets/wikipedia}{Wikipedia 20240401}} & \multirow{2}{*}{CC BY-SA 3.0, GFDL} & \multirow{2}{*}{1,008,186} & \multirow{2}{*}{891} & 
            dev  & 350 & 23 & 817 & 1,481 \\
            \cline{10-14}
             &  &  &  &  &  &  &  &  &
            test & 1,404 & 22 & 3,411 & 5,881 \\
            \cline{3-14}
            % qa wiki es
             &  & \multirow{2}{*}{Spanish} & \multirow{2}{*}{es} & \multirow{2}{*}{default} & \multirow{2}{*}{\href{https://huggingface.co/datasets/wikipedia}{Wikipedia 20240401}} & \multirow{2}{*}{CC BY-SA 3.0, GFDL} & \multirow{2}{*}{1,008,147} & \multirow{2}{*}{801} & 
            dev  & 345 & 22 & 879 & 1,451 \\
            \cline{10-14}
             &  &  &  &  &  &  &  &  &
            test & 1,380 & 23 & 3,531 & 5,767 \\
            \cline{3-14}
            % qa wiki fa
             &  & \multirow{2}{*}{Persian} & \multirow{2}{*}{fa} & \multirow{2}{*}{default} & \multirow{2}{*}{\href{https://huggingface.co/datasets/wikipedia}{Wikipedia 20240401}} & \multirow{2}{*}{CC BY-SA 3.0, GFDL} & \multirow{2}{*}{999,223} & \multirow{2}{*}{627} & 
            dev  & 332 & 41 & 1,179 & 1,444 \\
            \cline{10-14}
             &  &  &  &  &  &  &  &  &
            test & 1,328 & 45 & 4,538 & 5,581 \\
            \cline{3-14}
            % qa wiki fr
             &  & \multirow{2}{*}{French} & \multirow{2}{*}{fr} & \multirow{2}{*}{default} & \multirow{2}{*}{\href{https://huggingface.co/datasets/wikipedia}{Wikipedia 20240401}} & \multirow{2}{*}{CC BY-SA 3.0, GFDL} & \multirow{2}{*}{1,008,270} & \multirow{2}{*}{779} & 
            dev  & 356 & 20 & 768 & 1,496 \\
            \cline{10-14}
             &  &  &  &  &  &  &  &  &
            test & 1,424 & 21 & 3,429 & 5,989 \\
            \cline{3-14}
            % qa wiki hi
             &  & \multirow{2}{*}{Hindi} & \multirow{2}{*}{hi} & \multirow{2}{*}{default} & \multirow{2}{*}{\href{https://huggingface.co/datasets/wikipedia}{Wikipedia 20240401}} & \multirow{2}{*}{CC BY-SA 3.0, GFDL} & \multirow{2}{*}{162,188} & \multirow{2}{*}{997} & 
            dev  & 340 & 59 & 995 & 1,324 \\
            \cline{10-14}
             &  &  &  &  &  &  &  &  &
            test & 1,360 & 59 & 3,911 & 5,225 \\
            \cline{3-14}
            % qa wiki id
             &  & \multirow{2}{*}{Indonesian} & \multirow{2}{*}{id} & \multirow{2}{*}{default} & \multirow{2}{*}{\href{https://huggingface.co/datasets/wikipedia}{Wikipedia 20240401}} & \multirow{2}{*}{CC BY-SA 3.0, GFDL} & \multirow{2}{*}{687,513} & \multirow{2}{*}{451} & 
            dev  & 343 & 22 & 1,003 & 1,365 \\
            \cline{10-14}
             &  &  &  &  &  &  &  &  &
            test & 1,373 & 21 & 4,089 & 5,486 \\
            \cline{3-14}
            % qa wiki ja
             &  & \multirow{2}{*}{Japanese} & \multirow{2}{*}{ja} & \multirow{2}{*}{default} & \multirow{2}{*}{\href{https://huggingface.co/datasets/wikipedia}{Wikipedia 20240401}} & \multirow{2}{*}{CC BY-SA 3.0, GFDL} & \multirow{2}{*}{1,008,365} & \multirow{2}{*}{1,470} & 
            dev  & 358 & 30 & 1,099 & 1,537 \\
            \cline{10-14}
             &  &  &  &  &  &  &  &  &
            test & 1,432 & 32 & 4,303 & 6,101 \\
            \cline{3-14}
            % qa wiki ko
             &  & \multirow{2}{*}{Korean} & \multirow{2}{*}{ko} & \multirow{2}{*}{default} & \multirow{2}{*}{\href{https://huggingface.co/datasets/wikipedia}{Wikipedia 20240401}} & \multirow{2}{*}{CC BY-SA 3.0, GFDL} & \multirow{2}{*}{665,227} & \multirow{2}{*}{775} & 
            dev  & 346 & 39 & 1,109 & 1,423 \\
            \cline{10-14}
             &  &  &  &  &  &  &  &  &
            test & 1,384 & 35 & 4,604 & 5,810 \\
            \cline{3-14}
            % qa wiki ru
             &  & \multirow{2}{*}{Russian} & \multirow{2}{*}{ru} & \multirow{2}{*}{default} & \multirow{2}{*}{\href{https://huggingface.co/datasets/wikipedia}{Wikipedia 20240401}} & \multirow{2}{*}{CC BY-SA 3.0, GFDL} & \multirow{2}{*}{1,008,405} & \multirow{2}{*}{1,211} & 
            dev  & 365 & 30 & 1,154 & 1,505 \\
            \cline{10-14}
             &  &  &  &  &  &  &  &  &
            test & 1,463 & 30 & 4,516 & 6,283 \\
            \cline{3-14}
            % qa wiki zh
             &  & \multirow{2}{*}{Chinese} & \multirow{2}{*}{zh} & \multirow{2}{*}{default} & \multirow{2}{*}{\href{https://huggingface.co/datasets/wikipedia}{Wikipedia 20240401}} & \multirow{2}{*}{CC BY-SA 3.0, GFDL} & \multirow{2}{*}{1,011,604} & \multirow{2}{*}{557} & 
            dev  & 335 & 30 & 952 & 1,301 \\
            \cline{10-14}
             &  &  &  &  &  &  &  &  &
            test & 1,344 & 30 & 3,793 & 5,662 \\
        \bottomrule
        \end{tabular}
    }
    \vspace{-5pt}
    \caption{Statistics of all datasets in \airbench 24.05 (Part 4).}
    \label{tab:air-bench_datasets_2405_4}
\end{sidewaystable*}

% ====================== 24.05 Part 5 ==========================
\begin{sidewaystable*}[t]
    \centering
    \setlength{\extrarowheight}{2pt}
    \resizebox{\textwidth}{!}{
        \begin{tabular}{|c|c|c|c|c|c|c|c|c|c|c|c|c|c|}
        \toprule
            \multirow{2}{*}{\textbf{Task}} & \multirow{2}{*}{\textbf{Domain}} & \multicolumn{2}{c|}{\textbf{Language}} & \multirow{2}{*}{\textbf{Dataset Name}} & \multicolumn{2}{c|}{\textbf{Source of Corpus}} & \multirow{2}{*}{\textbf{\#corpus}} & \multirow{2}{*}{\begin{tabular}[c]{@{}c@{}} \textbf{Avg \#token} \\ \textbf{of corpus} \end{tabular}} & \multirow{2}{*}{\textbf{Split}} & \multirow{2}{*}{\begin{tabular}[c]{@{}c@{}} \textbf{\# of} \\ \textbf{queries} \end{tabular}} & \multirow{2}{*}{\begin{tabular}[c]{@{}c@{}} \textbf{Avg \#token} \\ \textbf{of queries} \end{tabular}} & \multirow{2}{*}{\begin{tabular}[c]{@{}c@{}} \textbf{\# of} \\ \textbf{positives} \end{tabular}} & \multirow{2}{*}{\begin{tabular}[c]{@{}c@{}} \textbf{\# of hard} \\ \textbf{negatives} \end{tabular}} \\
            \cline{3-4} \cline{6-7}
             &  & \textbf{Name} & \textbf{ISO Code} &  & \textbf{Link \& Citation} & \textbf{License} &  &  &  &  &  &  & \\
            \midrule
            \midrule
            
            \multirow{15}{*}{long-doc} & \multirow{4}{*}{arxiv} & \multirow{4}{*}{English} & \multirow{4}{*}{en} & gemini & \href{https://arxiv.org/pdf/2312.11805.pdf}{Paper of Gemini} & CC BY 4.0 & 276 & 136 & 
            test  & 249 & 18 & 249 & 0 \\
            \cline{5-14}
             &  &  &  & gpt3 & \href{https://arxiv.org/pdf/2005.14165.pdf}{Paper of GPT-3} & \begin{tabular}[c]{@{}c@{}} arXiv.org perpetual, \\ non-exclusive license 1.0 \end{tabular} & 515 & 137 & 
            test  & 337 & 16 & 496 & 0 \\
            \cline{5-14}
             &  &  &  & llama2 & \href{https://arxiv.org/pdf/2307.09288.pdf}{Paper of Llama 2} & \begin{tabular}[c]{@{}c@{}} arXiv.org perpetual, \\ non-exclusive license 1.0 \end{tabular} & 566 & 136 & 
            dev  & 326 & 18 & 635 & 0 \\
            \cline{5-14}
             &  &  &  & llm-survey & \href{https://arxiv.org/pdf/2303.18223.pdf}{Survey of LLM} & \begin{tabular}[c]{@{}c@{}} arXiv.org perpetual, \\ non-exclusive license 1.0 \end{tabular} & 1,144 & 135 & 
            test  & 357 & 17 & 924 & 0 \\
            \cline{2-14}
            % long-doc book en
             & \multirow{2}{*}{book} & \multirow{2}{*}{English} & \multirow{2}{*}{en} & a-brief-history-of-time\_stephen-hawking & \href{https://www.docdroid.net/GCLN82v/stephen-hawking-a-brief-history-of-time-pdf}{\textit{A Brief History of Time}} & Unknown & 778 & 127 & 
            dev  & 370 & 16 & 876 & 0 \\
            \cline{5-14}
             &  &  &  & origin-of-species\_darwin & \href{https://www.vliz.be/docs/Zeecijfers/Origin_of_Species.pdf}{\textit{On the Origin of Species}} & Unknown & 1,758 & 126 & 
            test  & 366 & 16 & 1,145 & 0 \\
            \cline{2-14}
            % long-doc healthcare en
             & \multirow{5}{*}{healthcare} & \multirow{5}{*}{English} & \multirow{5}{*}{en} & pubmed\_100K-200K\_1 & \href{https://github.com/armancohan/long-summarization}{long-summarization}~\cite{cohan-etal-2018-discourse} & Apache 2.0 & 899 & 133 & 
            test  & 372 & 20 & 1,008 & 0 \\
            \cline{5-14}
             &  &  &  & pubmed\_100K-200K\_2 & \href{https://github.com/armancohan/long-summarization}{long-summarization}~\cite{cohan-etal-2018-discourse} & Apache 2.0 & 872 & 136 & 
            test  & 355 & 18 & 980 & 0 \\
            \cline{5-14}
             &  &  &  & pubmed\_100K-200K\_3 & \href{https://github.com/armancohan/long-summarization}{long-summarization}~\cite{cohan-etal-2018-discourse} & Apache 2.0 & 873 & 133 & 
            dev  & 357 & 19 & 978 & 0 \\
            \cline{5-14}
             &  &  &  & pubmed\_30K-40K\_10-merged & \href{https://github.com/armancohan/long-summarization}{long-summarization}~\cite{cohan-etal-2018-discourse} & Apache 2.0 & 2,154 & 133 & 
            test  & 368 & 18 & 1,485 & 0 \\
            \cline{5-14}
             &  &  &  & pubmed\_40K-50K\_5-merged & \href{https://github.com/armancohan/long-summarization}{long-summarization}~\cite{cohan-etal-2018-discourse} & Apache 2.0 & 1,731 & 136 & 
            test  & 336 & 21 & 1,046 & 0 \\
            \cline{2-14}
            % long-doc law en
             & \multirow{4}{*}{law} & \multirow{4}{*}{English} & \multirow{4}{*}{en} & lex\_files\_300K-400K & \href{https://huggingface.co/datasets/lexlms/lex_files}{LexFiles}~\cite{chalkidis-etal-2023-lexfiles} & CC BY-NC-SA 4.0 & 2,797 & 137 & 
            dev  & 339 & 15 & 1,307 & 0 \\
            \cline{5-14}
             &  &  &  & lex\_files\_400K-500K & \href{https://huggingface.co/datasets/lexlms/lex_files}{LexFiles}~\cite{chalkidis-etal-2023-lexfiles} & CC BY-NC-SA 4.0 & 3,320 & 137 & 
            test  & 333 & 17 & 1,427 & 0 \\
            \cline{5-14}
             &  &  &  & lex\_files\_500K-600K & \href{https://huggingface.co/datasets/lexlms/lex_files}{LexFiles}~\cite{chalkidis-etal-2023-lexfiles} & CC BY-NC-SA 4.0 & 4,087 & 136 &
            test  & 346 & 17 & 1,324 & 0 \\
            \cline{5-14}
             &  &  &  & lex\_files\_600K-700K & \href{https://huggingface.co/datasets/lexlms/lex_files}{LexFiles}~\cite{chalkidis-etal-2023-lexfiles} & CC BY-NC-SA 4.0 & 5,049 & 138 & 
            test  & 338 & 18 & 1,442 & 0 \\
        \bottomrule
        \end{tabular}
    }
    \vspace{-5pt}
    \caption{Statistics of all datasets in \airbench 24.05 (Part 5).}
    \label{tab:air-bench_datasets_2405_5}
\end{sidewaystable*}

% Detailed Evaluation Results

% English: All Models
% - BM25, BM25+bge-reranker-v2-m3
% - bge-{small, base, large}-en-v1.5, gte-{small, base, large}-en, e5-{small, base, large}-v2
% - SFR-Embedding-Mistral, repllama-v1-7b-lora-passage, bge-en-icl

% Multilingual: All Models
% - BM25, BM25+bge-reranker-v2-m3
% - multilingual-e5-{small, base, large}, gte-Qwen2-{1.5B, 7B}-instruct, bge-m3, multilingual-e5-large-instruct
% - e5-mistral-7b-instruct, bge-multilingual-gemma2
% - gte-multilingual-base, bce-embedding-base_v1

% ================================ English Models ================================

\begin{sidewaystable*}[t]
    \centering
    \setlength{\extrarowheight}{2pt}
    \resizebox{\textwidth}{!}{
        \begin{tabular}{l|c|c|c|c|c|c|c|c|c|c|c|c|c|c|c|c|c}
        \toprule
            \multirow{2}{*}{\textbf{Dataset ($\downarrow$)}} & \multirow{2}{*}{\begin{tabular}[c]{@{}c@{}} bge-en-icl \\ (zero-shot) \end{tabular}} & \multirow{2}{*}{\begin{tabular}[c]{@{}c@{}} bge-en-icl-e5data \\ (zero-shot) \end{tabular}} & \multicolumn{3}{c|}{bge-*-en-v1.5} & \multicolumn{3}{c|}{e5-*-v2} & \multicolumn{3}{c|}{gte-*} & \multicolumn{2}{c|}{NV-Embed-*} & \multirow{2}{*}{\begin{tabular}[c]{@{}c@{}} gte-large- \\ en-v1.5 \end{tabular}} & \multirow{2}{*}{\begin{tabular}[c]{@{}c@{}} SFR- \\ Embedding-Mistral \end{tabular}} & \multirow{2}{*}{\begin{tabular}[c]{@{}c@{}} SFR- \\ Embedding-2\_R \end{tabular}} & \multirow{2}{*}{\begin{tabular}[c]{@{}c@{}} repllama-v1-7b- \\ lora-passage \end{tabular}} \\
            \cline{4-14}
             &  &  & small & base & large & small & base & large & small & base & large & v1 & v2 &  &  &  &  \\
            \midrule
            \midrule
            \multicolumn{18}{l}{\textbf{QA Task (English, 7 Datasets)}} \\
            \midrule
            arxiv\_en & 50.43 & 50.43 & 35.56 & 36.44 & 38.68 & 35.44 & 37.56 & 38.31 & 36.25 & 38.09 & 39.20 & 47.39 & 49.88 & 40.00 & 48.10 & 43.58 & 47.20 \\
            finance\_en & 55.48 & 55.48 & 44.68 & 47.35 & 45.59 & 49.44 & 47.33 & 51.34 & 47.60 & 49.08 & 47.91 & 51.21 & 54.09 & 54.33 & 58.34 & 56.02 & 54.14 \\
            healthcare\_en & 57.20 & 57.20 & 47.23 & 49.50 & 50.93 & 46.48 & 48.93 & 52.45 & 47.48 & 49.20 & 49.02 & 58.98 & 59.85 & 48.34 & 59.09 & 55.83 & 56.25 \\
            law\_en & 28.92 & 28.92 & 17.12 & 16.71 & 23.53 & 15.67 & 16.34 & 21.28 & 15.89 & 15.63 & 17.57 & 20.77 & 25.43 & 18.73 & 22.52 & 21.22 & 17.98 \\
            news\_en & 54.92 & 54.92 & 41.38 & 40.59 & 43.70 & 43.37 & 42.00 & 45.26 & 40.70 & 43.64 & 43.70 & 50.93 & 52.35 & 45.74 & 51.20 & 50.57 & 51.58 \\
            web\_en & 60.80 & 60.80 & 47.04 & 46.77 & 48.92 & 45.27 & 46.12 & 48.60 & 49.55 & 50.41 & 51.47 & 56.46 & 58.76 & 50.02 & 57.44 & 55.57 & 58.06 \\
            wiki\_en & 73.44 & 73.44 & 62.15 & 62.53 & 63.05 & 65.94 & 66.61 & 66.54 & 65.07 & 65.92 & 65.96 & 71.08 & 73.06 & 66.59 & 72.78 & 72.77 & 70.84 \\
            \midrule
            \multicolumn{18}{l}{\textbf{Long-Doc Task (English, 11 Datasets)}} \\
            \midrule
            arxiv\_en\_gemini & 87.55 & 87.55 & 69.08 & 69.88 & 75.50 & 75.50 & 72.29 & 76.31 & 72.69 & 71.89 & 74.70 & 83.94 & 85.54 & 72.29 & 79.92 & 78.72 & 82.73 \\
            arxiv\_en\_gpt3 & 81.65 & 81.65 & 65.23 & 64.91 & 64.37 & 65.83 & 69.44 & 75.07 & 66.79 & 67.41 & 70.62 & 83.31 & 83.28 & 68.30 & 76.88 & 72.28 & 76.81 \\
            arxiv\_en\_llm\_survey & 71.05 & 71.05 & 46.25 & 49.80 & 51.32 & 52.71 & 54.30 & 58.07 & 49.54 & 55.85 & 56.58 & 67.62 & 69.33 & 55.94 & 63.51 & 60.15 & 64.71 \\
            book\_en\_origin\_of\_species\_darwin & 71.07 & 71.07 & 52.50 & 56.09 & 60.22 & 53.03 & 54.07 & 61.51 & 58.37 & 59.76 & 63.32 & 68.21 & 70.77 & 51.35 & 66.18 & 62.49 & 66.64 \\
            healthcare\_en\_pubmed\_100k\_200k\_1 & 71.23 & 71.23 & 57.24 & 60.50 & 61.68 & 58.76 & 62.33 & 64.62 & 56.94 & 57.59 & 58.76 & 68.51 & 69.17 & 57.78 & 63.77 & 65.13 & 68.17 \\
            healthcare\_en\_pubmed\_100k\_200k\_2 & 75.59 & 75.59 & 57.23 & 59.39 & 62.54 & 62.14 & 66.00 & 69.10 & 58.30 & 61.70 & 64.49 & 75.53 & 77.96 & 63.91 & 73.15 & 70.78 & 69.78 \\
            healthcare\_en\_pubmed\_30k\_40k\_10\_merged & 79.00 & 79.00 & 65.76 & 67.33 & 69.36 & 68.32 & 69.74 & 74.16 & 67.03 & 68.50 & 68.34 & 78.45 & 79.98 & 64.79 & 74.45 & 71.97 & 77.24 \\
            healthcare\_en\_pubmed\_40k\_50k\_5\_merged & 65.57 & 65.57 & 52.03 & 57.15 & 57.28 & 57.94 & 57.83 & 61.35 & 50.79 & 54.53 & 53.57 & 64.89 & 64.57 & 55.14 & 61.95 & 61.19 & 60.55 \\
            law\_en\_lex\_files\_400k\_500k & 67.73 & 67.73 & 54.46 & 58.07 & 60.25 & 53.20 & 58.06 & 62.17 & 55.56 & 59.03 & 62.66 & 66.54 & 66.06 & 58.80 & 61.76 & 60.35 & 64.77 \\
            law\_en\_lex\_files\_500k\_600k & 68.52 & 68.52 & 53.03 & 60.23 & 61.68 & 57.98 & 60.08 & 65.89 & 58.75 & 61.63 & 63.54 & 68.14 & 71.56 & 63.36 & 66.31 & 63.16 & 67.55 \\
            law\_en\_lex\_files\_600k\_700k & 68.73 & 68.73 & 49.98 & 53.58 & 56.30 & 51.79 & 57.61 & 59.56 & 53.84 & 55.29 & 56.53 & 67.76 & 69.68 & 56.18 & 61.18 & 57.90 & 63.97 \\
        \bottomrule
        \end{tabular}
    }
    \vspace{-5pt}
    \caption{Detailed evaluation results of English IR models on QA (English, test) datasets and Long-Doc (English, test) datasets of \airbench 24.05.}
    \label{tab:detailed_en_results}
\end{sidewaystable*}

% ================================ Multilingual Models ================================

\begin{table*}[t]
    \centering
    \setlength{\extrarowheight}{2pt}
    \resizebox{\linewidth}{!}{
        \begin{tabular}{l|c|c|c|c|c|c|c|c|c|c|c|c|c|c}
        \toprule
            \multirow{2}{*}{\textbf{Dataset ($\downarrow$)}} & \multirow{2}{*}{BM25} & \multirow{2}{*}{\begin{tabular}[c]{@{}c@{}} BM25 + \\ bge-reranker-v2-m3  \end{tabular}} & \multicolumn{3}{c|}{multilingual-e5-*} & \multicolumn{2}{c|}{gte-Qwen2-*-instruct} & \multirow{2}{*}{bge-m3} & \multirow{2}{*}{\begin{tabular}[c]{@{}c@{}} multilingual-e5- \\ large-instruct \end{tabular}} & \multirow{2}{*}{\begin{tabular}[c]{@{}c@{}} e5-mistral- \\ 7b-instruct \end{tabular}} & \multirow{2}{*}{\begin{tabular}[c]{@{}c@{}} bge-multilingual- \\ gemma2 \end{tabular}} & \multirow{2}{*}{\begin{tabular}[c]{@{}c@{}} gte-multilingual- \\ base \end{tabular}} & \multirow{2}{*}{\begin{tabular}[c]{@{}c@{}} bce-embedding- \\ base\_v1 \end{tabular}} & \multirow{2}{*}{\begin{tabular}[c]{@{}c@{}} jina- \\ embeddings-v3 \end{tabular}} \\
            \cline{4-8}
             &  &  & small & base & large & 1.5B & 7B &  &  &  &  &  &  & \\
            \midrule
            \midrule
            \multicolumn{15}{l}{\textbf{QA Task (Multilingual, 53 Datasets)}} \\
            \midrule
            arxiv\_en & 33.18 & 50.30 & 32.98 & 34.17 & 37.84 & 42.15 & 41.33 & 41.64 & 39.52 & 46.06 & 24.00 & 41.28 & 22.60 & 39.65 \\
            finance\_ar & 35.78 & 51.78 & 36.17 & 43.82 & 45.34 & 44.12 & 43.55 & 45.76 & 48.95 & 44.59 & 50.25 & 45.59 & 25.00 & 46.32 \\
            finance\_en & 45.13 & 58.04 & 47.32 & 50.29 & 49.05 & 55.21 & 59.23 & 52.92 & 52.79 & 55.90 & 50.08 & 53.24 & 41.67 & 51.70 \\
            finance\_fr & 27.63 & 52.08 & 25.90 & 33.83 & 36.41 & 36.52 & 39.57 & 41.44 & 42.73 & 38.98 & 51.10 & 35.47 & 19.27 & 37.14 \\
            finance\_zh & 22.43 & 42.35 & 30.46 & 32.07 & 34.71 & 34.28 & 34.61 & 40.23 & 37.72 & 33.10 & 39.23 & 36.84 & 25.72 & 33.96 \\
            healthcare\_de & 50.02 & 63.43 & 43.90 & 47.34 & 46.14 & 46.34 & 53.91 & 49.00 & 52.06 & 53.12 & 55.40 & 50.68 & 25.55 & 49.86 \\
            healthcare\_en & 34.84 & 53.76 & 44.21 & 49.16 & 50.63 & 52.11 & 54.46 & 49.05 & 54.02 & 56.24 & 47.48 & 47.48 & 29.89 & 49.42 \\
            healthcare\_es & 31.25 & 50.85 & 45.67 & 50.30 & 54.91 & 49.49 & 53.78 & 53.05 & 51.74 & 47.67 & 63.13 & 46.35 & 29.90 & 52.75 \\
            healthcare\_fr & 28.02 & 50.99 & 19.75 & 28.53 & 32.40 & 33.86 & 30.29 & 39.29 & 36.64 & 37.28 & 45.13 & 34.92 & 6.39 & 32.68 \\
            healthcare\_zh & 18.10 & 43.58 & 28.97 & 28.08 & 33.62 & 39.13 & 38.66 & 42.31 & 39.76 & 36.05 & 42.35 & 37.94 & 25.84 & 38.91 \\
            law\_de & 12.33 & 22.95 & 11.93 & 13.35 & 13.56 & 12.81 & 13.18 & 20.11 & 15.59 & 14.77 & 15.75 & 12.65 & 5.72 & 11.71 \\
            law\_en & 19.50 & 34.17 & 14.61 & 15.76 & 19.71 & 20.19 & 22.75 & 26.95 & 16.90 & 19.61 & 22.60 & 11.44 & 8.67 & 16.78 \\
            law\_fr & 13.16 & 23.19 & 7.34 & 10.30 & 9.94 & 12.72 & 13.15 & 20.20 & 15.12 & 14.38 & 14.29 & 11.68 & 2.64 & 9.76 \\
            news\_ar & 26.54 & 50.17 & 32.16 & 37.49 & 40.64 & 35.93 & 37.63 & 44.93 & 48.20 & 38.95 & 48.41 & 39.13 & 13.43 & 44.04 \\
            news\_bn & 29.33 & 41.60 & 44.97 & 46.48 & 52.17 & 20.27 & 61.31 & 59.03 & 49.31 & 25.50 & 58.77 & 56.00 & 17.90 & 53.73 \\
            news\_de & 38.52 & 55.11 & 39.06 & 43.70 & 43.34 & 43.08 & 44.89 & 47.87 & 47.84 & 46.48 & 52.05 & 43.93 & 21.15 & 46.39 \\
            news\_en & 39.72 & 57.63 & 38.70 & 43.05 & 43.48 & 47.44 & 52.74 & 47.34 & 44.27 & 47.89 & 50.29 & 47.55 & 30.74 & 45.61 \\
            news\_es & 33.09 & 54.65 & 36.14 & 38.88 & 40.41 & 39.90 & 45.21 & 44.70 & 45.99 & 45.34 & 49.90 & 40.47 & 19.76 & 42.94 \\
            news\_fa & 24.95 & 52.02 & 33.07 & 36.70 & 40.03 & 26.40 & 30.09 & 43.81 & 45.59 & 29.72 & 43.40 & 39.05 & 15.79 & 37.90 \\
            news\_fr & 41.20 & 60.79 & 28.56 & 40.51 & 36.59 & 45.60 & 49.76 & 49.52 & 50.59 & 49.61 & 56.80 & 45.86 & 22.75 & 46.56 \\
            news\_hi & 31.93 & 54.95 & 32.96 & 32.85 & 39.12 & 23.39 & 30.28 & 42.12 & 39.66 & 29.82 & 44.89 & 36.64 & 14.02 & 40.02 \\
            news\_id & 42.82 & 66.52 & 35.87 & 41.26 & 41.03 & 34.82 & 46.44 & 47.45 & 48.59 & 45.93 & 50.65 & 41.27 & 19.20 & 44.86 \\
            news\_ja & 38.12 & 57.95 & 37.42 & 39.06 & 45.24 & 41.95 & 44.13 & 47.09 & 47.60 & 43.47 & 51.51 & 42.62 & 21.44 & 41.96 \\
            news\_ko & 34.79 & 59.22 & 40.05 & 43.16 & 47.79 & 44.55 & 47.19 & 48.13 & 50.52 & 46.47 & 51.64 & 40.39 & 20.70 & 45.18 \\
            news\_ru & 31.67 & 55.72 & 37.90 & 42.06 & 43.24 & 43.09 & 46.55 & 48.31 & 48.81 & 46.59 & 51.48 & 42.93 & 20.56 & 46.65 \\
            news\_zh & 15.22 & 30.61 & 27.34 & 36.24 & 39.67 & 36.43 & 43.17 & 41.00 & 35.46 & 35.98 & 43.42 & 36.20 & 27.56 & 40.56 \\
            science\_ru & 39.78 & 62.84 & 43.70 & 47.01 & 51.87 & 54.04 & 45.21 & 55.18 & 56.86 & 53.07 & 44.13 & 48.69 & 26.06 & 50.24 \\
            web\_ar & 39.15 & 60.85 & 41.15 & 46.78 & 47.74 & 48.85 & 55.56 & 52.53 & 56.40 & 49.56 & 59.97 & 47.12 & 18.89 & 53.40 \\
            web\_bn & 47.47 & 68.73 & 44.65 & 46.71 & 51.10 & 38.37 & 51.45 & 55.53 & 56.17 & 46.83 & 59.68 & 50.89 & 25.03 & 55.55 \\
            web\_de & 46.14 & 61.30 & 45.06 & 45.90 & 46.89 & 47.73 & 48.62 & 51.89 & 50.87 & 50.88 & 57.72 & 47.22 & 26.31 & 48.06 \\
            web\_en & 41.46 & 60.51 & 38.71 & 43.24 & 42.81 & 52.68 & 58.99 & 53.88 & 41.58 & 52.08 & 56.48 & 52.05 & 30.55 & 47.38 \\
            web\_es & 42.52 & 60.89 & 42.57 & 46.04 & 46.44 & 50.69 & 54.11 & 51.78 & 52.24 & 54.45 & 58.20 & 49.56 & 26.77 & 49.42 \\
            web\_fa & 42.61 & 64.98 & 45.91 & 48.44 & 50.45 & 41.71 & 49.55 & 55.81 & 58.68 & 45.86 & 62.43 & 49.70 & 21.83 & 52.84 \\
            web\_fr & 46.62 & 63.48 & 30.61 & 43.13 & 39.56 & 51.60 & 55.16 & 51.46 & 50.20 & 54.52 & 59.54 & 50.34 & 26.94 & 48.80 \\
            web\_hi & 50.70 & 71.06 & 50.53 & 51.50 & 56.44 & 40.53 & 53.06 & 57.06 & 56.32 & 49.43 & 64.50 & 56.30 & 25.22 & 58.79 \\
            web\_id & 48.80 & 67.23 & 39.52 & 46.37 & 44.80 & 48.32 & 55.51 & 53.14 & 54.49 & 55.17 & 60.00 & 50.50 & 21.02 & 52.76 \\
            web\_ja & 47.41 & 64.53 & 45.49 & 47.36 & 52.21 & 52.21 & 57.27 & 54.75 & 54.89 & 51.80 & 60.26 & 51.18 & 27.65 & 50.10 \\
            web\_ko & 44.73 & 61.51 & 45.07 & 46.53 & 53.59 & 52.48 & 57.54 & 55.28 & 55.81 & 54.22 & 59.64 & 47.41 & 23.53 & 51.87 \\
            web\_ru & 42.92 & 63.59 & 42.85 & 47.59 & 48.51 & 52.35 & 55.88 & 54.53 & 54.97 & 53.85 & 60.12 & 49.77 & 28.24 & 50.51 \\
            web\_zh & 33.69 & 52.96 & 42.14 & 44.27 & 48.17 & 47.48 & 51.66 & 50.20 & 47.06 & 45.68 & 53.04 & 46.75 & 35.66 & 47.66 \\
            wiki\_ar & 43.66 & 63.82 & 50.61 & 54.35 & 60.65 & 47.74 & 59.44 & 59.65 & 63.21 & 52.98 & 63.42 & 54.40 & 19.38 & 57.89 \\
            wiki\_bn & 55.80 & 72.97 & 53.57 & 53.13 & 60.33 & 51.35 & 58.17 & 64.33 & 64.45 & 56.84 & 69.48 & 58.12 & 25.81 & 62.81 \\
            wiki\_de & 61.20 & 73.32 & 56.58 & 57.89 & 59.70 & 56.30 & 63.97 & 64.68 & 65.81 & 65.40 & 67.91 & 62.83 & 30.17 & 62.08 \\
            wiki\_en & 60.27 & 75.46 & 61.89 & 62.78 & 63.85 & 66.45 & 73.59 & 69.70 & 68.62 & 71.38 & 72.80 & 69.12 & 30.97 & 64.96 \\
            wiki\_es & 57.24 & 73.70 & 59.53 & 59.41 & 61.61 & 60.94 & 67.62 & 65.40 & 68.10 & 69.49 & 71.79 & 63.42 & 34.99 & 63.65 \\
            wiki\_fa & 48.02 & 67.43 & 54.07 & 56.47 & 59.69 & 44.29 & 57.05 & 61.15 & 64.20 & 51.77 & 67.57 & 53.24 & 27.63 & 57.75 \\
            wiki\_fr & 62.71 & 76.51 & 50.94 & 59.04 & 60.71 & 61.90 & 70.32 & 66.04 & 69.72 & 69.29 & 71.28 & 66.69 & 33.14 & 64.67 \\
            wiki\_hi & 57.81 & 74.76 & 62.73 & 63.59 & 68.59 & 51.57 & 60.54 & 69.02 & 71.81 & 63.93 & 75.39 & 67.62 & 32.02 & 68.74 \\
            wiki\_id & 58.14 & 75.16 & 59.00 & 60.95 & 61.82 & 54.47 & 61.81 & 66.30 & 66.36 & 66.23 & 68.91 & 62.79 & 25.92 & 62.75 \\
            wiki\_ja & 56.43 & 72.90 & 54.32 & 51.31 & 61.07 & 55.97 & 62.88 & 60.86 & 64.12 & 57.72 & 68.29 & 57.62 & 20.26 & 58.26 \\
            wiki\_ko & 43.93 & 67.17 & 55.75 & 56.26 & 62.64 & 54.89 & 59.17 & 62.36 & 64.79 & 60.30 & 66.78 & 55.63 & 20.96 & 58.28 \\
            wiki\_ru & 53.99 & 68.60 & 53.80 & 52.96 & 58.16 & 53.45 & 62.95 & 60.18 & 62.57 & 58.70 & 64.15 & 57.03 & 28.08 & 59.41 \\
            wiki\_zh & 40.24 & 63.51 & 53.63 & 59.44 & 61.83 & 58.33 & 67.50 & 63.52 & 62.82 & 57.19 & 68.64 & 61.86 & 35.46 & 62.70 \\
            \midrule
            \multicolumn{15}{l}{\textbf{Long-Doc Task (English, 11 Datasets)}} \\
            \midrule
            arxiv\_en\_gemini & 63.85 & 82.33 & 75.10 & 74.70 & 76.71 & 75.10 & 73.09 & 82.33 & 76.71 & 77.51 & 81.53 & 74.70 & 71.89 & 72.69 \\
            arxiv\_en\_gpt3 & 56.13 & 74.56 & 67.21 & 70.23 & 73.71 & 73.39 & 71.61 & 71.93 & 69.12 & 76.85 & 75.12 & 72.13 & 69.98 & 71.39 \\
            arxiv\_en\_llm\_survey & 47.76 & 68.77 & 54.11 & 58.05 & 60.29 & 53.63 & 50.33 & 61.25 & 58.87 & 62.28 & 59.65 & 57.89 & 52.85 & 55.96 \\
            book\_en\_origin\_of\_species\_darwin & 42.07 & 65.42 & 50.12 & 55.93 & 59.39 & 63.02 & 58.39 & 59.41 & 61.94 & 64.50 & 67.08 & 57.78 & 48.53 & 62.20 \\
            healthcare\_en\_pubmed\_100k\_200k\_1 & 60.21 & 78.17 & 58.56 & 63.94 & 63.47 & 60.54 & 62.06 & 65.64 & 62.97 & 64.40 & 71.48 & 64.40 & 48.08 & 58.33 \\
            healthcare\_en\_pubmed\_100k\_200k\_2 & 61.78 & 82.29 & 59.41 & 61.79 & 63.94 & 66.71 & 69.05 & 67.31 & 64.42 & 71.40 & 79.21 & 68.32 & 50.88 & 57.63 \\
            healthcare\_en\_pubmed\_30k\_40k\_10\_merged & 65.45 & 84.12 & 66.44 & 70.90 & 70.36 & 70.47 & 70.75 & 70.98 & 72.13 & 74.65 & 79.78 & 73.03 & 58.36 & 67.06 \\
            healthcare\_en\_pubmed\_40k\_50k\_5\_merged & 53.90 & 72.44 & 55.84 & 57.31 & 60.13 & 56.65 & 59.22 & 56.45 & 59.07 & 62.91 & 66.72 & 60.10 & 45.09 & 53.67 \\
            law\_en\_lex\_files\_400k\_500k & 42.75 & 68.51 & 51.85 & 58.70 & 64.56 & 58.70 & 61.84 & 63.59 & 59.82 & 58.61 & 66.14 & 59.56 & 47.08 & 60.64 \\
            law\_en\_lex\_files\_500k\_600k & 42.99 & 67.93 & 56.03 & 61.41 & 67.72 & 60.78 & 64.33 & 64.12 & 60.90 & 62.73 & 69.83 & 64.56 & 49.92 & 60.82 \\
            law\_en\_lex\_files\_600k\_700k & 47.12 & 71.70 & 52.92 & 57.33 & 62.50 & 58.61 & 63.04 & 60.60 & 57.67 & 60.15 & 69.03 & 60.17 & 42.74 & 56.06 \\
        \bottomrule
        \end{tabular}
    }
    \vspace{-5pt}
    \caption{Detailed evaluation results of multilingual IR models on QA (Multilingual, test) datasets and Long-Doc (English, test) datasets of \airbench 24.05.}
    \label{tab:detailed_multilingual_results}
\end{table*}

% \begin{tabular}[c]{@{}c@{}}  \\  \end{tabular}

\end{document}